\documentclass[twocolumn,10pt,amsmath,amssymb,aps,pra,superscriptaddress,nofootinbib]{revtex4-2}
\usepackage{graphicx}
\usepackage{dcolumn}
\usepackage{bm}

\usepackage{siunitx}
\usepackage[T1]{fontenc}
\usepackage{multirow}

\usepackage{titlesec}

\begin{document}

\title{Automated All-RF Tuning for Spin Qubit Readout and Control}

\author{Cornelius Carlsson}
\thanks{These authors contributed equally to this work}
\affiliation{Department of Engineering Science, University of Oxford, Oxford OX1 3PJ, United Kingdom}
\email[]{cornelius.carlsson@spc.ox.ac.uk}

\author{Jaime Saez-Mollejo}
\thanks{These authors contributed equally to this work}
\affiliation{Institute of Science and Technology Austria, Am Campus 1, 3400 Klosterneuburg, Austria}
\email[]{jaime.saezmollejo@ist.ac.at}

\author{Federico Fedele}
\affiliation{Department of Engineering Science, University of Oxford, Oxford OX1 3PJ, United Kingdom}

\author{Stefano Calcaterra}
\affiliation{L-NESS, Physics Department, Politecnico di Milano, Como, Italy}

\author{Daniel Chrastina}
\affiliation{L-NESS, Physics Department, Politecnico di Milano, Como, Italy}

\author{Giovanni Isella}
\affiliation{L-NESS, Physics Department, Politecnico di Milano, Como, Italy}

\author{Georgios Katsaros}
\affiliation{Institute of Science and Technology Austria, Am Campus 1, 3400 Klosterneuburg, Austria}

\author{Natalia Ares}
\email[]{natalia.ares@eng.ox.ac.uk}
\affiliation{Department of Engineering Science, University of Oxford, Oxford OX1 3PJ, United Kingdom}

\date{\today}

\begin{abstract}

Efficient tuning of spin qubits remains a major bottleneck in scaling semiconductor quantum dot-based quantum processors. A key challenge is the rapid identification of gate voltage regimes suitable for qubit initialisation, control, and readout. Here, we leverage radio-frequency charge sensing to automate spin qubit tuning, achieving a median tuning time of approximately 15 minutes. In a single continuous run, our routine identifies spin qubits at 12 distinct charge transitions in under 17 hours. Beyond tuning, our routine autonomously acquires data revealing the gate-voltage dependence of the exchange interaction, dephasing time, and quality factor -- quantities that vary substantially between charge configurations. These results represent a step change in high-throughput spin qubit tuning and provide a foundation for a systematic and automated exploration of semiconductor quantum circuits.

\end{abstract}

\maketitle

\titleformat{\section}[block]
  {\normalfont\small\bfseries\centering}
  {\thesection.}
  {0.75em}
  {\uppercase}

\section{\label{sec:intro}Introduction}
Spins in gate-defined semiconductor quantum dots are strong contenders for realising dense qubit architectures and for running high-fidelity quantum circuits \cite{zwerver2022qubits, neyens2024probing, george202412, noiri2022fast, xue2022quantum, lawrie2023simultaneous, wang2024operating, huang2024high}.
However, reaching an operating condition for spin qubits entails a careful and often tedious calibration of gate voltages -- a task quickly becoming manually intractable as devices scale.

To address this, machine learning methods have emerged as valuable tuning agents, relieving the need for heuristic measurements \cite{kalantre2019machine, zwolak2020autotuning, durrer2020automated, zwolak2021ray, ziegler2023tuning, alexeev2024artificial} and providing significant speed-ups over human experts \cite{moon2020machine, severin2024, schuff2024fully}. With the decision-making process delegated to machines, the primary bottleneck becomes the time spent acquiring data. Approaches have been developed to reduce data requirements \cite{lennon2019efficiently, nguyen2021deep}, yet the potential time savings offered by employing high-bandwidth measurement techniques, such as radio-frequency (rf) reflectometry, are considerably greater \cite{vigneau2023probing}. Only recently have autonomous tuning methods been extended to the rf domain, with efforts focused on quantum dot tuning \cite{van2022all}, gate virtualisation \cite{hickie2024, rao2024mavis} and qubit optimisation \cite{katiraee2025unified, berritta2024real}. This leaves the critical task of autonomously tuning quantum dots into spin qubits, using rf reflectometry, yet to be implemented.

\begin{figure*}[t]
\includegraphics[width=\linewidth]{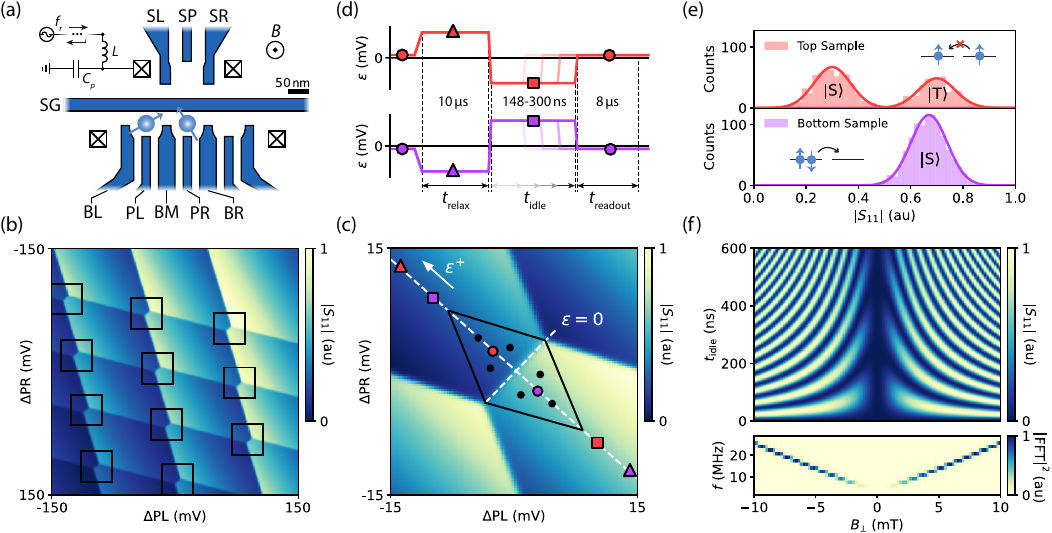}
\caption{\label{fig:fig_1}\textbf{Device layout and algorithm flow on simulated data.} \textbf{(a)} A hole double quantum dot is confined by applying positive voltages on plunger gates PL and PR, and barrier gates BL, BM and BR. A proximal sensing dot, separated by a splitter gate SG, is confined by SP, SL and SR, and couples to a readout tank circuit via its left ohmic with $L=470$\! nH and $f_r=249.5$\! MHz. Details on the Ge/SiGe heterostructure and fabrication can be found in \cite{saez-mollejo_exchange_2025}. \textbf{(b)} Charge stability diagram with automatically located interdot charge transitions (IDTs) outlined in black. Visiting each IDT is done in an autonomous loop, verifying the presence, or absence, of qubits at each. \textbf{(c)} Narrow-range voltage scan centred about an IDT with automatically extracted potential meta-stable regions (black lines), within which pairs of readout voltage points are generated (circular markers). An example pair is shown in red/purple, along with additional pulse points depicted by square and triangular markers. \textbf{(d)} Pulse sequences for identifying Pauli spin blockade (PSB) colour-coded by pulsing direction -- red for top samples, purple for bottom samples. Markers and the detuning axis ($\varepsilon$) match those in c), with marker shapes indicating the role of each voltage pulse point: triangles for relaxation periods ($t_{\mathrm{relax}}$) where the spin system returns to its ground state; squares for idling periods ($t_{\mathrm{idle}}$) where spin rotations can occur; circles for readout periods ($t_{\mathrm{readout}}$) where the rf tone is applied and the reflected signal is measured. \textbf{(e)} Example histograms of readout signals following the scheme in (d) with a clear PSB signature due to a triplet signal emerging for only a single pulsing direction (top). \textbf{(f)} Dependence of singlet-triplet oscillations on the sample's out-of-plane magnetic field, accompanied by a Fourier transform along $t_{\mathrm{idle}}$. All simulations emulate the response in amplitude of the reflected rf signal ($S_{11}$).}
\end{figure*}

In this work, we leverage machine learning to develop a first-ever automated routine for tuning spin qubits using rf charge sensing. The backbone of our routine is built on neural networks that enable navigation through two-dimensional voltage spaces, gate virtualisation, extraction of transition line features, and setting of voltage pulse points. The versatility of these networks opens the path to large-scale qubit characterisation and $in$-$situ$ engineering of relevant qubit parameters, going beyond an initial qubit tune-up. We showcase this in real-time on a depletion-mode Ge/SiGe planar heterostructure device, measuring singlet-triplet oscillations autonomously at 12 different charge occupations, in addition to acquiring measurements for characterising the gate-voltage tunability of the exchange interaction, dephasing time, and Q factor. Starting from a double quantum dot charge stability diagram, all measurements presented in this paper are launched automatically, with no active user input.

\section{\label{sec:tuning_method}Tuning Method}
Central to the tuning of spin qubits is the identification of a regime that exhibits Pauli spin blockade (PSB), a spin-to-charge conversion mechanism commonly used for qubit readout. However, unknown charge occupations, low signal-to-noise ratios, and complex blockade lifting mechanisms \cite{lundberg2024non} can all obscure PSB signatures. This is further complicated in systems with a strong spin-orbit interaction, where site-dependent $g$-tensors and the orientation of quantisation axes can reduce readout visibility \cite{sen2023classification, mutter2020g, hendrickx2024sweet, jin2024probing, saez-mollejo_exchange_2025}. Fortunately, automated rf measurements enable these unfavourable regimes to be discarded quickly, allowing the PSB mechanism to be exploited effectively for spin qubit tuning. This approach overcomes the longer measurement times and reliance on ohmic reservoir couplings typical of automated PSB searches that use charge transport measurements \cite{schuff2023identifying}.

Our routine follows a well-established sequence of steps. A schematic of our device is shown in Fig.~\ref{fig:fig_1}a, and a 2.5 \unit{\milli\tesla} magnetic field is applied in the sample's out-of-plane direction throughout the routine, unless otherwise stated. The routine begins with a wide voltage range charge stability diagram (CSD), obtained by sweeping the two plunger gates, PL and PR. A navigation stage (Sec.~\ref{sec:navigation}) automatically detects each interdot charge transition (IDT) in the CSD, as shown in Fig.~\ref{fig:fig_1}b. Next, each IDT is visited and the search for qubits is streamlined by iterating through different barrier gate voltage configurations in an autonomous tuning loop. 

When pulsing across an effective (1,1)--(2,0) interdot charge transition of a double quantum dot, PSB is manifested by the extension of a meta-stable (1,1) spin triplet state into the (2,0) region of charge stability. In the above, $(n_L, n_R)$ refers to the charge occupation on the left and right dot, respectively. At each barrier voltage iteration, an image segmentation stage (Sec.~\ref{sec:segmentation}) creates bounds for the two potential meta-stable regions, as shown in Fig.~\ref{fig:fig_1}c. For the general case where the singlet-triplet splitting in an effective (2,0) charge region is larger than all other relevant energy scales, these regions are triangular \cite{petta2005pulsed}. Readout voltage points are sampled in each triangle, with points enclosed by the top-left triangle denoted as `top' samples and those enclosed by the bottom-right triangle denoted as `bottom' samples. Top and bottom samples located parallel to the detuning axis ($\varepsilon$), and that lay equidistant from the interdot line ($\varepsilon = 0$), constitute one sample pair. Each sample pair participates in the pulsing and readout scheme shown in Fig.~\ref{fig:fig_1}d. Alternating pulsing directions across the IDT enables our routine to detect PSB independently of the charge occupation parity. An evaluation stage (Sec. \ref{sec:evaluation}) scores the presence of PSB for each sample pair based on the prominence of a blocked, single-shot, readout signal. A prototypical example of a high-scoring iteration for the case where a top sample shows PSB is provided in Fig.~\ref{fig:fig_1}e. If PSB exists, our routine confirms the presence of a qubit by probing singlet-triplet oscillations. A modulation of the oscillation frequency by the magnetic field, as shown in Fig.~\ref{fig:fig_1}f, verifies singlet-triplet dynamics. If a maximum number of iterations is reached and oscillations are still absent, a new IDT is explored until all IDTs have been visited. In the following sections, we provide detail on each of these steps in our routine. A flow chart is provided in the Supplementary Materials.

\section{Automation Modules}
\subsection{\label{sec:navigation}Navigation}
Computer vision techniques, including filtering, gradient extraction, thresholding, and the Hough transform, are standard tools for line detection in images. However, when applied to CSDs \cite{lapointe2020algorithm, mills2019computer}, these methods fail if transition lines become obscured by noise or charge latching. To overcome this, we train a convolutional neural network (CNN) for detecting IDTs, called the `Interdot CNN', using $5 \times 10^5$ simulated CSD images. These are generated using the GPU-accelerated constant-capacitance model solver QArray \cite{van2024qarray, van2024qarraycodebase}, and incorporate realistic levels of white noise, telegraphic noise, thermal broadening, and latching. Network architectures and performance benchmarks can be found in the Supplementary Material.

\begin{figure}[b]
\includegraphics[width=\linewidth]{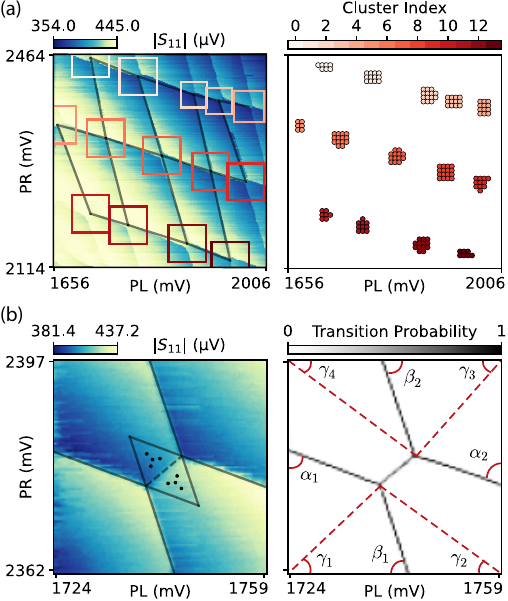}
\caption{\label{fig:fig_2}\textbf{Navigation and image segmentation stages of our routine applied to example experimental data}. \textbf{(a)} Wide voltage range charge stability diagram with automatically placed interdot windows and connections (left). Sliding the `Interdot CNN' yields positive classification clusters (right), whose centroids mark interdot charge transition locations. \textbf{(b)} Narrow range voltage scan of an interdot charge transition with an automatic reconstruction of transition lines, potential meta-stable triangles, and placement of readout voltage points (left). The output of the `Line CNN' gives pixel-wise probabilities of transitions (right). These probabilities are passed onto the `Angle CNN', which returns eight angles used to parametrise the transition lines.}
\end{figure}

A sliding window of $20 \times 20$ pixels feeds patches of the CSD to the `Interdot CNN', which assigns binary classifications to patch centres: `1' if an IDT is detected (circles in Fig.~\ref{fig:fig_2}a right) and `0' otherwise. Our patch resolution of $\sim$1\! mV per pixel is large enough to fully enclose IDTs for typical CSD scan ranges and resolutions \cite{liu2022automated}, yet small enough to become blind to CSD features not captured by the training data. These include transition line curvature, caused by tunnel coupling, and non-uniform transition line periodicities, caused by shell filling \cite{van2021two}, spurious dots, and/or disorder potentials \cite{craig2024bridging}.

Positive classifications made by the `Interdot CNN' surround true IDT locations in the voltage space spanned by PL and PR, as illustrated in Fig.~\ref{fig:fig_2}a. A $k$-means algorithm \cite{scikit-learn} is used to cluster these positive classifications, and IDT voltage locations are defined by the cluster centroids. The boundaries of charge stability regions are approximated using the location of each IDT's nearest neighbours, producing the connections shown in black in Fig.~\ref{fig:fig_2}a left. Using this spatial information, IDTs are indexed and assigned unique voltage ranges and detuning amplitudes for their respective image segmentation scans (next section) and pulsed measurements (Sec.\! \ref{subsec:measurement}). Further details on these sub-routines are provided in the Supplementary Materials.

\subsection{\label{sec:segmentation}Image Segmentation}
To define potential meta-stable regions, all dot-reservoir transition lines around an IDT must be specified. To achieve this, our routine uses two neural networks in series. The first network, called `Line CNN', de-noises $100\times100$ pixel images of IDT voltage scans and returns high-intensity pixels only where charge transition lines are present. Input images span $\sim$20--35\! mV, adjusted automatically to accommodate different IDT lengths (Appendix \ref{app:virtualisation}). The second network, called `Angle CNN', accepts the output of the `Line CNN' and returns 8 angles that fully parametrise the transition lines, enabling their reconstruction. The results of this threshold-free procedure are shown in Fig.~\ref{fig:fig_2}b.

Encoding transition line information using angles, rather than coordinates, prevents our networks from learning line locations using pixels at exclusively triple points and image edges \cite{muto2024visual}. By encouraging the incorporation of all information in the IDT image, our line parametrisations become robust to pixel noise and charge latching. Moreover, our two-stage framework allows the intermediate output from the `Line CNN' to be used for detecting poor IDT scans, as discussed in the Supplementary Materials.

Black markers in Fig.~\ref{fig:fig_2}b left indicate the readout voltage points corresponding to `top' and `bottom' samples, used for PSB evaluation. Their quantity can vary, as desired by the user. We find four sample pairs sufficient to detect PSB while considerably reducing measurement time from previous works, which often evaluate >50 sampling points inside potential metastable triangles \cite{jirovec2021singlet, saez-mollejo_exchange_2025, botzem2018tuning, blumoff2022fast}.

\subsection{\label{sec:evaluation}PSB Evaluation}

\subsubsection{\label{subsec:measurement}Measurement outcomes}
To probe the existence of PSB, a three-stage pulse sequence is performed on each sample of each pair, following the scheme in Fig.~\ref{fig:fig_1}d. First, spins are relaxed for 10\! \unit{\micro\second} by pulsing deep into the same charge occupation as the readout point. Then, the system is initialised using a 16\! \unit{\nano\second} ramp across the inerdot line ($\varepsilon = 0$). After an idling period $t_\mathrm{idle}$, another 16\! \unit{\nano\second} ramp returns the system to the readout point where the rf readout tone is applied and the reflected signal is integrated over 8\! \unit{\micro\second}. This sequence is repeated for 1000 cycles with $t_\mathrm{idle}$ uniformly distributed between 148\! \unit{\nano\second} and 300\! \unit{\nano\second} in 4\! \unit{\nano\second} increments. These pulse times are typical for baseband-controlled qubits in such Ge/SiGe heterostructures \cite{jirovec2021singlet, zhang_universal_2024, kelly2025identifying}.

\begin{figure}[b]
\includegraphics[width=\linewidth]{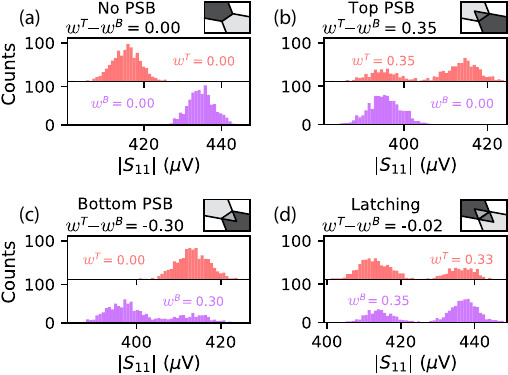}
\caption{\label{fig:fig_3}\textbf{Experimental histograms for each pulsed measurement outcome}: \textbf{(a)} `No PSB', \textbf{(b)} `Top PSB', \textbf{(c)} `Bottom PSB', \textbf{(d)} `Latching'. Red/purple histograms in each panel correspond to data acquired for the top/bottom sample of a pair. Cartoons in the top right of each panel serve as visual guides for each outcome, depicting how an averaged measurement would appear if our pulse scheme were applied at each pixel in an IDT voltage scan.}
\end{figure}

In general, readout signals for each sample pair can be mapped to one of four possible outcomes, each represented by a panel in Fig.~\ref{fig:fig_3}. In the absence of spin blockade, both pulsing directions yield purely unblocked signals. This outcome, depicted in Fig.~\ref{fig:fig_3}a, we call `No PSB'. Conversely, under a suitable electrostatic configuration, if the system is initialised in an effective (1,1) charge region, an admixing between singlet and triplet states will induce spin rotations during $t_\mathrm{idle}$ \cite{jirovec2022dynamics}. This introduces a blocked signal component for one of the pulsing directions, as seen in Fig.~\ref{fig:fig_3}(b-c). We call this the `Top PSB' or `Bottom PSB' outcome, depending on the sample for which spin blockade occurs. Since the evolution frequency is not known \textit{a priori}, we vary $t_\mathrm{idle}$ over the pulse sequence cycles to encourage the measurement of blocked and unblocked spin states in equal share, thereby balancing the counts of each component in the signal distribution. Finally, if the inverse of the interdot tunelling rate exceeds the readout time, both samples may yield blocked signals regardless of their charge occupation, as shown in Fig.~\ref{fig:fig_3}d. This regime, which we call `Latching', is not suitable for spin qubit readout.

\subsubsection{\label{subsec:score}Score function}
The signal distribution $p(x^j)$ produced by each sample is assumed to follow the Gaussian mixture model defined in Eq.~\ref{eq:GMM}. We denote the means of the model by $\tilde{\mu}_i^j$ and standard deviations by $\tilde{\sigma}_i^j$, where $i \in \{\mathrm{b, u}\}$ denotes the component (blocked or unblocked) and $j \in \{\mathrm{T, B}\}$ denotes the sample's associated readout position (top or bottom).
\begin{equation}
p(x^j) = w^j\mathcal{N}(x^j|\tilde{\mu}_b^j, \tilde{\sigma}_b^j) + \\ (1-w^j)\mathcal{N}(x^j|\tilde{\mu}_u^j, \tilde{\sigma}_u^j),\label{eq:GMM}
\end{equation}
\begin{equation}
\mathcal{N}(x^j|\tilde{\mu}_i^j, \tilde{\sigma}_i^j) = \frac{1}{\tilde{\sigma}_i^j\sqrt{2\pi}} e^{-\frac{1}{2} \left(\frac{x^j-\tilde{\mu}_i^j}{\tilde{\sigma}_i^j}\right)^2},
\end{equation}

The weight assigned to the blocked component for sample $j$ is $0 \leq w^j \leq 1$, and $(1-w^j)$, the unblocked weight. If $w^T = w^B = 0$, we recover two unimodal Gaussian distributions characteristic of `No PSB'. Conversely, if $w^T \sim 0.5$ and $w^B\sim 0$, or $w^B \sim 0.5$ and $w^T\sim 0$, either `Top PSB' or `Bottom PSB' outcomes are satisfied, respectively. Our PSB score function is therefore,
\begin{equation}
\mathrm{score} = \frac{1}{N} \left\lvert\sum_n^N (w_n^T - w_n^B)\right\rvert,\label{eq:score_func}
\end{equation}
where $n$ runs over all sample pairs and $N=4$ is the total number of pairs. Crucially, the score is independent of the pulsing direction for which spin blockade emerges, and will be low if both directions contain similar weights, i.e. `Latching'.

The parameters of Eq.~\ref{eq:GMM} are calculated using maximum likelihood estimation (MLE) within bounds imposed by reference distributions. The acquisition of reference distributions is discussed in Appendix \ref{app:references} and MLE calculations in Appendix \ref{app:MLE_methods}.

\subsubsection{\label{subsec:criteria}Stopping criteria}
At each tuning iteration, singlet-triplet oscillations are probed at the readout point of the highest scoring sample. If this measurement contains a Fourier component with normalised amplitude above 0.5 and prominence greater than 0.2, an automatic fitting procedure computes the $R^2$ coefficient at the detected frequency. If $R^2 > 0.75$, a qubit is deemed present and the tuning loop exits with a ``succes'' statement. If $0.55 < R^2 \leq 0.75$, the routine proceeds with two checks and exits the tuning loop successfully if any one check is satisfied.

The first check is that the PSB score is greater than 0.1. Although ideal blocked weights of 0.5 are expected, state preparation and measurement errors diminish readout visibility. The second check requires more than half of all sample pairs to share the same outcome, which must be either `Top PSB' or `Bottom PSB'. A sample pair's outcome is determined using the Bayesian Information Criterion \cite{schwarz1978estimating}, detailed in Appendix \ref{app:outcome_methods}.

Our routine's three complimentary criteria: the prominence of singlet-triplet oscillations, the PSB score, and sample pair outcomes, enable reliable detection of qubits when dealing with imperfect or noisy measurements.

\section{\label{sec:experiment}Experimental Results}
We now turn to an experimental demonstration of our routine. We begin by presenting the barrier voltage tuning path around a single IDT, then show results from an autonomous qubit search across 26 IDTs, and conclude with a summary of qubit properties across the 12 charge occupations where qubits were autonomously found. All gates are virtualised against the sensor plunger (SP) before running our routine using the method described in \cite{hickie2024}. At the start of each IDT's voltage tuning loop, barrier gates are automatically virtualised against PL and PR (see Appendix \ref{app:virtualisation}), and with every barrier voltage adjustment, SP is also fine-tuned, the IDT is re-centred, and readout points are automatically re-sampled. 

\subsection{Single IDT tuning}
Since the score function landscape is flat in most regions where PSB is absent, tuning is done primarily through an exploration of the barrier voltage search space. Latin hypercube sampling \cite{scipy} is used to generate 40 candidate barrier voltage configurations within user-defined voltage bounds, and a path is followed through them that minimises large voltage jumps. Throughout this work, we use bounds of $[-20,+90]$/ $[-80,+90]$/$[-60,+80]$\! mV on BL/BM/BR, based on where double quantum dot confinement was maintained during the initial device tune-up. If the stopping criteria are not met during this exploration but a promising candidate exists, our routine enters into an optimisation loop for up to 20 additional iterations within a refined voltage subspace. Candidate selection criteria and optimisation details are provided in Appendix \ref{app:optimisation}. For optimisation, we use the Nelder-Mead algorithm due to its gradient-free optimisation and suitability to non-linear objective function landscapes \cite{nelder1965simplex}. 

\begin{figure}[t]
\includegraphics[width=\linewidth]{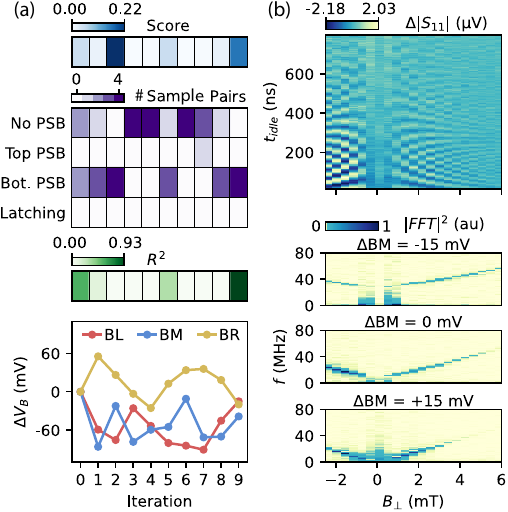}
\caption{\label{fig:fig_4}\textbf{Example tuning path taken by our routine}. \textbf{(a)} Moving downward: the PSB score, counts of sample pair outcomes, fitted $R^2$ coefficient, and barrier gate voltages, for each tuning iteration. \textbf{(b)} Magnetic field sweep at the barrier voltage configuration of iteration 9, accompanied by Fourier transforms at two additional BM voltages of $\pm$15\! mV. The mean of each time trace has been subtracted.}
\end{figure}

Figure \ref{fig:fig_4}a shows the interplay of our three stopping criteria over an example tuning loop that ends in success. Prominent oscillations are found at iteration 9 ($R^2 = 0.929$), accompanied by a high PSB score (0.159) and unanimous `Bottom PSB' outcomes across sample pairs. The modulation of the oscillation frequency $f$ by the magnetic field, shown in Fig.~\ref{fig:fig_4}b, is consistent with singlet-triplet transitions. At $B_{\perp} = 2.5$\! \unit{\milli\tesla}, we report a $T_2^\star$ of 410\! \unit{\nano\second} by fitting the decay envelope to $e^{-t/T_2^\star}$. 

Fast Fourier transforms (FFTs) enable the extraction of the exchange energy $J$ by extrapolating the oscillation frequency to zero field. The effect of individually changing BM by $\pm15$\! mV is shown in top and bottom FFTs of Fig.~\ref{fig:fig_4}b. For increasing BM, $J / h$ is tuned from $\sim0$\! \unit{\mega\hertz} to $\sim30$\! \unit{\mega\hertz}, consistent with the expected increase in tunnel coupling. A more comprehensive comparison of qubit properties is discussed in the next section.

\begin{figure*}[t]
\includegraphics[width=\linewidth]{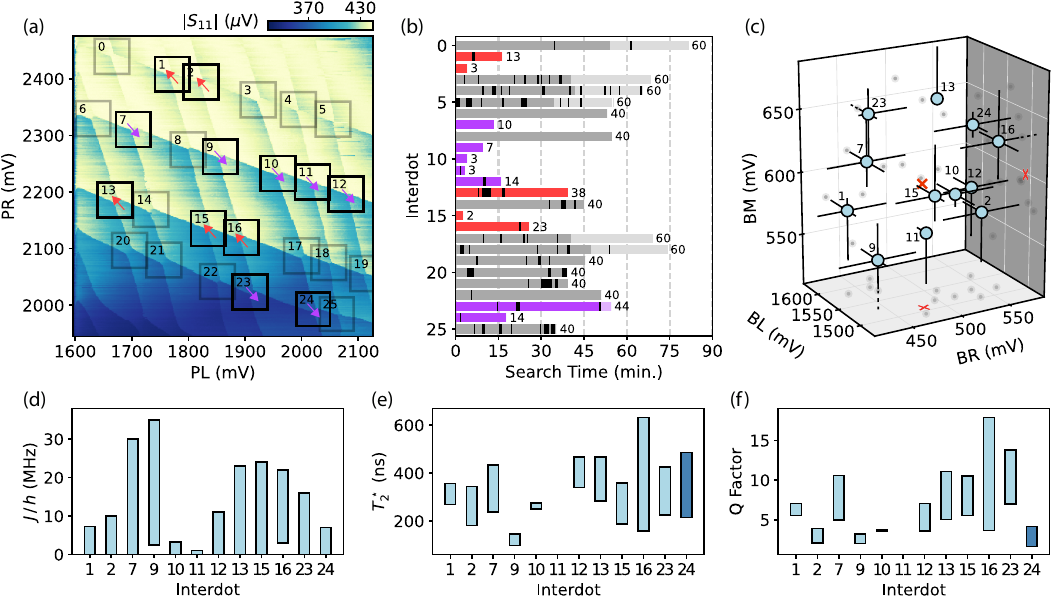}
\caption{\label{fig:fig_5}\textbf{Summary of results from an autonomous run of our routine}. \textbf{(a)} Charge stability diagram with automatically placed boxes about detected interdot charge transitions (IDTs), each enumerated. Arrows, where present, indicate the direction of the readout pulse for which PSB and singlet-triple oscillations were found. \textbf{(b)} Bar plot of the time spent searching for qubits at each IDT. Light-grey segments correspond to optimisation iterations, occurring only if a promising barrier voltage candidate (but no qubit) was found within
40 iterations (Appendix \ref{app:optimisation}). Black segments correspond to configurations that did not yield IDT scans of sufficient quality to proceed to our routine's evaluation stage (see Supplementary Materials). Values at the end of each bar are the number of tested voltage configurations. \textbf{(c)} Barrier voltage configurations at which qubits were found, following the IDT numbering in (a). Grey markers are projections of these configurations and the starting configuration is marked with a red cross. Line extensions indicate barrier voltage ranges over which oscillations persist ($R^2 > 0.55$), with an upper bound of $\pm 40$\! mV, and lower bound of $\pm 5$\! mV (not visible for IDT 13). Dashed lines indicate voltages that go beyond the axes plane. \textbf{(d--f)} Overall variability in qubit properties over the barrier voltage ranges in (c): (d) exchange interaction strength $J / h$, (e) dephasing time $T_2^\star$, and (f) Q factor. Data fits are provided in the Supplementary Materials. All values of $T_2^\star$ and the Q factor are reported at $B=2.5$\! \unit{\milli\tesla}, except for interdot 24 (shown in dark blue), which is at $B=1.0$\! \unit{\milli\tesla}. Data with multiple oscillation frequencies were omitted from this analysis.}
\end{figure*}

\subsection{Multi-IDT tuning}
\indent The main results are summarised in Fig.~\ref{fig:fig_5}. Out of 26 IDTs, our autonomous routine finds PSB and singlet-triplet oscillations at 12 of them. These are depicted using black boxes in Fig.~\ref{fig:fig_5}a. 

Conventionally, PSB follows a checkerboard pattern across IDTs \cite{johnson2005singlet}, which is not obeyed by our device within the voltage ranges our routine operated. We speculate this is due to a spurious dot coupled to PL, as suggested by the two different vertical transition line slopes and varying gate cross-capacitances (Appendix \ref{app:virtualisation}). Alternatively, spin-orbit effects could also be responsible for the breakdown of the checkerboard pattern. This highlights our routine's capacity to operate independently of human bias, which often restricts experimenters to investigate only a handful of transitions. 

A benchmark of the tuning time at each IDT is shown in Fig.~\ref{fig:fig_5}b. After acquiring the charge stability diagram in (a), our routine takes $<17$ hours to find qubits among all IDTs. We note that the majority of this time (13 hours, or $\sim78\%$) was devoted to tuning unsuccessful IDTs, whereas successful IDTs took between 3 and 55 minutes, with a median tuning time of just 15 minutes. The additional time spent virtualising barrier gates is under 2.5 minutes per IDT.

Figure \ref{fig:fig_5}c maps the barrier voltage configurations at which our stopping criteria were met. The variety in distances from the starting barrier configuration (red cross) highlights the appropriacy of our exploratory tuning approach (see Supplementary Materials). Furthermore, the spread in configurations underscores the challenge of finding PSB manually, as no single configuration leads to success across all the IDTs.

At each successful IDT, our routine automatically finds barrier voltage bounds that sustain singlet-triplet oscillations ($R^2 > 0.55$). These are depicted using lines in Fig.~\ref{fig:fig_5}c, with an upper-bound set to $\pm40$\! mV. We embed magnetic field sweeps between these barrier bounds autonomously into our routine, thereby opening the path to a systematic characterisation of qubit properties, and their tunability.

In Fig~\ref{fig:fig_5}(d-f) we report $J/h$, $T_2^\star$, and the quality factor $Q = f \times T_2^\star$ across different IDTs, and their ranges within the aforementioned barrier bounds. Notably, our routine uncovers a factor-of-six variation in $T_2^\star$ and nearly an order of magnitude variation in the Q factor despite not explicitly optimising over these qubit properties. Based on these findings, IDT 16 emerges as a potential best candidate for operating a singlet-triplet qubit with the highest fidelity among all tested IDTs \cite{stano2022review}. For other qubit modalities, such as exchange-only and Loss-DiVincenzo, good control over the exchange interaction is desirable \cite{ha2025two, lawrie2023simultaneous}, making IDT 7 compelling for further experiments. Due to an ambiguity in the nature of the found oscillations (see Supplementary Materials), we refrain from fitting g-factors in this work, but note that they may be accessed by varying ramp durations and magnetic field angles \cite{jirovec2022dynamics, liles2024singlet, kelly2025identifying}.

\section{Conclusion \& Outlook}
We have developed a machine learning-assisted routine capable of efficiently finding coherent singlet-triplet oscillations over many interdot charge transitions. We have trained neural networks that, provided a large charge stability diagram, can locate interdot charge transitions autonomously, virtualise barrier gates, and extract potential meta-stable regions in the presence of both experimental noise and latching. The pulsing scheme, score function and stopping criteria we have developed for detecting PSB, allow our routine to operate without prior knowledge of charge occupations. Furthermore, we have shown how our routine can enable a systematic characterisation of the exchange interaction, dephasing time, and Q factor, which show large spreads between interdot charge transitions.

The impact of our routine lies in the speed advantage that it offers, enabling prolonged experiments to be run independently and without manual decision-making. This brings $in$-$situ$ engineering of qubit properties, without being constrained to a single charge transition, within reach. For example, our routine, in conjunction with magnetic spectroscopy at different voltage configurations, could be used for $g$-factor tuning \cite{ares2013nature, liles2021electrical}, exploring magnetic field sweet lines \cite{mauro2024geometry, bassi2024optimal, sen2023classification}, and studying the anisotropic exchange interaction \cite{geyer2024anisotropic, saez-mollejo_exchange_2025} in systems with strong spin-orbit interaction. In addition, one could use our routine to verify the presence of spurious dots by considering conventional PSB patterns, as well as studying PSB lifting mechanisms and their susceptibility to gate voltage \cite{froning2021strong}. Given the vast amount of data generated by our routine, one could also envisage its use in training meta-learning models to extract qubit properties and Hamiltonian parameters `on-the-fly' \cite{schorling2025meta}. 

We anticipate a prompt integration of automated double quantum dot tuning into our routine \cite{gebhart2023learning}, unlocking efficient tuning of spin qubit arrays and statistical characterisations of qubit uniformity across a variety of spin qubit architectures \cite{tosato2025qarpet}.

\begin{acknowledgments}
The authors would like to thank Barnaby van Straaten, Jonas Schuff, Daniel Jirovec and Hanifa Tidjani for fruitful discussions. 

This research was supported by the Scientific Service Units of ISTA through resources provided by the MIBA Machine Shop and the Nanofabrication Facility. G.K. acknowledges support from the NOMIS Foundation, the HORIZON-RIA (project no. 101069515) and the FWF Projects (DOIs: 10.55776/F86 and 10.55776/I5060). N.A. acknowledges support from the European Research Council (grant agreement 948932), and the Royal Society (grant no. URF/R1/191150). This project received support from the US Army Research Office (ARO) under Award No. W911NF-24-2-0043. C.C. acknowledges support from the UKRI Doctoral Training Partnership related to EP/W524311/1 (project ref. 2887634).
\end{acknowledgments}

\section*{Author Contributions}
C.C. and J.SM. conducted the experiments and analysis equally. C.C. developed the machine learning models, and mathematical methods with support from J.SM., J.SM. fabricated the device and identified its operational regime. S.C., D.C. and G.I. supplied the heterostructure. C.C. and J.SM. wrote the manuscript with input from G.K. and N.A., F.F., G.K. and N.A. co-supervised the project. N.A. conceived the project.

\section*{Competing Interests}
Natalia Ares declares a competing interest as a founder of QuantrolOx, which develops machine learning-based software for quantum control.

\section*{Data \& Code Availability}
All data included in this work will be accessible at the Institute of Science and Technology Austria repository. All machine learning models and example code for running our routine will become available on GitHub upon final publication.

\bibliography{refs.bib}

\begin{thebibliography}{69}%
\makeatletter
\providecommand \@ifxundefined [1]{%
 \@ifx{#1\undefined}
}%
\providecommand \@ifnum [1]{%
 \ifnum #1\expandafter \@firstoftwo
 \else \expandafter \@secondoftwo
 \fi
}%
\providecommand \@ifx [1]{%
 \ifx #1\expandafter \@firstoftwo
 \else \expandafter \@secondoftwo
 \fi
}%
\providecommand \natexlab [1]{#1}%
\providecommand \enquote  [1]{``#1''}%
\providecommand \bibnamefont  [1]{#1}%
\providecommand \bibfnamefont [1]{#1}%
\providecommand \citenamefont [1]{#1}%
\providecommand \href@noop [0]{\@secondoftwo}%
\providecommand \href [0]{\begingroup \@sanitize@url \@href}%
\providecommand \@href[1]{\@@startlink{#1}\@@href}%
\providecommand \@@href[1]{\endgroup#1\@@endlink}%
\providecommand \@sanitize@url [0]{\catcode `\\12\catcode `\$12\catcode `\&12\catcode `\#12\catcode `\^12\catcode `\_12\catcode `\%12\relax}%
\providecommand \@@startlink[1]{}%
\providecommand \@@endlink[0]{}%
\providecommand \url  [0]{\begingroup\@sanitize@url \@url }%
\providecommand \@url [1]{\endgroup\@href {#1}{\urlprefix }}%
\providecommand \urlprefix  [0]{URL }%
\providecommand \Eprint [0]{\href }%
\providecommand \doibase [0]{https://doi.org/}%
\providecommand \selectlanguage [0]{\@gobble}%
\providecommand \bibinfo  [0]{\@secondoftwo}%
\providecommand \bibfield  [0]{\@secondoftwo}%
\providecommand \translation [1]{[#1]}%
\providecommand \BibitemOpen [0]{}%
\providecommand \bibitemStop [0]{}%
\providecommand \bibitemNoStop [0]{.\EOS\space}%
\providecommand \EOS [0]{\spacefactor3000\relax}%
\providecommand \BibitemShut  [1]{\csname bibitem#1\endcsname}%
\let\auto@bib@innerbib\@empty
%</preamble>
\bibitem [{\citenamefont {Zwerver}\ \emph {et~al.}(2022)\citenamefont {Zwerver}, \citenamefont {Kr{\"a}henmann}, \citenamefont {Watson}, \citenamefont {Lampert}, \citenamefont {George}, \citenamefont {Pillarisetty}, \citenamefont {Bojarski}, \citenamefont {Amin}, \citenamefont {Amitonov}, \citenamefont {Boter} \emph {et~al.}}]{zwerver2022qubits}%
  \BibitemOpen
  \bibfield  {author} {\bibinfo {author} {\bibfnamefont {A.}~\bibnamefont {Zwerver}}, \bibinfo {author} {\bibfnamefont {T.}~\bibnamefont {Kr{\"a}henmann}}, \bibinfo {author} {\bibfnamefont {T.}~\bibnamefont {Watson}}, \bibinfo {author} {\bibfnamefont {L.}~\bibnamefont {Lampert}}, \bibinfo {author} {\bibfnamefont {H.~C.}\ \bibnamefont {George}}, \bibinfo {author} {\bibfnamefont {R.}~\bibnamefont {Pillarisetty}}, \bibinfo {author} {\bibfnamefont {S.}~\bibnamefont {Bojarski}}, \bibinfo {author} {\bibfnamefont {P.}~\bibnamefont {Amin}}, \bibinfo {author} {\bibfnamefont {S.}~\bibnamefont {Amitonov}}, \bibinfo {author} {\bibfnamefont {J.}~\bibnamefont {Boter}}, \emph {et~al.},\ }\bibfield  {title} {\bibinfo {title} {Qubits made by advanced semiconductor manufacturing},\ }\href@noop {} {\bibfield  {journal} {\bibinfo  {journal} {Nature Electronics}\ }\textbf {\bibinfo {volume} {5}},\ \bibinfo {pages} {184} (\bibinfo {year} {2022})}\BibitemShut {NoStop}%
\bibitem [{\citenamefont {Neyens}\ \emph {et~al.}(2024)\citenamefont {Neyens}, \citenamefont {Zietz}, \citenamefont {Watson}, \citenamefont {Luthi}, \citenamefont {Nethwewala}, \citenamefont {George}, \citenamefont {Henry}, \citenamefont {Islam}, \citenamefont {Wagner}, \citenamefont {Borjans} \emph {et~al.}}]{neyens2024probing}%
  \BibitemOpen
  \bibfield  {author} {\bibinfo {author} {\bibfnamefont {S.}~\bibnamefont {Neyens}}, \bibinfo {author} {\bibfnamefont {O.~K.}\ \bibnamefont {Zietz}}, \bibinfo {author} {\bibfnamefont {T.~F.}\ \bibnamefont {Watson}}, \bibinfo {author} {\bibfnamefont {F.}~\bibnamefont {Luthi}}, \bibinfo {author} {\bibfnamefont {A.}~\bibnamefont {Nethwewala}}, \bibinfo {author} {\bibfnamefont {H.~C.}\ \bibnamefont {George}}, \bibinfo {author} {\bibfnamefont {E.}~\bibnamefont {Henry}}, \bibinfo {author} {\bibfnamefont {M.}~\bibnamefont {Islam}}, \bibinfo {author} {\bibfnamefont {A.~J.}\ \bibnamefont {Wagner}}, \bibinfo {author} {\bibfnamefont {F.}~\bibnamefont {Borjans}}, \emph {et~al.},\ }\bibfield  {title} {\bibinfo {title} {Probing single electrons across 300-mm spin qubit wafers},\ }\href@noop {} {\bibfield  {journal} {\bibinfo  {journal} {Nature}\ }\textbf {\bibinfo {volume} {629}},\ \bibinfo {pages} {80} (\bibinfo {year} {2024})}\BibitemShut {NoStop}%
\bibitem [{\citenamefont {George}\ \emph {et~al.}(2024)\citenamefont {George}, \citenamefont {M{\k{a}}dzik}, \citenamefont {Henry}, \citenamefont {Wagner}, \citenamefont {Islam}, \citenamefont {Borjans}, \citenamefont {Connors}, \citenamefont {Corrigan}, \citenamefont {Curry}, \citenamefont {Harper} \emph {et~al.}}]{george202412}%
  \BibitemOpen
  \bibfield  {author} {\bibinfo {author} {\bibfnamefont {H.~C.}\ \bibnamefont {George}}, \bibinfo {author} {\bibfnamefont {M.~T.}\ \bibnamefont {M{\k{a}}dzik}}, \bibinfo {author} {\bibfnamefont {E.~M.}\ \bibnamefont {Henry}}, \bibinfo {author} {\bibfnamefont {A.~J.}\ \bibnamefont {Wagner}}, \bibinfo {author} {\bibfnamefont {M.~M.}\ \bibnamefont {Islam}}, \bibinfo {author} {\bibfnamefont {F.}~\bibnamefont {Borjans}}, \bibinfo {author} {\bibfnamefont {E.~J.}\ \bibnamefont {Connors}}, \bibinfo {author} {\bibfnamefont {J.}~\bibnamefont {Corrigan}}, \bibinfo {author} {\bibfnamefont {M.}~\bibnamefont {Curry}}, \bibinfo {author} {\bibfnamefont {M.~K.}\ \bibnamefont {Harper}}, \emph {et~al.},\ }\bibfield  {title} {\bibinfo {title} {12-spin-qubit arrays fabricated on a 300 mm semiconductor manufacturing line},\ }\href@noop {} {\bibfield  {journal} {\bibinfo  {journal} {Nano Letters}\ }\textbf {\bibinfo {volume} {25}},\ \bibinfo {pages} {793} (\bibinfo {year} {2024})}\BibitemShut {NoStop}%
\bibitem [{\citenamefont {Noiri}\ \emph {et~al.}(2022)\citenamefont {Noiri}, \citenamefont {Takeda}, \citenamefont {Nakajima}, \citenamefont {Kobayashi}, \citenamefont {Sammak}, \citenamefont {Scappucci},\ and\ \citenamefont {Tarucha}}]{noiri2022fast}%
  \BibitemOpen
  \bibfield  {author} {\bibinfo {author} {\bibfnamefont {A.}~\bibnamefont {Noiri}}, \bibinfo {author} {\bibfnamefont {K.}~\bibnamefont {Takeda}}, \bibinfo {author} {\bibfnamefont {T.}~\bibnamefont {Nakajima}}, \bibinfo {author} {\bibfnamefont {T.}~\bibnamefont {Kobayashi}}, \bibinfo {author} {\bibfnamefont {A.}~\bibnamefont {Sammak}}, \bibinfo {author} {\bibfnamefont {G.}~\bibnamefont {Scappucci}},\ and\ \bibinfo {author} {\bibfnamefont {S.}~\bibnamefont {Tarucha}},\ }\bibfield  {title} {\bibinfo {title} {Fast universal quantum gate above the fault-tolerance threshold in silicon},\ }\href@noop {} {\bibfield  {journal} {\bibinfo  {journal} {Nature}\ }\textbf {\bibinfo {volume} {601}},\ \bibinfo {pages} {338} (\bibinfo {year} {2022})}\BibitemShut {NoStop}%
\bibitem [{\citenamefont {Xue}\ \emph {et~al.}(2022)\citenamefont {Xue}, \citenamefont {Russ}, \citenamefont {Samkharadze}, \citenamefont {Undseth}, \citenamefont {Sammak}, \citenamefont {Scappucci},\ and\ \citenamefont {Vandersypen}}]{xue2022quantum}%
  \BibitemOpen
  \bibfield  {author} {\bibinfo {author} {\bibfnamefont {X.}~\bibnamefont {Xue}}, \bibinfo {author} {\bibfnamefont {M.}~\bibnamefont {Russ}}, \bibinfo {author} {\bibfnamefont {N.}~\bibnamefont {Samkharadze}}, \bibinfo {author} {\bibfnamefont {B.}~\bibnamefont {Undseth}}, \bibinfo {author} {\bibfnamefont {A.}~\bibnamefont {Sammak}}, \bibinfo {author} {\bibfnamefont {G.}~\bibnamefont {Scappucci}},\ and\ \bibinfo {author} {\bibfnamefont {L.~M.}\ \bibnamefont {Vandersypen}},\ }\bibfield  {title} {\bibinfo {title} {Quantum logic with spin qubits crossing the surface code threshold},\ }\href@noop {} {\bibfield  {journal} {\bibinfo  {journal} {Nature}\ }\textbf {\bibinfo {volume} {601}},\ \bibinfo {pages} {343} (\bibinfo {year} {2022})}\BibitemShut {NoStop}%
\bibitem [{\citenamefont {Lawrie}\ \emph {et~al.}(2023)\citenamefont {Lawrie}, \citenamefont {Rimbach-Russ}, \citenamefont {Riggelen}, \citenamefont {Hendrickx}, \citenamefont {Snoo}, \citenamefont {Sammak}, \citenamefont {Scappucci}, \citenamefont {Helsen},\ and\ \citenamefont {Veldhorst}}]{lawrie2023simultaneous}%
  \BibitemOpen
  \bibfield  {author} {\bibinfo {author} {\bibfnamefont {W.}~\bibnamefont {Lawrie}}, \bibinfo {author} {\bibfnamefont {M.}~\bibnamefont {Rimbach-Russ}}, \bibinfo {author} {\bibfnamefont {F.~v.}\ \bibnamefont {Riggelen}}, \bibinfo {author} {\bibfnamefont {N.}~\bibnamefont {Hendrickx}}, \bibinfo {author} {\bibfnamefont {S.~d.}\ \bibnamefont {Snoo}}, \bibinfo {author} {\bibfnamefont {A.}~\bibnamefont {Sammak}}, \bibinfo {author} {\bibfnamefont {G.}~\bibnamefont {Scappucci}}, \bibinfo {author} {\bibfnamefont {J.}~\bibnamefont {Helsen}},\ and\ \bibinfo {author} {\bibfnamefont {M.}~\bibnamefont {Veldhorst}},\ }\bibfield  {title} {\bibinfo {title} {Simultaneous single-qubit driving of semiconductor spin qubits at the fault-tolerant threshold},\ }\href@noop {} {\bibfield  {journal} {\bibinfo  {journal} {Nature Communications}\ }\textbf {\bibinfo {volume} {14}},\ \bibinfo {pages} {3617} (\bibinfo {year} {2023})}\BibitemShut {NoStop}%
\bibitem [{\citenamefont {Wang}\ \emph {et~al.}(2024)\citenamefont {Wang}, \citenamefont {John}, \citenamefont {Tidjani}, \citenamefont {Yu}, \citenamefont {Ivlev}, \citenamefont {D{\'e}prez}, \citenamefont {van Riggelen-Doelman}, \citenamefont {Woods}, \citenamefont {Hendrickx}, \citenamefont {Lawrie} \emph {et~al.}}]{wang2024operating}%
  \BibitemOpen
  \bibfield  {author} {\bibinfo {author} {\bibfnamefont {C.-A.}\ \bibnamefont {Wang}}, \bibinfo {author} {\bibfnamefont {V.}~\bibnamefont {John}}, \bibinfo {author} {\bibfnamefont {H.}~\bibnamefont {Tidjani}}, \bibinfo {author} {\bibfnamefont {C.~X.}\ \bibnamefont {Yu}}, \bibinfo {author} {\bibfnamefont {A.~S.}\ \bibnamefont {Ivlev}}, \bibinfo {author} {\bibfnamefont {C.}~\bibnamefont {D{\'e}prez}}, \bibinfo {author} {\bibfnamefont {F.}~\bibnamefont {van Riggelen-Doelman}}, \bibinfo {author} {\bibfnamefont {B.~D.}\ \bibnamefont {Woods}}, \bibinfo {author} {\bibfnamefont {N.~W.}\ \bibnamefont {Hendrickx}}, \bibinfo {author} {\bibfnamefont {W.~I.}\ \bibnamefont {Lawrie}}, \emph {et~al.},\ }\bibfield  {title} {\bibinfo {title} {Operating semiconductor quantum processors with hopping spins},\ }\href@noop {} {\bibfield  {journal} {\bibinfo  {journal} {Science}\ }\textbf {\bibinfo {volume} {385}},\ \bibinfo {pages} {447} (\bibinfo {year} {2024})}\BibitemShut {NoStop}%
\bibitem [{\citenamefont {Huang}\ \emph {et~al.}(2024)\citenamefont {Huang}, \citenamefont {Su}, \citenamefont {Lim}, \citenamefont {Feng}, \citenamefont {van Straaten}, \citenamefont {Severin}, \citenamefont {Gilbert}, \citenamefont {Dumoulin~Stuyck}, \citenamefont {Tanttu}, \citenamefont {Serrano} \emph {et~al.}}]{huang2024high}%
  \BibitemOpen
  \bibfield  {author} {\bibinfo {author} {\bibfnamefont {J.~Y.}\ \bibnamefont {Huang}}, \bibinfo {author} {\bibfnamefont {R.~Y.}\ \bibnamefont {Su}}, \bibinfo {author} {\bibfnamefont {W.~H.}\ \bibnamefont {Lim}}, \bibinfo {author} {\bibfnamefont {M.}~\bibnamefont {Feng}}, \bibinfo {author} {\bibfnamefont {B.}~\bibnamefont {van Straaten}}, \bibinfo {author} {\bibfnamefont {B.}~\bibnamefont {Severin}}, \bibinfo {author} {\bibfnamefont {W.}~\bibnamefont {Gilbert}}, \bibinfo {author} {\bibfnamefont {N.}~\bibnamefont {Dumoulin~Stuyck}}, \bibinfo {author} {\bibfnamefont {T.}~\bibnamefont {Tanttu}}, \bibinfo {author} {\bibfnamefont {S.}~\bibnamefont {Serrano}}, \emph {et~al.},\ }\bibfield  {title} {\bibinfo {title} {High-fidelity spin qubit operation and algorithmic initialization above 1 k},\ }\href@noop {} {\bibfield  {journal} {\bibinfo  {journal} {Nature}\ }\textbf {\bibinfo {volume} {627}},\ \bibinfo {pages} {772} (\bibinfo {year} {2024})}\BibitemShut {NoStop}%
\bibitem [{\citenamefont {Kalantre}\ \emph {et~al.}(2019)\citenamefont {Kalantre}, \citenamefont {Zwolak}, \citenamefont {Ragole}, \citenamefont {Wu}, \citenamefont {Zimmerman}, \citenamefont {Stewart~Jr},\ and\ \citenamefont {Taylor}}]{kalantre2019machine}%
  \BibitemOpen
  \bibfield  {author} {\bibinfo {author} {\bibfnamefont {S.~S.}\ \bibnamefont {Kalantre}}, \bibinfo {author} {\bibfnamefont {J.~P.}\ \bibnamefont {Zwolak}}, \bibinfo {author} {\bibfnamefont {S.}~\bibnamefont {Ragole}}, \bibinfo {author} {\bibfnamefont {X.}~\bibnamefont {Wu}}, \bibinfo {author} {\bibfnamefont {N.~M.}\ \bibnamefont {Zimmerman}}, \bibinfo {author} {\bibfnamefont {M.}~\bibnamefont {Stewart~Jr}},\ and\ \bibinfo {author} {\bibfnamefont {J.~M.}\ \bibnamefont {Taylor}},\ }\bibfield  {title} {\bibinfo {title} {Machine learning techniques for state recognition and auto-tuning in quantum dots},\ }\href@noop {} {\bibfield  {journal} {\bibinfo  {journal} {npj Quantum Information}\ }\textbf {\bibinfo {volume} {5}},\ \bibinfo {pages} {6} (\bibinfo {year} {2019})}\BibitemShut {NoStop}%
\bibitem [{\citenamefont {Zwolak}\ \emph {et~al.}(2020)\citenamefont {Zwolak}, \citenamefont {McJunkin}, \citenamefont {Kalantre}, \citenamefont {Dodson}, \citenamefont {MacQuarrie}, \citenamefont {Savage}, \citenamefont {Lagally}, \citenamefont {Coppersmith}, \citenamefont {Eriksson},\ and\ \citenamefont {Taylor}}]{zwolak2020autotuning}%
  \BibitemOpen
  \bibfield  {author} {\bibinfo {author} {\bibfnamefont {J.~P.}\ \bibnamefont {Zwolak}}, \bibinfo {author} {\bibfnamefont {T.}~\bibnamefont {McJunkin}}, \bibinfo {author} {\bibfnamefont {S.~S.}\ \bibnamefont {Kalantre}}, \bibinfo {author} {\bibfnamefont {J.}~\bibnamefont {Dodson}}, \bibinfo {author} {\bibfnamefont {E.}~\bibnamefont {MacQuarrie}}, \bibinfo {author} {\bibfnamefont {D.}~\bibnamefont {Savage}}, \bibinfo {author} {\bibfnamefont {M.}~\bibnamefont {Lagally}}, \bibinfo {author} {\bibfnamefont {S.}~\bibnamefont {Coppersmith}}, \bibinfo {author} {\bibfnamefont {M.~A.}\ \bibnamefont {Eriksson}},\ and\ \bibinfo {author} {\bibfnamefont {J.~M.}\ \bibnamefont {Taylor}},\ }\bibfield  {title} {\bibinfo {title} {Autotuning of double-dot devices in situ with machine learning},\ }\href@noop {} {\bibfield  {journal} {\bibinfo  {journal} {Physical review applied}\ }\textbf {\bibinfo {volume} {13}},\ \bibinfo {pages} {034075} (\bibinfo {year} {2020})}\BibitemShut {NoStop}%
\bibitem [{\citenamefont {Durrer}\ \emph {et~al.}(2020)\citenamefont {Durrer}, \citenamefont {Kratochwil}, \citenamefont {Koski}, \citenamefont {Landig}, \citenamefont {Reichl}, \citenamefont {Wegscheider}, \citenamefont {Ihn},\ and\ \citenamefont {Greplova}}]{durrer2020automated}%
  \BibitemOpen
  \bibfield  {author} {\bibinfo {author} {\bibfnamefont {R.}~\bibnamefont {Durrer}}, \bibinfo {author} {\bibfnamefont {B.}~\bibnamefont {Kratochwil}}, \bibinfo {author} {\bibfnamefont {J.~V.}\ \bibnamefont {Koski}}, \bibinfo {author} {\bibfnamefont {A.~J.}\ \bibnamefont {Landig}}, \bibinfo {author} {\bibfnamefont {C.}~\bibnamefont {Reichl}}, \bibinfo {author} {\bibfnamefont {W.}~\bibnamefont {Wegscheider}}, \bibinfo {author} {\bibfnamefont {T.}~\bibnamefont {Ihn}},\ and\ \bibinfo {author} {\bibfnamefont {E.}~\bibnamefont {Greplova}},\ }\bibfield  {title} {\bibinfo {title} {Automated tuning of double quantum dots into specific charge states using neural networks},\ }\href@noop {} {\bibfield  {journal} {\bibinfo  {journal} {Physical Review Applied}\ }\textbf {\bibinfo {volume} {13}},\ \bibinfo {pages} {054019} (\bibinfo {year} {2020})}\BibitemShut {NoStop}%
\bibitem [{\citenamefont {Zwolak}\ \emph {et~al.}(2021)\citenamefont {Zwolak}, \citenamefont {McJunkin}, \citenamefont {Kalantre}, \citenamefont {Neyens}, \citenamefont {MacQuarrie}, \citenamefont {Eriksson},\ and\ \citenamefont {Taylor}}]{zwolak2021ray}%
  \BibitemOpen
  \bibfield  {author} {\bibinfo {author} {\bibfnamefont {J.~P.}\ \bibnamefont {Zwolak}}, \bibinfo {author} {\bibfnamefont {T.}~\bibnamefont {McJunkin}}, \bibinfo {author} {\bibfnamefont {S.~S.}\ \bibnamefont {Kalantre}}, \bibinfo {author} {\bibfnamefont {S.~F.}\ \bibnamefont {Neyens}}, \bibinfo {author} {\bibfnamefont {E.}~\bibnamefont {MacQuarrie}}, \bibinfo {author} {\bibfnamefont {M.~A.}\ \bibnamefont {Eriksson}},\ and\ \bibinfo {author} {\bibfnamefont {J.~M.}\ \bibnamefont {Taylor}},\ }\bibfield  {title} {\bibinfo {title} {Ray-based framework for state identification in quantum dot devices},\ }\href@noop {} {\bibfield  {journal} {\bibinfo  {journal} {PRX Quantum}\ }\textbf {\bibinfo {volume} {2}},\ \bibinfo {pages} {020335} (\bibinfo {year} {2021})}\BibitemShut {NoStop}%
\bibitem [{\citenamefont {Ziegler}\ \emph {et~al.}(2023{\natexlab{a}})\citenamefont {Ziegler}, \citenamefont {Luthi}, \citenamefont {Ramsey}, \citenamefont {Borjans}, \citenamefont {Zheng},\ and\ \citenamefont {Zwolak}}]{ziegler2023tuning}%
  \BibitemOpen
  \bibfield  {author} {\bibinfo {author} {\bibfnamefont {J.}~\bibnamefont {Ziegler}}, \bibinfo {author} {\bibfnamefont {F.}~\bibnamefont {Luthi}}, \bibinfo {author} {\bibfnamefont {M.}~\bibnamefont {Ramsey}}, \bibinfo {author} {\bibfnamefont {F.}~\bibnamefont {Borjans}}, \bibinfo {author} {\bibfnamefont {G.}~\bibnamefont {Zheng}},\ and\ \bibinfo {author} {\bibfnamefont {J.~P.}\ \bibnamefont {Zwolak}},\ }\bibfield  {title} {\bibinfo {title} {Tuning arrays with rays: Physics-informed tuning of quantum dot charge states},\ }\href@noop {} {\bibfield  {journal} {\bibinfo  {journal} {Physical Review Applied}\ }\textbf {\bibinfo {volume} {20}},\ \bibinfo {pages} {034067} (\bibinfo {year} {2023}{\natexlab{a}})}\BibitemShut {NoStop}%
\bibitem [{\citenamefont {Alexeev}\ \emph {et~al.}(2024)\citenamefont {Alexeev}, \citenamefont {Farag}, \citenamefont {Patti}, \citenamefont {Wolf}, \citenamefont {Ares}, \citenamefont {Aspuru-Guzik}, \citenamefont {Benjamin}, \citenamefont {Cai}, \citenamefont {Chandani}, \citenamefont {Fedele} \emph {et~al.}}]{alexeev2024artificial}%
  \BibitemOpen
  \bibfield  {author} {\bibinfo {author} {\bibfnamefont {Y.}~\bibnamefont {Alexeev}}, \bibinfo {author} {\bibfnamefont {M.~H.}\ \bibnamefont {Farag}}, \bibinfo {author} {\bibfnamefont {T.~L.}\ \bibnamefont {Patti}}, \bibinfo {author} {\bibfnamefont {M.~E.}\ \bibnamefont {Wolf}}, \bibinfo {author} {\bibfnamefont {N.}~\bibnamefont {Ares}}, \bibinfo {author} {\bibfnamefont {A.}~\bibnamefont {Aspuru-Guzik}}, \bibinfo {author} {\bibfnamefont {S.~C.}\ \bibnamefont {Benjamin}}, \bibinfo {author} {\bibfnamefont {Z.}~\bibnamefont {Cai}}, \bibinfo {author} {\bibfnamefont {Z.}~\bibnamefont {Chandani}}, \bibinfo {author} {\bibfnamefont {F.}~\bibnamefont {Fedele}}, \emph {et~al.},\ }\bibfield  {title} {\bibinfo {title} {Artificial intelligence for quantum computing},\ }\href@noop {} {\bibfield  {journal} {\bibinfo  {journal} {arXiv preprint arXiv:2411.09131}\ } (\bibinfo {year} {2024})}\BibitemShut {NoStop}%
\bibitem [{\citenamefont {Moon}\ \emph {et~al.}(2020)\citenamefont {Moon}, \citenamefont {Lennon}, \citenamefont {Kirkpatrick}, \citenamefont {van Esbroeck}, \citenamefont {Camenzind}, \citenamefont {Yu}, \citenamefont {Vigneau}, \citenamefont {Zumb{\"u}hl}, \citenamefont {Briggs}, \citenamefont {Osborne} \emph {et~al.}}]{moon2020machine}%
  \BibitemOpen
  \bibfield  {author} {\bibinfo {author} {\bibfnamefont {H.}~\bibnamefont {Moon}}, \bibinfo {author} {\bibfnamefont {D.~T.}\ \bibnamefont {Lennon}}, \bibinfo {author} {\bibfnamefont {J.}~\bibnamefont {Kirkpatrick}}, \bibinfo {author} {\bibfnamefont {N.~M.}\ \bibnamefont {van Esbroeck}}, \bibinfo {author} {\bibfnamefont {L.~C.}\ \bibnamefont {Camenzind}}, \bibinfo {author} {\bibfnamefont {L.}~\bibnamefont {Yu}}, \bibinfo {author} {\bibfnamefont {F.}~\bibnamefont {Vigneau}}, \bibinfo {author} {\bibfnamefont {D.~M.}\ \bibnamefont {Zumb{\"u}hl}}, \bibinfo {author} {\bibfnamefont {G.~A.~D.}\ \bibnamefont {Briggs}}, \bibinfo {author} {\bibfnamefont {M.~A.}\ \bibnamefont {Osborne}}, \emph {et~al.},\ }\bibfield  {title} {\bibinfo {title} {Machine learning enables completely automatic tuning of a quantum device faster than human experts},\ }\href@noop {} {\bibfield  {journal} {\bibinfo  {journal} {Nature communications}\ }\textbf {\bibinfo {volume} {11}},\ \bibinfo {pages} {4161} (\bibinfo {year} {2020})}\BibitemShut
  {NoStop}%
\bibitem [{\citenamefont {Severin}\ \emph {et~al.}(2024)\citenamefont {Severin}, \citenamefont {Lennon}, \citenamefont {Camenzind}, \citenamefont {Vigneau}, \citenamefont {Fedele}, \citenamefont {Jirovec}, \citenamefont {Ballabio}, \citenamefont {Chrastina}, \citenamefont {Isella}, \citenamefont {de~Kruijf}, \citenamefont {Carballido}, \citenamefont {Svab}, \citenamefont {Kuhlmann}, \citenamefont {Geyer}, \citenamefont {Froning}, \citenamefont {Moon}, \citenamefont {Osborne}, \citenamefont {Sejdinovic}, \citenamefont {Katsaros}, \citenamefont {Zumb{\"u}hl}, \citenamefont {Briggs},\ and\ \citenamefont {Ares}}]{severin2024}%
  \BibitemOpen
  \bibfield  {author} {\bibinfo {author} {\bibfnamefont {B.}~\bibnamefont {Severin}}, \bibinfo {author} {\bibfnamefont {D.~T.}\ \bibnamefont {Lennon}}, \bibinfo {author} {\bibfnamefont {L.~C.}\ \bibnamefont {Camenzind}}, \bibinfo {author} {\bibfnamefont {F.}~\bibnamefont {Vigneau}}, \bibinfo {author} {\bibfnamefont {F.}~\bibnamefont {Fedele}}, \bibinfo {author} {\bibfnamefont {D.}~\bibnamefont {Jirovec}}, \bibinfo {author} {\bibfnamefont {A.}~\bibnamefont {Ballabio}}, \bibinfo {author} {\bibfnamefont {D.}~\bibnamefont {Chrastina}}, \bibinfo {author} {\bibfnamefont {G.}~\bibnamefont {Isella}}, \bibinfo {author} {\bibfnamefont {M.}~\bibnamefont {de~Kruijf}}, \bibinfo {author} {\bibfnamefont {M.~J.}\ \bibnamefont {Carballido}}, \bibinfo {author} {\bibfnamefont {S.}~\bibnamefont {Svab}}, \bibinfo {author} {\bibfnamefont {A.~V.}\ \bibnamefont {Kuhlmann}}, \bibinfo {author} {\bibfnamefont {S.}~\bibnamefont {Geyer}}, \bibinfo {author} {\bibfnamefont {F.~N.~M.}\ \bibnamefont {Froning}}, \bibinfo {author}
  {\bibfnamefont {H.}~\bibnamefont {Moon}}, \bibinfo {author} {\bibfnamefont {M.~A.}\ \bibnamefont {Osborne}}, \bibinfo {author} {\bibfnamefont {D.}~\bibnamefont {Sejdinovic}}, \bibinfo {author} {\bibfnamefont {G.}~\bibnamefont {Katsaros}}, \bibinfo {author} {\bibfnamefont {D.~M.}\ \bibnamefont {Zumb{\"u}hl}}, \bibinfo {author} {\bibfnamefont {G.~A.~D.}\ \bibnamefont {Briggs}},\ and\ \bibinfo {author} {\bibfnamefont {N.}~\bibnamefont {Ares}},\ }\bibfield  {title} {\bibinfo {title} {Cross-architecture tuning of silicon and sige-based quantum devices using machine learning},\ }\href@noop {} {\bibfield  {journal} {\bibinfo  {journal} {Scientific Reports}\ }\textbf {\bibinfo {volume} {14}},\ \bibinfo {pages} {17281} (\bibinfo {year} {2024})}\BibitemShut {NoStop}%
\bibitem [{\citenamefont {Schuff}\ \emph {et~al.}(2024)\citenamefont {Schuff}, \citenamefont {Carballido}, \citenamefont {Kotzagiannidis}, \citenamefont {Calvo}, \citenamefont {Caselli}, \citenamefont {Rawling}, \citenamefont {Craig}, \citenamefont {van Straaten}, \citenamefont {Severin}, \citenamefont {Fedele} \emph {et~al.}}]{schuff2024fully}%
  \BibitemOpen
  \bibfield  {author} {\bibinfo {author} {\bibfnamefont {J.}~\bibnamefont {Schuff}}, \bibinfo {author} {\bibfnamefont {M.~J.}\ \bibnamefont {Carballido}}, \bibinfo {author} {\bibfnamefont {M.}~\bibnamefont {Kotzagiannidis}}, \bibinfo {author} {\bibfnamefont {J.~C.}\ \bibnamefont {Calvo}}, \bibinfo {author} {\bibfnamefont {M.}~\bibnamefont {Caselli}}, \bibinfo {author} {\bibfnamefont {J.}~\bibnamefont {Rawling}}, \bibinfo {author} {\bibfnamefont {D.~L.}\ \bibnamefont {Craig}}, \bibinfo {author} {\bibfnamefont {B.}~\bibnamefont {van Straaten}}, \bibinfo {author} {\bibfnamefont {B.}~\bibnamefont {Severin}}, \bibinfo {author} {\bibfnamefont {F.}~\bibnamefont {Fedele}}, \emph {et~al.},\ }\bibfield  {title} {\bibinfo {title} {Fully autonomous tuning of a spin qubit},\ }\href@noop {} {\bibfield  {journal} {\bibinfo  {journal} {arXiv preprint arXiv:2402.03931}\ } (\bibinfo {year} {2024})}\BibitemShut {NoStop}%
\bibitem [{\citenamefont {Lennon}\ \emph {et~al.}(2019)\citenamefont {Lennon}, \citenamefont {Moon}, \citenamefont {Camenzind}, \citenamefont {Yu}, \citenamefont {Zumb{\"u}hl}, \citenamefont {Briggs}, \citenamefont {Osborne}, \citenamefont {Laird},\ and\ \citenamefont {Ares}}]{lennon2019efficiently}%
  \BibitemOpen
  \bibfield  {author} {\bibinfo {author} {\bibfnamefont {D.~T.}\ \bibnamefont {Lennon}}, \bibinfo {author} {\bibfnamefont {H.}~\bibnamefont {Moon}}, \bibinfo {author} {\bibfnamefont {L.~C.}\ \bibnamefont {Camenzind}}, \bibinfo {author} {\bibfnamefont {L.}~\bibnamefont {Yu}}, \bibinfo {author} {\bibfnamefont {D.~M.}\ \bibnamefont {Zumb{\"u}hl}}, \bibinfo {author} {\bibfnamefont {G.~A.~D.}\ \bibnamefont {Briggs}}, \bibinfo {author} {\bibfnamefont {M.~A.}\ \bibnamefont {Osborne}}, \bibinfo {author} {\bibfnamefont {E.~A.}\ \bibnamefont {Laird}},\ and\ \bibinfo {author} {\bibfnamefont {N.}~\bibnamefont {Ares}},\ }\bibfield  {title} {\bibinfo {title} {Efficiently measuring a quantum device using machine learning},\ }\href@noop {} {\bibfield  {journal} {\bibinfo  {journal} {npj Quantum Information}\ }\textbf {\bibinfo {volume} {5}},\ \bibinfo {pages} {79} (\bibinfo {year} {2019})}\BibitemShut {NoStop}%
\bibitem [{\citenamefont {Nguyen}\ \emph {et~al.}(2021)\citenamefont {Nguyen}, \citenamefont {Orbell}, \citenamefont {Lennon}, \citenamefont {Moon}, \citenamefont {Vigneau}, \citenamefont {Camenzind}, \citenamefont {Yu}, \citenamefont {Zumb{\"u}hl}, \citenamefont {Briggs}, \citenamefont {Osborne} \emph {et~al.}}]{nguyen2021deep}%
  \BibitemOpen
  \bibfield  {author} {\bibinfo {author} {\bibfnamefont {V.}~\bibnamefont {Nguyen}}, \bibinfo {author} {\bibfnamefont {S.}~\bibnamefont {Orbell}}, \bibinfo {author} {\bibfnamefont {D.~T.}\ \bibnamefont {Lennon}}, \bibinfo {author} {\bibfnamefont {H.}~\bibnamefont {Moon}}, \bibinfo {author} {\bibfnamefont {F.}~\bibnamefont {Vigneau}}, \bibinfo {author} {\bibfnamefont {L.~C.}\ \bibnamefont {Camenzind}}, \bibinfo {author} {\bibfnamefont {L.}~\bibnamefont {Yu}}, \bibinfo {author} {\bibfnamefont {D.~M.}\ \bibnamefont {Zumb{\"u}hl}}, \bibinfo {author} {\bibfnamefont {G.~A.~D.}\ \bibnamefont {Briggs}}, \bibinfo {author} {\bibfnamefont {M.~A.}\ \bibnamefont {Osborne}}, \emph {et~al.},\ }\bibfield  {title} {\bibinfo {title} {Deep reinforcement learning for efficient measurement of quantum devices},\ }\href@noop {} {\bibfield  {journal} {\bibinfo  {journal} {npj Quantum Information}\ }\textbf {\bibinfo {volume} {7}},\ \bibinfo {pages} {100} (\bibinfo {year} {2021})}\BibitemShut {NoStop}%
\bibitem [{\citenamefont {Vigneau}\ \emph {et~al.}(2023)\citenamefont {Vigneau}, \citenamefont {Fedele}, \citenamefont {Chatterjee}, \citenamefont {Reilly}, \citenamefont {Kuemmeth}, \citenamefont {Gonzalez-Zalba}, \citenamefont {Laird},\ and\ \citenamefont {Ares}}]{vigneau2023probing}%
  \BibitemOpen
  \bibfield  {author} {\bibinfo {author} {\bibfnamefont {F.}~\bibnamefont {Vigneau}}, \bibinfo {author} {\bibfnamefont {F.}~\bibnamefont {Fedele}}, \bibinfo {author} {\bibfnamefont {A.}~\bibnamefont {Chatterjee}}, \bibinfo {author} {\bibfnamefont {D.}~\bibnamefont {Reilly}}, \bibinfo {author} {\bibfnamefont {F.}~\bibnamefont {Kuemmeth}}, \bibinfo {author} {\bibfnamefont {M.~F.}\ \bibnamefont {Gonzalez-Zalba}}, \bibinfo {author} {\bibfnamefont {E.}~\bibnamefont {Laird}},\ and\ \bibinfo {author} {\bibfnamefont {N.}~\bibnamefont {Ares}},\ }\bibfield  {title} {\bibinfo {title} {Probing quantum devices with radio-frequency reflectometry},\ }\href@noop {} {\bibfield  {journal} {\bibinfo  {journal} {Applied Physics Reviews}\ }\textbf {\bibinfo {volume} {10}} (\bibinfo {year} {2023})}\BibitemShut {NoStop}%
\bibitem [{\citenamefont {van Straaten}\ \emph {et~al.}(2022)\citenamefont {van Straaten}, \citenamefont {Fedele}, \citenamefont {Vigneau}, \citenamefont {Hickie}, \citenamefont {Jirovec}, \citenamefont {Ballabio}, \citenamefont {Chrastina}, \citenamefont {Isella}, \citenamefont {Katsaros},\ and\ \citenamefont {Ares}}]{van2022all}%
  \BibitemOpen
  \bibfield  {author} {\bibinfo {author} {\bibfnamefont {B.}~\bibnamefont {van Straaten}}, \bibinfo {author} {\bibfnamefont {F.}~\bibnamefont {Fedele}}, \bibinfo {author} {\bibfnamefont {F.}~\bibnamefont {Vigneau}}, \bibinfo {author} {\bibfnamefont {J.}~\bibnamefont {Hickie}}, \bibinfo {author} {\bibfnamefont {D.}~\bibnamefont {Jirovec}}, \bibinfo {author} {\bibfnamefont {A.}~\bibnamefont {Ballabio}}, \bibinfo {author} {\bibfnamefont {D.}~\bibnamefont {Chrastina}}, \bibinfo {author} {\bibfnamefont {G.}~\bibnamefont {Isella}}, \bibinfo {author} {\bibfnamefont {G.}~\bibnamefont {Katsaros}},\ and\ \bibinfo {author} {\bibfnamefont {N.}~\bibnamefont {Ares}},\ }\bibfield  {title} {\bibinfo {title} {All rf-based tuning algorithm for quantum devices using machine learning},\ }\href@noop {} {\bibfield  {journal} {\bibinfo  {journal} {arXiv preprint arXiv:2211.04504}\ } (\bibinfo {year} {2022})}\BibitemShut {NoStop}%
\bibitem [{\citenamefont {Hickie}\ \emph {et~al.}(2024)\citenamefont {Hickie}, \citenamefont {van Straaten}, \citenamefont {Fedele}, \citenamefont {Jirovec}, \citenamefont {Ballabio}, \citenamefont {Chrastina}, \citenamefont {Isella}, \citenamefont {Katsaros},\ and\ \citenamefont {Ares}}]{hickie2024}%
  \BibitemOpen
  \bibfield  {author} {\bibinfo {author} {\bibfnamefont {J.}~\bibnamefont {Hickie}}, \bibinfo {author} {\bibfnamefont {B.}~\bibnamefont {van Straaten}}, \bibinfo {author} {\bibfnamefont {F.}~\bibnamefont {Fedele}}, \bibinfo {author} {\bibfnamefont {D.}~\bibnamefont {Jirovec}}, \bibinfo {author} {\bibfnamefont {A.}~\bibnamefont {Ballabio}}, \bibinfo {author} {\bibfnamefont {D.}~\bibnamefont {Chrastina}}, \bibinfo {author} {\bibfnamefont {G.}~\bibnamefont {Isella}}, \bibinfo {author} {\bibfnamefont {G.}~\bibnamefont {Katsaros}},\ and\ \bibinfo {author} {\bibfnamefont {N.}~\bibnamefont {Ares}},\ }\bibfield  {title} {\bibinfo {title} {Automated long-range compensation of an rf quantum dot sensor},\ }\href@noop {} {\bibfield  {journal} {\bibinfo  {journal} {Physical Review Applied}\ }\textbf {\bibinfo {volume} {22}},\ \bibinfo {pages} {064026} (\bibinfo {year} {2024})}\BibitemShut {NoStop}%
\bibitem [{\citenamefont {Rao}\ \emph {et~al.}(2024)\citenamefont {Rao}, \citenamefont {Buterakos}, \citenamefont {van Straaten}, \citenamefont {John}, \citenamefont {Yu}, \citenamefont {Oosterhout}, \citenamefont {Stehouwer}, \citenamefont {Scappucci}, \citenamefont {Veldhorst}, \citenamefont {Borsoi} \emph {et~al.}}]{rao2024mavis}%
  \BibitemOpen
  \bibfield  {author} {\bibinfo {author} {\bibfnamefont {A.~S.}\ \bibnamefont {Rao}}, \bibinfo {author} {\bibfnamefont {D.}~\bibnamefont {Buterakos}}, \bibinfo {author} {\bibfnamefont {B.}~\bibnamefont {van Straaten}}, \bibinfo {author} {\bibfnamefont {V.}~\bibnamefont {John}}, \bibinfo {author} {\bibfnamefont {C.~X.}\ \bibnamefont {Yu}}, \bibinfo {author} {\bibfnamefont {S.~D.}\ \bibnamefont {Oosterhout}}, \bibinfo {author} {\bibfnamefont {L.}~\bibnamefont {Stehouwer}}, \bibinfo {author} {\bibfnamefont {G.}~\bibnamefont {Scappucci}}, \bibinfo {author} {\bibfnamefont {M.}~\bibnamefont {Veldhorst}}, \bibinfo {author} {\bibfnamefont {F.}~\bibnamefont {Borsoi}}, \emph {et~al.},\ }\bibfield  {title} {\bibinfo {title} {Mavis: Modular autonomous virtualization system for two-dimensional semiconductor quantum dot arrays},\ }\href@noop {} {\bibfield  {journal} {\bibinfo  {journal} {arXiv preprint arXiv:2411.12516}\ } (\bibinfo {year} {2024})}\BibitemShut {NoStop}%
\bibitem [{\citenamefont {Katiraee-Far}\ \emph {et~al.}(2025)\citenamefont {Katiraee-Far}, \citenamefont {Matsumoto}, \citenamefont {Undseth}, \citenamefont {De~Smet}, \citenamefont {Gualtieri}, \citenamefont {Meinersen}, \citenamefont {de~Fuentes}, \citenamefont {Capannelli}, \citenamefont {Rimbach-Russ}, \citenamefont {Scappucci} \emph {et~al.}}]{katiraee2025unified}%
  \BibitemOpen
  \bibfield  {author} {\bibinfo {author} {\bibfnamefont {S.~R.}\ \bibnamefont {Katiraee-Far}}, \bibinfo {author} {\bibfnamefont {Y.}~\bibnamefont {Matsumoto}}, \bibinfo {author} {\bibfnamefont {B.}~\bibnamefont {Undseth}}, \bibinfo {author} {\bibfnamefont {M.}~\bibnamefont {De~Smet}}, \bibinfo {author} {\bibfnamefont {V.}~\bibnamefont {Gualtieri}}, \bibinfo {author} {\bibfnamefont {C.~V.}\ \bibnamefont {Meinersen}}, \bibinfo {author} {\bibfnamefont {I.~F.}\ \bibnamefont {de~Fuentes}}, \bibinfo {author} {\bibfnamefont {K.}~\bibnamefont {Capannelli}}, \bibinfo {author} {\bibfnamefont {M.}~\bibnamefont {Rimbach-Russ}}, \bibinfo {author} {\bibfnamefont {G.}~\bibnamefont {Scappucci}}, \emph {et~al.},\ }\bibfield  {title} {\bibinfo {title} {Unified evolutionary optimization for high-fidelity spin qubit operations},\ }\href@noop {} {\bibfield  {journal} {\bibinfo  {journal} {arXiv preprint arXiv:2503.12256}\ } (\bibinfo {year} {2025})}\BibitemShut {NoStop}%
\bibitem [{\citenamefont {Berritta}\ \emph {et~al.}(2024)\citenamefont {Berritta}, \citenamefont {Rasmussen}, \citenamefont {Krzywda}, \citenamefont {Van Der~Heijden}, \citenamefont {Fedele}, \citenamefont {Fallahi}, \citenamefont {Gardner}, \citenamefont {Manfra}, \citenamefont {Van~Nieuwenburg}, \citenamefont {Danon} \emph {et~al.}}]{berritta2024real}%
  \BibitemOpen
  \bibfield  {author} {\bibinfo {author} {\bibfnamefont {F.}~\bibnamefont {Berritta}}, \bibinfo {author} {\bibfnamefont {T.}~\bibnamefont {Rasmussen}}, \bibinfo {author} {\bibfnamefont {J.~A.}\ \bibnamefont {Krzywda}}, \bibinfo {author} {\bibfnamefont {J.}~\bibnamefont {Van Der~Heijden}}, \bibinfo {author} {\bibfnamefont {F.}~\bibnamefont {Fedele}}, \bibinfo {author} {\bibfnamefont {S.}~\bibnamefont {Fallahi}}, \bibinfo {author} {\bibfnamefont {G.~C.}\ \bibnamefont {Gardner}}, \bibinfo {author} {\bibfnamefont {M.~J.}\ \bibnamefont {Manfra}}, \bibinfo {author} {\bibfnamefont {E.}~\bibnamefont {Van~Nieuwenburg}}, \bibinfo {author} {\bibfnamefont {J.}~\bibnamefont {Danon}}, \emph {et~al.},\ }\bibfield  {title} {\bibinfo {title} {Real-time two-axis control of a spin qubit},\ }\href@noop {} {\bibfield  {journal} {\bibinfo  {journal} {Nature Communications}\ }\textbf {\bibinfo {volume} {15}},\ \bibinfo {pages} {1676} (\bibinfo {year} {2024})}\BibitemShut {NoStop}%
\bibitem [{\citenamefont {Saez-Mollejo}\ \emph {et~al.}(2025)\citenamefont {Saez-Mollejo}, \citenamefont {Jirovec}, \citenamefont {Schell}, \citenamefont {Kukucka}, \citenamefont {Calcaterra}, \citenamefont {Chrastina}, \citenamefont {Isella}, \citenamefont {Rimbach-Russ}, \citenamefont {Bosco},\ and\ \citenamefont {Katsaros}}]{saez-mollejo_exchange_2025}%
  \BibitemOpen
  \bibfield  {author} {\bibinfo {author} {\bibfnamefont {J.}~\bibnamefont {Saez-Mollejo}}, \bibinfo {author} {\bibfnamefont {D.}~\bibnamefont {Jirovec}}, \bibinfo {author} {\bibfnamefont {Y.}~\bibnamefont {Schell}}, \bibinfo {author} {\bibfnamefont {J.}~\bibnamefont {Kukucka}}, \bibinfo {author} {\bibfnamefont {S.}~\bibnamefont {Calcaterra}}, \bibinfo {author} {\bibfnamefont {D.}~\bibnamefont {Chrastina}}, \bibinfo {author} {\bibfnamefont {G.}~\bibnamefont {Isella}}, \bibinfo {author} {\bibfnamefont {M.}~\bibnamefont {Rimbach-Russ}}, \bibinfo {author} {\bibfnamefont {S.}~\bibnamefont {Bosco}},\ and\ \bibinfo {author} {\bibfnamefont {G.}~\bibnamefont {Katsaros}},\ }\bibfield  {title} {\bibinfo {title} {Exchange anisotropies in microwave-driven singlet-triplet qubits},\ }\href {https://doi.org/10.1038/s41467-025-58969-y} {\bibfield  {journal} {\bibinfo  {journal} {Nature Communications}\ }\textbf {\bibinfo {volume} {16}},\ \bibinfo {pages} {3862} (\bibinfo {year} {2025})}\BibitemShut {NoStop}%
\bibitem [{\citenamefont {Lundberg}\ \emph {et~al.}(2024)\citenamefont {Lundberg}, \citenamefont {Ibberson}, \citenamefont {Li}, \citenamefont {Hutin}, \citenamefont {Abadillo-Uriel}, \citenamefont {Filippone}, \citenamefont {Bertrand}, \citenamefont {Nunnenkamp}, \citenamefont {Lee}, \citenamefont {Stelmashenko} \emph {et~al.}}]{lundberg2024non}%
  \BibitemOpen
  \bibfield  {author} {\bibinfo {author} {\bibfnamefont {T.}~\bibnamefont {Lundberg}}, \bibinfo {author} {\bibfnamefont {D.~J.}\ \bibnamefont {Ibberson}}, \bibinfo {author} {\bibfnamefont {J.}~\bibnamefont {Li}}, \bibinfo {author} {\bibfnamefont {L.}~\bibnamefont {Hutin}}, \bibinfo {author} {\bibfnamefont {J.~C.}\ \bibnamefont {Abadillo-Uriel}}, \bibinfo {author} {\bibfnamefont {M.}~\bibnamefont {Filippone}}, \bibinfo {author} {\bibfnamefont {B.}~\bibnamefont {Bertrand}}, \bibinfo {author} {\bibfnamefont {A.}~\bibnamefont {Nunnenkamp}}, \bibinfo {author} {\bibfnamefont {C.-M.}\ \bibnamefont {Lee}}, \bibinfo {author} {\bibfnamefont {N.}~\bibnamefont {Stelmashenko}}, \emph {et~al.},\ }\bibfield  {title} {\bibinfo {title} {Non-symmetric pauli spin blockade in a silicon double quantum dot},\ }\href@noop {} {\bibfield  {journal} {\bibinfo  {journal} {npj Quantum Information}\ }\textbf {\bibinfo {volume} {10}},\ \bibinfo {pages} {28} (\bibinfo {year} {2024})}\BibitemShut {NoStop}%
\bibitem [{\citenamefont {Sen}\ \emph {et~al.}(2023)\citenamefont {Sen}, \citenamefont {Frank}, \citenamefont {Kolok}, \citenamefont {Danon},\ and\ \citenamefont {P{\'a}lyi}}]{sen2023classification}%
  \BibitemOpen
  \bibfield  {author} {\bibinfo {author} {\bibfnamefont {A.}~\bibnamefont {Sen}}, \bibinfo {author} {\bibfnamefont {G.}~\bibnamefont {Frank}}, \bibinfo {author} {\bibfnamefont {B.}~\bibnamefont {Kolok}}, \bibinfo {author} {\bibfnamefont {J.}~\bibnamefont {Danon}},\ and\ \bibinfo {author} {\bibfnamefont {A.}~\bibnamefont {P{\'a}lyi}},\ }\bibfield  {title} {\bibinfo {title} {Classification and magic magnetic field directions for spin-orbit-coupled double quantum dots},\ }\href@noop {} {\bibfield  {journal} {\bibinfo  {journal} {Physical Review B}\ }\textbf {\bibinfo {volume} {108}},\ \bibinfo {pages} {245406} (\bibinfo {year} {2023})}\BibitemShut {NoStop}%
\bibitem [{\citenamefont {Mutter}\ and\ \citenamefont {Burkard}(2020)}]{mutter2020g}%
  \BibitemOpen
  \bibfield  {author} {\bibinfo {author} {\bibfnamefont {P.~M.}\ \bibnamefont {Mutter}}\ and\ \bibinfo {author} {\bibfnamefont {G.}~\bibnamefont {Burkard}},\ }\bibfield  {title} {\bibinfo {title} {g-tensor resonance in double quantum dots with site-dependent g-tensors},\ }\href@noop {} {\bibfield  {journal} {\bibinfo  {journal} {Materials for Quantum Technology}\ }\textbf {\bibinfo {volume} {1}},\ \bibinfo {pages} {015003} (\bibinfo {year} {2020})}\BibitemShut {NoStop}%
\bibitem [{\citenamefont {Hendrickx}\ \emph {et~al.}(2024)\citenamefont {Hendrickx}, \citenamefont {Massai}, \citenamefont {Mergenthaler}, \citenamefont {Schupp}, \citenamefont {Paredes}, \citenamefont {Bedell}, \citenamefont {Salis},\ and\ \citenamefont {Fuhrer}}]{hendrickx2024sweet}%
  \BibitemOpen
  \bibfield  {author} {\bibinfo {author} {\bibfnamefont {N.}~\bibnamefont {Hendrickx}}, \bibinfo {author} {\bibfnamefont {L.}~\bibnamefont {Massai}}, \bibinfo {author} {\bibfnamefont {M.}~\bibnamefont {Mergenthaler}}, \bibinfo {author} {\bibfnamefont {F.~J.}\ \bibnamefont {Schupp}}, \bibinfo {author} {\bibfnamefont {S.}~\bibnamefont {Paredes}}, \bibinfo {author} {\bibfnamefont {S.}~\bibnamefont {Bedell}}, \bibinfo {author} {\bibfnamefont {G.}~\bibnamefont {Salis}},\ and\ \bibinfo {author} {\bibfnamefont {A.}~\bibnamefont {Fuhrer}},\ }\bibfield  {title} {\bibinfo {title} {Sweet-spot operation of a germanium hole spin qubit with highly anisotropic noise sensitivity},\ }\href@noop {} {\bibfield  {journal} {\bibinfo  {journal} {Nature Materials}\ }\textbf {\bibinfo {volume} {23}},\ \bibinfo {pages} {920} (\bibinfo {year} {2024})}\BibitemShut {NoStop}%
\bibitem [{\citenamefont {Jin}\ \emph {et~al.}(2024)\citenamefont {Jin}, \citenamefont {Hillier}, \citenamefont {Liles}, \citenamefont {Wang}, \citenamefont {Shamim}, \citenamefont {Vorreiter}, \citenamefont {Li}, \citenamefont {Godfrin}, \citenamefont {Kubicek}, \citenamefont {De~Greve} \emph {et~al.}}]{jin2024probing}%
  \BibitemOpen
  \bibfield  {author} {\bibinfo {author} {\bibfnamefont {I.~K.}\ \bibnamefont {Jin}}, \bibinfo {author} {\bibfnamefont {J.}~\bibnamefont {Hillier}}, \bibinfo {author} {\bibfnamefont {S.~D.}\ \bibnamefont {Liles}}, \bibinfo {author} {\bibfnamefont {Z.}~\bibnamefont {Wang}}, \bibinfo {author} {\bibfnamefont {A.}~\bibnamefont {Shamim}}, \bibinfo {author} {\bibfnamefont {I.}~\bibnamefont {Vorreiter}}, \bibinfo {author} {\bibfnamefont {R.}~\bibnamefont {Li}}, \bibinfo {author} {\bibfnamefont {C.}~\bibnamefont {Godfrin}}, \bibinfo {author} {\bibfnamefont {S.}~\bibnamefont {Kubicek}}, \bibinfo {author} {\bibfnamefont {K.}~\bibnamefont {De~Greve}}, \emph {et~al.},\ }\bibfield  {title} {\bibinfo {title} {Probing g-tensor reproducibility and spin-orbit effects in planar silicon hole quantum dots},\ }\href@noop {} {\bibfield  {journal} {\bibinfo  {journal} {arXiv preprint arXiv:2411.06016}\ } (\bibinfo {year} {2024})}\BibitemShut {NoStop}%
\bibitem [{\citenamefont {Schuff}\ \emph {et~al.}(2023)\citenamefont {Schuff}, \citenamefont {Lennon}, \citenamefont {Geyer}, \citenamefont {Craig}, \citenamefont {Fedele}, \citenamefont {Vigneau}, \citenamefont {Camenzind}, \citenamefont {Kuhlmann}, \citenamefont {Briggs}, \citenamefont {Zumb{\"u}hl} \emph {et~al.}}]{schuff2023identifying}%
  \BibitemOpen
  \bibfield  {author} {\bibinfo {author} {\bibfnamefont {J.}~\bibnamefont {Schuff}}, \bibinfo {author} {\bibfnamefont {D.~T.}\ \bibnamefont {Lennon}}, \bibinfo {author} {\bibfnamefont {S.}~\bibnamefont {Geyer}}, \bibinfo {author} {\bibfnamefont {D.~L.}\ \bibnamefont {Craig}}, \bibinfo {author} {\bibfnamefont {F.}~\bibnamefont {Fedele}}, \bibinfo {author} {\bibfnamefont {F.}~\bibnamefont {Vigneau}}, \bibinfo {author} {\bibfnamefont {L.~C.}\ \bibnamefont {Camenzind}}, \bibinfo {author} {\bibfnamefont {A.~V.}\ \bibnamefont {Kuhlmann}}, \bibinfo {author} {\bibfnamefont {G.~A.~D.}\ \bibnamefont {Briggs}}, \bibinfo {author} {\bibfnamefont {D.~M.}\ \bibnamefont {Zumb{\"u}hl}}, \emph {et~al.},\ }\bibfield  {title} {\bibinfo {title} {Identifying pauli spin blockade using deep learning},\ }\href@noop {} {\bibfield  {journal} {\bibinfo  {journal} {Quantum}\ }\textbf {\bibinfo {volume} {7}},\ \bibinfo {pages} {1077} (\bibinfo {year} {2023})}\BibitemShut {NoStop}%
\bibitem [{\citenamefont {Petta}\ \emph {et~al.}(2005)\citenamefont {Petta}, \citenamefont {Johnson}, \citenamefont {Yacoby}, \citenamefont {Marcus}, \citenamefont {Hanson},\ and\ \citenamefont {Gossard}}]{petta2005pulsed}%
  \BibitemOpen
  \bibfield  {author} {\bibinfo {author} {\bibfnamefont {J.}~\bibnamefont {Petta}}, \bibinfo {author} {\bibfnamefont {A.}~\bibnamefont {Johnson}}, \bibinfo {author} {\bibfnamefont {A.}~\bibnamefont {Yacoby}}, \bibinfo {author} {\bibfnamefont {C.}~\bibnamefont {Marcus}}, \bibinfo {author} {\bibfnamefont {M.}~\bibnamefont {Hanson}},\ and\ \bibinfo {author} {\bibfnamefont {A.}~\bibnamefont {Gossard}},\ }\bibfield  {title} {\bibinfo {title} {Pulsed-gate measurements of the singlet-triplet relaxation time in a two-electron double quantum dot},\ }\href@noop {} {\bibfield  {journal} {\bibinfo  {journal} {Physical Review B—Condensed Matter and Materials Physics}\ }\textbf {\bibinfo {volume} {72}},\ \bibinfo {pages} {161301} (\bibinfo {year} {2005})}\BibitemShut {NoStop}%
\bibitem [{\citenamefont {Lapointe-Major}\ \emph {et~al.}(2020)\citenamefont {Lapointe-Major}, \citenamefont {Germain}, \citenamefont {Camirand~Lemyre}, \citenamefont {Lachance-Quirion}, \citenamefont {Rochette}, \citenamefont {Camirand~Lemyre},\ and\ \citenamefont {Pioro-Ladri{\`e}re}}]{lapointe2020algorithm}%
  \BibitemOpen
  \bibfield  {author} {\bibinfo {author} {\bibfnamefont {M.}~\bibnamefont {Lapointe-Major}}, \bibinfo {author} {\bibfnamefont {O.}~\bibnamefont {Germain}}, \bibinfo {author} {\bibfnamefont {J.}~\bibnamefont {Camirand~Lemyre}}, \bibinfo {author} {\bibfnamefont {D.}~\bibnamefont {Lachance-Quirion}}, \bibinfo {author} {\bibfnamefont {S.}~\bibnamefont {Rochette}}, \bibinfo {author} {\bibfnamefont {F.}~\bibnamefont {Camirand~Lemyre}},\ and\ \bibinfo {author} {\bibfnamefont {M.}~\bibnamefont {Pioro-Ladri{\`e}re}},\ }\bibfield  {title} {\bibinfo {title} {Algorithm for automated tuning of a quantum dot into the single-electron regime},\ }\href@noop {} {\bibfield  {journal} {\bibinfo  {journal} {Physical Review B}\ }\textbf {\bibinfo {volume} {102}},\ \bibinfo {pages} {085301} (\bibinfo {year} {2020})}\BibitemShut {NoStop}%
\bibitem [{\citenamefont {Mills}\ \emph {et~al.}(2019)\citenamefont {Mills}, \citenamefont {Feldman}, \citenamefont {Monical}, \citenamefont {Lewis}, \citenamefont {Larson}, \citenamefont {Mounce},\ and\ \citenamefont {Petta}}]{mills2019computer}%
  \BibitemOpen
  \bibfield  {author} {\bibinfo {author} {\bibfnamefont {A.}~\bibnamefont {Mills}}, \bibinfo {author} {\bibfnamefont {M.}~\bibnamefont {Feldman}}, \bibinfo {author} {\bibfnamefont {C.}~\bibnamefont {Monical}}, \bibinfo {author} {\bibfnamefont {P.}~\bibnamefont {Lewis}}, \bibinfo {author} {\bibfnamefont {K.}~\bibnamefont {Larson}}, \bibinfo {author} {\bibfnamefont {A.}~\bibnamefont {Mounce}},\ and\ \bibinfo {author} {\bibfnamefont {J.~R.}\ \bibnamefont {Petta}},\ }\bibfield  {title} {\bibinfo {title} {Computer-automated tuning procedures for semiconductor quantum dot arrays},\ }\href@noop {} {\bibfield  {journal} {\bibinfo  {journal} {Applied Physics Letters}\ }\textbf {\bibinfo {volume} {115}} (\bibinfo {year} {2019})}\BibitemShut {NoStop}%
\bibitem [{\citenamefont {van Straaten}\ \emph {et~al.}(2024{\natexlab{a}})\citenamefont {van Straaten}, \citenamefont {Hickie}, \citenamefont {Schorling}, \citenamefont {Schuff}, \citenamefont {Fedele},\ and\ \citenamefont {Ares}}]{van2024qarray}%
  \BibitemOpen
  \bibfield  {author} {\bibinfo {author} {\bibfnamefont {B.}~\bibnamefont {van Straaten}}, \bibinfo {author} {\bibfnamefont {J.}~\bibnamefont {Hickie}}, \bibinfo {author} {\bibfnamefont {L.}~\bibnamefont {Schorling}}, \bibinfo {author} {\bibfnamefont {J.}~\bibnamefont {Schuff}}, \bibinfo {author} {\bibfnamefont {F.}~\bibnamefont {Fedele}},\ and\ \bibinfo {author} {\bibfnamefont {N.}~\bibnamefont {Ares}},\ }\bibfield  {title} {\bibinfo {title} {{QArray: A GPU-accelerated constant capacitance model simulator for large quantum dot arrays}},\ }\href {https://doi.org/10.21468/SciPostPhysCodeb.35} {\bibfield  {journal} {\bibinfo  {journal} {SciPost Phys. Codebases}\ ,\ \bibinfo {pages} {35}} (\bibinfo {year} {2024}{\natexlab{a}})}\BibitemShut {NoStop}%
\bibitem [{\citenamefont {van Straaten}\ \emph {et~al.}(2024{\natexlab{b}})\citenamefont {van Straaten}, \citenamefont {Hickie}, \citenamefont {Schorling}, \citenamefont {Schuff}, \citenamefont {Fedele},\ and\ \citenamefont {Ares}}]{van2024qarraycodebase}%
  \BibitemOpen
  \bibfield  {author} {\bibinfo {author} {\bibfnamefont {B.}~\bibnamefont {van Straaten}}, \bibinfo {author} {\bibfnamefont {J.}~\bibnamefont {Hickie}}, \bibinfo {author} {\bibfnamefont {L.}~\bibnamefont {Schorling}}, \bibinfo {author} {\bibfnamefont {J.}~\bibnamefont {Schuff}}, \bibinfo {author} {\bibfnamefont {F.}~\bibnamefont {Fedele}},\ and\ \bibinfo {author} {\bibfnamefont {N.}~\bibnamefont {Ares}},\ }\bibfield  {title} {\bibinfo {title} {{Codebase release 1.3 for QArray}},\ }\href {https://doi.org/10.21468/SciPostPhysCodeb.35-r1.3} {\bibfield  {journal} {\bibinfo  {journal} {SciPost Phys. Codebases}\ ,\ \bibinfo {pages} {35}} (\bibinfo {year} {2024}{\natexlab{b}})}\BibitemShut {NoStop}%
\bibitem [{\citenamefont {Liu}\ \emph {et~al.}(2022)\citenamefont {Liu}, \citenamefont {Wang}, \citenamefont {Wang}, \citenamefont {Sun}, \citenamefont {Yin}, \citenamefont {Li}, \citenamefont {Cao},\ and\ \citenamefont {Guo}}]{liu2022automated}%
  \BibitemOpen
  \bibfield  {author} {\bibinfo {author} {\bibfnamefont {H.}~\bibnamefont {Liu}}, \bibinfo {author} {\bibfnamefont {B.}~\bibnamefont {Wang}}, \bibinfo {author} {\bibfnamefont {N.}~\bibnamefont {Wang}}, \bibinfo {author} {\bibfnamefont {Z.}~\bibnamefont {Sun}}, \bibinfo {author} {\bibfnamefont {H.}~\bibnamefont {Yin}}, \bibinfo {author} {\bibfnamefont {H.}~\bibnamefont {Li}}, \bibinfo {author} {\bibfnamefont {G.}~\bibnamefont {Cao}},\ and\ \bibinfo {author} {\bibfnamefont {G.}~\bibnamefont {Guo}},\ }\bibfield  {title} {\bibinfo {title} {An automated approach for consecutive tuning of quantum dot arrays},\ }\href@noop {} {\bibfield  {journal} {\bibinfo  {journal} {Applied Physics Letters}\ }\textbf {\bibinfo {volume} {121}} (\bibinfo {year} {2022})}\BibitemShut {NoStop}%
\bibitem [{\citenamefont {Van~Riggelen}\ \emph {et~al.}(2021)\citenamefont {Van~Riggelen}, \citenamefont {Hendrickx}, \citenamefont {Lawrie}, \citenamefont {Russ}, \citenamefont {Sammak}, \citenamefont {Scappucci},\ and\ \citenamefont {Veldhorst}}]{van2021two}%
  \BibitemOpen
  \bibfield  {author} {\bibinfo {author} {\bibfnamefont {F.}~\bibnamefont {Van~Riggelen}}, \bibinfo {author} {\bibfnamefont {N.}~\bibnamefont {Hendrickx}}, \bibinfo {author} {\bibfnamefont {W.}~\bibnamefont {Lawrie}}, \bibinfo {author} {\bibfnamefont {M.}~\bibnamefont {Russ}}, \bibinfo {author} {\bibfnamefont {A.}~\bibnamefont {Sammak}}, \bibinfo {author} {\bibfnamefont {G.}~\bibnamefont {Scappucci}},\ and\ \bibinfo {author} {\bibfnamefont {M.}~\bibnamefont {Veldhorst}},\ }\bibfield  {title} {\bibinfo {title} {A two-dimensional array of single-hole quantum dots},\ }\href@noop {} {\bibfield  {journal} {\bibinfo  {journal} {Applied Physics Letters}\ }\textbf {\bibinfo {volume} {118}} (\bibinfo {year} {2021})}\BibitemShut {NoStop}%
\bibitem [{\citenamefont {Craig}\ \emph {et~al.}(2024)\citenamefont {Craig}, \citenamefont {Moon}, \citenamefont {Fedele}, \citenamefont {Lennon}, \citenamefont {van Straaten}, \citenamefont {Vigneau}, \citenamefont {Camenzind}, \citenamefont {Zumb{\"u}hl}, \citenamefont {Briggs}, \citenamefont {Osborne} \emph {et~al.}}]{craig2024bridging}%
  \BibitemOpen
  \bibfield  {author} {\bibinfo {author} {\bibfnamefont {D.~L.}\ \bibnamefont {Craig}}, \bibinfo {author} {\bibfnamefont {H.}~\bibnamefont {Moon}}, \bibinfo {author} {\bibfnamefont {F.}~\bibnamefont {Fedele}}, \bibinfo {author} {\bibfnamefont {D.~T.}\ \bibnamefont {Lennon}}, \bibinfo {author} {\bibfnamefont {B.}~\bibnamefont {van Straaten}}, \bibinfo {author} {\bibfnamefont {F.}~\bibnamefont {Vigneau}}, \bibinfo {author} {\bibfnamefont {L.~C.}\ \bibnamefont {Camenzind}}, \bibinfo {author} {\bibfnamefont {D.~M.}\ \bibnamefont {Zumb{\"u}hl}}, \bibinfo {author} {\bibfnamefont {G.~A.~D.}\ \bibnamefont {Briggs}}, \bibinfo {author} {\bibfnamefont {M.~A.}\ \bibnamefont {Osborne}}, \emph {et~al.},\ }\bibfield  {title} {\bibinfo {title} {Bridging the reality gap in quantum devices with physics-aware machine learning},\ }\href@noop {} {\bibfield  {journal} {\bibinfo  {journal} {Physical Review X}\ }\textbf {\bibinfo {volume} {14}},\ \bibinfo {pages} {011001} (\bibinfo {year} {2024})}\BibitemShut {NoStop}%
\bibitem [{\citenamefont {Pedregosa}\ \emph {et~al.}(2011)\citenamefont {Pedregosa}, \citenamefont {Varoquaux}, \citenamefont {Gramfort}, \citenamefont {Michel}, \citenamefont {Thirion}, \citenamefont {Grisel}, \citenamefont {Blondel}, \citenamefont {Prettenhofer}, \citenamefont {Weiss}, \citenamefont {Dubourg}, \citenamefont {Vanderplas}, \citenamefont {Passos}, \citenamefont {Cournapeau}, \citenamefont {Brucher}, \citenamefont {Perrot},\ and\ \citenamefont {Duchesnay}}]{scikit-learn}%
  \BibitemOpen
  \bibfield  {author} {\bibinfo {author} {\bibfnamefont {F.}~\bibnamefont {Pedregosa}}, \bibinfo {author} {\bibfnamefont {G.}~\bibnamefont {Varoquaux}}, \bibinfo {author} {\bibfnamefont {A.}~\bibnamefont {Gramfort}}, \bibinfo {author} {\bibfnamefont {V.}~\bibnamefont {Michel}}, \bibinfo {author} {\bibfnamefont {B.}~\bibnamefont {Thirion}}, \bibinfo {author} {\bibfnamefont {O.}~\bibnamefont {Grisel}}, \bibinfo {author} {\bibfnamefont {M.}~\bibnamefont {Blondel}}, \bibinfo {author} {\bibfnamefont {P.}~\bibnamefont {Prettenhofer}}, \bibinfo {author} {\bibfnamefont {R.}~\bibnamefont {Weiss}}, \bibinfo {author} {\bibfnamefont {V.}~\bibnamefont {Dubourg}}, \bibinfo {author} {\bibfnamefont {J.}~\bibnamefont {Vanderplas}}, \bibinfo {author} {\bibfnamefont {A.}~\bibnamefont {Passos}}, \bibinfo {author} {\bibfnamefont {D.}~\bibnamefont {Cournapeau}}, \bibinfo {author} {\bibfnamefont {M.}~\bibnamefont {Brucher}}, \bibinfo {author} {\bibfnamefont {M.}~\bibnamefont {Perrot}},\ and\ \bibinfo {author} {\bibfnamefont
  {E.}~\bibnamefont {Duchesnay}},\ }\bibfield  {title} {\bibinfo {title} {Scikit-learn: Machine learning in {P}ython},\ }\href@noop {} {\bibfield  {journal} {\bibinfo  {journal} {Journal of Machine Learning Research}\ }\textbf {\bibinfo {volume} {12}},\ \bibinfo {pages} {2825} (\bibinfo {year} {2011})}\BibitemShut {NoStop}%
\bibitem [{\citenamefont {Muto}\ \emph {et~al.}(2024)\citenamefont {Muto}, \citenamefont {Nakaso}, \citenamefont {Shinozaki}, \citenamefont {Aizawa}, \citenamefont {Kitada}, \citenamefont {Nakajima}, \citenamefont {Delbecq}, \citenamefont {Yoneda}, \citenamefont {Takeda}, \citenamefont {Noiri} \emph {et~al.}}]{muto2024visual}%
  \BibitemOpen
  \bibfield  {author} {\bibinfo {author} {\bibfnamefont {Y.}~\bibnamefont {Muto}}, \bibinfo {author} {\bibfnamefont {T.}~\bibnamefont {Nakaso}}, \bibinfo {author} {\bibfnamefont {M.}~\bibnamefont {Shinozaki}}, \bibinfo {author} {\bibfnamefont {T.}~\bibnamefont {Aizawa}}, \bibinfo {author} {\bibfnamefont {T.}~\bibnamefont {Kitada}}, \bibinfo {author} {\bibfnamefont {T.}~\bibnamefont {Nakajima}}, \bibinfo {author} {\bibfnamefont {M.~R.}\ \bibnamefont {Delbecq}}, \bibinfo {author} {\bibfnamefont {J.}~\bibnamefont {Yoneda}}, \bibinfo {author} {\bibfnamefont {K.}~\bibnamefont {Takeda}}, \bibinfo {author} {\bibfnamefont {A.}~\bibnamefont {Noiri}}, \emph {et~al.},\ }\bibfield  {title} {\bibinfo {title} {Visual explanations of machine learning model estimating charge states in quantum dots},\ }\href@noop {} {\bibfield  {journal} {\bibinfo  {journal} {APL Machine Learning}\ }\textbf {\bibinfo {volume} {2}} (\bibinfo {year} {2024})}\BibitemShut {NoStop}%
\bibitem [{\citenamefont {Jirovec}\ \emph {et~al.}(2021)\citenamefont {Jirovec}, \citenamefont {Hofmann}, \citenamefont {Ballabio}, \citenamefont {Mutter}, \citenamefont {Tavani}, \citenamefont {Botifoll}, \citenamefont {Crippa}, \citenamefont {Kukucka}, \citenamefont {Sagi}, \citenamefont {Martins} \emph {et~al.}}]{jirovec2021singlet}%
  \BibitemOpen
  \bibfield  {author} {\bibinfo {author} {\bibfnamefont {D.}~\bibnamefont {Jirovec}}, \bibinfo {author} {\bibfnamefont {A.}~\bibnamefont {Hofmann}}, \bibinfo {author} {\bibfnamefont {A.}~\bibnamefont {Ballabio}}, \bibinfo {author} {\bibfnamefont {P.~M.}\ \bibnamefont {Mutter}}, \bibinfo {author} {\bibfnamefont {G.}~\bibnamefont {Tavani}}, \bibinfo {author} {\bibfnamefont {M.}~\bibnamefont {Botifoll}}, \bibinfo {author} {\bibfnamefont {A.}~\bibnamefont {Crippa}}, \bibinfo {author} {\bibfnamefont {J.}~\bibnamefont {Kukucka}}, \bibinfo {author} {\bibfnamefont {O.}~\bibnamefont {Sagi}}, \bibinfo {author} {\bibfnamefont {F.}~\bibnamefont {Martins}}, \emph {et~al.},\ }\bibfield  {title} {\bibinfo {title} {A singlet-triplet hole spin qubit in planar ge},\ }\href@noop {} {\bibfield  {journal} {\bibinfo  {journal} {Nature Materials}\ }\textbf {\bibinfo {volume} {20}},\ \bibinfo {pages} {1106} (\bibinfo {year} {2021})}\BibitemShut {NoStop}%
\bibitem [{\citenamefont {Botzem}\ \emph {et~al.}(2018)\citenamefont {Botzem}, \citenamefont {Shulman}, \citenamefont {Foletti}, \citenamefont {Harvey}, \citenamefont {Dial}, \citenamefont {Bethke}, \citenamefont {Cerfontaine}, \citenamefont {McNeil}, \citenamefont {Mahalu}, \citenamefont {Umansky} \emph {et~al.}}]{botzem2018tuning}%
  \BibitemOpen
  \bibfield  {author} {\bibinfo {author} {\bibfnamefont {T.}~\bibnamefont {Botzem}}, \bibinfo {author} {\bibfnamefont {M.~D.}\ \bibnamefont {Shulman}}, \bibinfo {author} {\bibfnamefont {S.}~\bibnamefont {Foletti}}, \bibinfo {author} {\bibfnamefont {S.~P.}\ \bibnamefont {Harvey}}, \bibinfo {author} {\bibfnamefont {O.~E.}\ \bibnamefont {Dial}}, \bibinfo {author} {\bibfnamefont {P.}~\bibnamefont {Bethke}}, \bibinfo {author} {\bibfnamefont {P.}~\bibnamefont {Cerfontaine}}, \bibinfo {author} {\bibfnamefont {R.~P.}\ \bibnamefont {McNeil}}, \bibinfo {author} {\bibfnamefont {D.}~\bibnamefont {Mahalu}}, \bibinfo {author} {\bibfnamefont {V.}~\bibnamefont {Umansky}}, \emph {et~al.},\ }\bibfield  {title} {\bibinfo {title} {Tuning methods for semiconductor spin qubits},\ }\href@noop {} {\bibfield  {journal} {\bibinfo  {journal} {Physical Review Applied}\ }\textbf {\bibinfo {volume} {10}},\ \bibinfo {pages} {054026} (\bibinfo {year} {2018})}\BibitemShut {NoStop}%
\bibitem [{\citenamefont {Blumoff}\ \emph {et~al.}(2022)\citenamefont {Blumoff}, \citenamefont {Pan}, \citenamefont {Keating}, \citenamefont {Andrews}, \citenamefont {Barnes}, \citenamefont {Brecht}, \citenamefont {Croke}, \citenamefont {Euliss}, \citenamefont {Fast}, \citenamefont {Jackson} \emph {et~al.}}]{blumoff2022fast}%
  \BibitemOpen
  \bibfield  {author} {\bibinfo {author} {\bibfnamefont {J.~Z.}\ \bibnamefont {Blumoff}}, \bibinfo {author} {\bibfnamefont {A.~S.}\ \bibnamefont {Pan}}, \bibinfo {author} {\bibfnamefont {T.~E.}\ \bibnamefont {Keating}}, \bibinfo {author} {\bibfnamefont {R.~W.}\ \bibnamefont {Andrews}}, \bibinfo {author} {\bibfnamefont {D.~W.}\ \bibnamefont {Barnes}}, \bibinfo {author} {\bibfnamefont {T.~L.}\ \bibnamefont {Brecht}}, \bibinfo {author} {\bibfnamefont {E.~T.}\ \bibnamefont {Croke}}, \bibinfo {author} {\bibfnamefont {L.~E.}\ \bibnamefont {Euliss}}, \bibinfo {author} {\bibfnamefont {J.~A.}\ \bibnamefont {Fast}}, \bibinfo {author} {\bibfnamefont {C.~A.}\ \bibnamefont {Jackson}}, \emph {et~al.},\ }\bibfield  {title} {\bibinfo {title} {Fast and high-fidelity state preparation and measurement in triple-quantum-dot spin qubits},\ }\href@noop {} {\bibfield  {journal} {\bibinfo  {journal} {PRX Quantum}\ }\textbf {\bibinfo {volume} {3}},\ \bibinfo {pages} {010352} (\bibinfo {year} {2022})}\BibitemShut {NoStop}%
\bibitem [{\citenamefont {Zhang}\ \emph {et~al.}(2024)\citenamefont {Zhang}, \citenamefont {Morozova}, \citenamefont {Rimbach-Russ}, \citenamefont {Jirovec}, \citenamefont {Hsiao}, \citenamefont {Fariña}, \citenamefont {Wang}, \citenamefont {Oosterhout}, \citenamefont {Sammak}, \citenamefont {Scappucci}, \citenamefont {Veldhorst},\ and\ \citenamefont {Vandersypen}}]{zhang_universal_2024}%
  \BibitemOpen
  \bibfield  {author} {\bibinfo {author} {\bibfnamefont {X.}~\bibnamefont {Zhang}}, \bibinfo {author} {\bibfnamefont {E.}~\bibnamefont {Morozova}}, \bibinfo {author} {\bibfnamefont {M.}~\bibnamefont {Rimbach-Russ}}, \bibinfo {author} {\bibfnamefont {D.}~\bibnamefont {Jirovec}}, \bibinfo {author} {\bibfnamefont {T.-K.}\ \bibnamefont {Hsiao}}, \bibinfo {author} {\bibfnamefont {P.~C.}\ \bibnamefont {Fariña}}, \bibinfo {author} {\bibfnamefont {C.-A.}\ \bibnamefont {Wang}}, \bibinfo {author} {\bibfnamefont {S.~D.}\ \bibnamefont {Oosterhout}}, \bibinfo {author} {\bibfnamefont {A.}~\bibnamefont {Sammak}}, \bibinfo {author} {\bibfnamefont {G.}~\bibnamefont {Scappucci}}, \bibinfo {author} {\bibfnamefont {M.}~\bibnamefont {Veldhorst}},\ and\ \bibinfo {author} {\bibfnamefont {L.~M.~K.}\ \bibnamefont {Vandersypen}},\ }\bibfield  {title} {\bibinfo {title} {Universal control of four singlet–triplet qubits},\ }\href {https://doi.org/10.1038/s41565-024-01817-9} {\bibfield  {journal} {\bibinfo  {journal} {Nature
  Nanotechnology}\ ,\ \bibinfo {pages} {1}} (\bibinfo {year} {2024})}\BibitemShut {NoStop}%
\bibitem [{\citenamefont {Kelly}\ \emph {et~al.}(2025)\citenamefont {Kelly}, \citenamefont {Massai}, \citenamefont {Hetényi}, \citenamefont {Pita-Vidal}, \citenamefont {Orekhov}, \citenamefont {Carlsson}, \citenamefont {Seidler}, \citenamefont {Tsoukalas}, \citenamefont {Sommer}, \citenamefont {Aldeghi}, \citenamefont {Bedell}, \citenamefont {Paredes}, \citenamefont {Schupp}, \citenamefont {Mergenthaler}, \citenamefont {Fuhrer}, \citenamefont {Salis},\ and\ \citenamefont {Harvey-Collard}}]{kelly2025identifying}%
  \BibitemOpen
  \bibfield  {author} {\bibinfo {author} {\bibfnamefont {E.~G.}\ \bibnamefont {Kelly}}, \bibinfo {author} {\bibfnamefont {L.}~\bibnamefont {Massai}}, \bibinfo {author} {\bibfnamefont {B.}~\bibnamefont {Hetényi}}, \bibinfo {author} {\bibfnamefont {M.}~\bibnamefont {Pita-Vidal}}, \bibinfo {author} {\bibfnamefont {A.}~\bibnamefont {Orekhov}}, \bibinfo {author} {\bibfnamefont {C.}~\bibnamefont {Carlsson}}, \bibinfo {author} {\bibfnamefont {I.}~\bibnamefont {Seidler}}, \bibinfo {author} {\bibfnamefont {K.}~\bibnamefont {Tsoukalas}}, \bibinfo {author} {\bibfnamefont {L.}~\bibnamefont {Sommer}}, \bibinfo {author} {\bibfnamefont {M.}~\bibnamefont {Aldeghi}}, \bibinfo {author} {\bibfnamefont {S.~W.}\ \bibnamefont {Bedell}}, \bibinfo {author} {\bibfnamefont {S.}~\bibnamefont {Paredes}}, \bibinfo {author} {\bibfnamefont {F.~J.}\ \bibnamefont {Schupp}}, \bibinfo {author} {\bibfnamefont {M.}~\bibnamefont {Mergenthaler}}, \bibinfo {author} {\bibfnamefont {A.}~\bibnamefont {Fuhrer}}, \bibinfo {author} {\bibfnamefont
  {G.}~\bibnamefont {Salis}},\ and\ \bibinfo {author} {\bibfnamefont {P.}~\bibnamefont {Harvey-Collard}},\ }\bibfield  {title} {\bibinfo {title} {Identifying and mitigating errors in hole spin qubit readout},\ }\href@noop {} {\bibfield  {journal} {\bibinfo  {journal} {arXiv preprint arXiv:2504.06898}\ } (\bibinfo {year} {2025})}\BibitemShut {NoStop}%
\bibitem [{\citenamefont {Jirovec}\ \emph {et~al.}(2022)\citenamefont {Jirovec}, \citenamefont {Mutter}, \citenamefont {Hofmann}, \citenamefont {Crippa}, \citenamefont {Rychetsky}, \citenamefont {Craig}, \citenamefont {Kukucka}, \citenamefont {Martins}, \citenamefont {Ballabio}, \citenamefont {Ares} \emph {et~al.}}]{jirovec2022dynamics}%
  \BibitemOpen
  \bibfield  {author} {\bibinfo {author} {\bibfnamefont {D.}~\bibnamefont {Jirovec}}, \bibinfo {author} {\bibfnamefont {P.~M.}\ \bibnamefont {Mutter}}, \bibinfo {author} {\bibfnamefont {A.}~\bibnamefont {Hofmann}}, \bibinfo {author} {\bibfnamefont {A.}~\bibnamefont {Crippa}}, \bibinfo {author} {\bibfnamefont {M.}~\bibnamefont {Rychetsky}}, \bibinfo {author} {\bibfnamefont {D.~L.}\ \bibnamefont {Craig}}, \bibinfo {author} {\bibfnamefont {J.}~\bibnamefont {Kukucka}}, \bibinfo {author} {\bibfnamefont {F.}~\bibnamefont {Martins}}, \bibinfo {author} {\bibfnamefont {A.}~\bibnamefont {Ballabio}}, \bibinfo {author} {\bibfnamefont {N.}~\bibnamefont {Ares}}, \emph {et~al.},\ }\bibfield  {title} {\bibinfo {title} {Dynamics of hole singlet-triplet qubits with large g-factor differences},\ }\href@noop {} {\bibfield  {journal} {\bibinfo  {journal} {Physical review letters}\ }\textbf {\bibinfo {volume} {128}},\ \bibinfo {pages} {126803} (\bibinfo {year} {2022})}\BibitemShut {NoStop}%
\bibitem [{\citenamefont {Schwarz}(1978)}]{schwarz1978estimating}%
  \BibitemOpen
  \bibfield  {author} {\bibinfo {author} {\bibfnamefont {G.}~\bibnamefont {Schwarz}},\ }\bibfield  {title} {\bibinfo {title} {Estimating the dimension of a model},\ }\href@noop {} {\bibfield  {journal} {\bibinfo  {journal} {The annals of statistics}\ ,\ \bibinfo {pages} {461}} (\bibinfo {year} {1978})}\BibitemShut {NoStop}%
\bibitem [{\citenamefont {Virtanen}\ \emph {et~al.}(2020)\citenamefont {Virtanen}, \citenamefont {Gommers}, \citenamefont {Oliphant}, \citenamefont {Haberland}, \citenamefont {Reddy}, \citenamefont {Cournapeau}, \citenamefont {Burovski}, \citenamefont {Peterson}, \citenamefont {Weckesser}, \citenamefont {Bright}, \citenamefont {{van der Walt}}, \citenamefont {Brett}, \citenamefont {Wilson}, \citenamefont {Millman}, \citenamefont {Mayorov},\ and\ \citenamefont {{others}}}]{scipy}%
  \BibitemOpen
  \bibfield  {author} {\bibinfo {author} {\bibfnamefont {P.}~\bibnamefont {Virtanen}}, \bibinfo {author} {\bibfnamefont {R.}~\bibnamefont {Gommers}}, \bibinfo {author} {\bibfnamefont {T.~E.}\ \bibnamefont {Oliphant}}, \bibinfo {author} {\bibfnamefont {M.}~\bibnamefont {Haberland}}, \bibinfo {author} {\bibfnamefont {T.}~\bibnamefont {Reddy}}, \bibinfo {author} {\bibfnamefont {D.}~\bibnamefont {Cournapeau}}, \bibinfo {author} {\bibfnamefont {E.}~\bibnamefont {Burovski}}, \bibinfo {author} {\bibfnamefont {P.}~\bibnamefont {Peterson}}, \bibinfo {author} {\bibfnamefont {W.}~\bibnamefont {Weckesser}}, \bibinfo {author} {\bibfnamefont {J.}~\bibnamefont {Bright}}, \bibinfo {author} {\bibfnamefont {S.~J.}\ \bibnamefont {{van der Walt}}}, \bibinfo {author} {\bibfnamefont {M.}~\bibnamefont {Brett}}, \bibinfo {author} {\bibfnamefont {J.}~\bibnamefont {Wilson}}, \bibinfo {author} {\bibfnamefont {K.~J.}\ \bibnamefont {Millman}}, \bibinfo {author} {\bibfnamefont {N.}~\bibnamefont {Mayorov}},\ and\ \bibinfo {author}
  {\bibnamefont {{others}}},\ }\bibfield  {title} {\bibinfo {title} {{{SciPy} 1.0: Fundamental Algorithms for Scientific Computing in Python}},\ }\href {https://doi.org/10.1038/s41592-019-0686-2} {\bibfield  {journal} {\bibinfo  {journal} {Nature Methods}\ }\textbf {\bibinfo {volume} {17}},\ \bibinfo {pages} {261} (\bibinfo {year} {2020})}\BibitemShut {NoStop}%
\bibitem [{\citenamefont {Nelder}\ and\ \citenamefont {Mead}(1965)}]{nelder1965simplex}%
  \BibitemOpen
  \bibfield  {author} {\bibinfo {author} {\bibfnamefont {J.~A.}\ \bibnamefont {Nelder}}\ and\ \bibinfo {author} {\bibfnamefont {R.}~\bibnamefont {Mead}},\ }\bibfield  {title} {\bibinfo {title} {A simplex method for function minimization},\ }\href@noop {} {\bibfield  {journal} {\bibinfo  {journal} {The computer journal}\ }\textbf {\bibinfo {volume} {7}},\ \bibinfo {pages} {308} (\bibinfo {year} {1965})}\BibitemShut {NoStop}%
\bibitem [{\citenamefont {Johnson}\ \emph {et~al.}(2005)\citenamefont {Johnson}, \citenamefont {Petta}, \citenamefont {Marcus}, \citenamefont {Hanson},\ and\ \citenamefont {Gossard}}]{johnson2005singlet}%
  \BibitemOpen
  \bibfield  {author} {\bibinfo {author} {\bibfnamefont {A.~C.}\ \bibnamefont {Johnson}}, \bibinfo {author} {\bibfnamefont {J.}~\bibnamefont {Petta}}, \bibinfo {author} {\bibfnamefont {C.}~\bibnamefont {Marcus}}, \bibinfo {author} {\bibfnamefont {M.}~\bibnamefont {Hanson}},\ and\ \bibinfo {author} {\bibfnamefont {A.}~\bibnamefont {Gossard}},\ }\bibfield  {title} {\bibinfo {title} {Singlet-triplet spin blockade and charge sensing in a few-electron double quantum dot},\ }\href@noop {} {\bibfield  {journal} {\bibinfo  {journal} {Physical Review B—Condensed Matter and Materials Physics}\ }\textbf {\bibinfo {volume} {72}},\ \bibinfo {pages} {165308} (\bibinfo {year} {2005})}\BibitemShut {NoStop}%
\bibitem [{\citenamefont {Stano}\ and\ \citenamefont {Loss}(2022)}]{stano2022review}%
  \BibitemOpen
  \bibfield  {author} {\bibinfo {author} {\bibfnamefont {P.}~\bibnamefont {Stano}}\ and\ \bibinfo {author} {\bibfnamefont {D.}~\bibnamefont {Loss}},\ }\bibfield  {title} {\bibinfo {title} {Review of performance metrics of spin qubits in gated semiconducting nanostructures},\ }\href@noop {} {\bibfield  {journal} {\bibinfo  {journal} {Nature Reviews Physics}\ }\textbf {\bibinfo {volume} {4}},\ \bibinfo {pages} {672} (\bibinfo {year} {2022})}\BibitemShut {NoStop}%
\bibitem [{\citenamefont {Ha}\ \emph {et~al.}(2025)\citenamefont {Ha}, \citenamefont {Acuna}, \citenamefont {Raach}, \citenamefont {Bloom}, \citenamefont {Brecht}, \citenamefont {Chappell}, \citenamefont {Choi}, \citenamefont {Christensen}, \citenamefont {Counts}, \citenamefont {Daprano} \emph {et~al.}}]{ha2025two}%
  \BibitemOpen
  \bibfield  {author} {\bibinfo {author} {\bibfnamefont {S.~D.}\ \bibnamefont {Ha}}, \bibinfo {author} {\bibfnamefont {E.}~\bibnamefont {Acuna}}, \bibinfo {author} {\bibfnamefont {K.}~\bibnamefont {Raach}}, \bibinfo {author} {\bibfnamefont {Z.~T.}\ \bibnamefont {Bloom}}, \bibinfo {author} {\bibfnamefont {T.~L.}\ \bibnamefont {Brecht}}, \bibinfo {author} {\bibfnamefont {J.~M.}\ \bibnamefont {Chappell}}, \bibinfo {author} {\bibfnamefont {M.~D.}\ \bibnamefont {Choi}}, \bibinfo {author} {\bibfnamefont {J.~E.}\ \bibnamefont {Christensen}}, \bibinfo {author} {\bibfnamefont {I.~T.}\ \bibnamefont {Counts}}, \bibinfo {author} {\bibfnamefont {D.}~\bibnamefont {Daprano}}, \emph {et~al.},\ }\bibfield  {title} {\bibinfo {title} {Two-dimensional si spin qubit arrays with multilevel interconnects},\ }\href@noop {} {\bibfield  {journal} {\bibinfo  {journal} {arXiv preprint arXiv:2502.08861}\ } (\bibinfo {year} {2025})}\BibitemShut {NoStop}%
\bibitem [{\citenamefont {Liles}\ \emph {et~al.}(2024)\citenamefont {Liles}, \citenamefont {Halverson}, \citenamefont {Wang}, \citenamefont {Shamim}, \citenamefont {Eggli}, \citenamefont {Jin}, \citenamefont {Hillier}, \citenamefont {Kumar}, \citenamefont {Vorreiter}, \citenamefont {Rendell} \emph {et~al.}}]{liles2024singlet}%
  \BibitemOpen
  \bibfield  {author} {\bibinfo {author} {\bibfnamefont {S.}~\bibnamefont {Liles}}, \bibinfo {author} {\bibfnamefont {D.}~\bibnamefont {Halverson}}, \bibinfo {author} {\bibfnamefont {Z.}~\bibnamefont {Wang}}, \bibinfo {author} {\bibfnamefont {A.}~\bibnamefont {Shamim}}, \bibinfo {author} {\bibfnamefont {R.}~\bibnamefont {Eggli}}, \bibinfo {author} {\bibfnamefont {I.~K.}\ \bibnamefont {Jin}}, \bibinfo {author} {\bibfnamefont {J.}~\bibnamefont {Hillier}}, \bibinfo {author} {\bibfnamefont {K.}~\bibnamefont {Kumar}}, \bibinfo {author} {\bibfnamefont {I.}~\bibnamefont {Vorreiter}}, \bibinfo {author} {\bibfnamefont {M.}~\bibnamefont {Rendell}}, \emph {et~al.},\ }\bibfield  {title} {\bibinfo {title} {A singlet-triplet hole-spin qubit in mos silicon},\ }\href@noop {} {\bibfield  {journal} {\bibinfo  {journal} {Nature Communications}\ }\textbf {\bibinfo {volume} {15}},\ \bibinfo {pages} {7690} (\bibinfo {year} {2024})}\BibitemShut {NoStop}%
\bibitem [{\citenamefont {Ares}\ \emph {et~al.}(2013)\citenamefont {Ares}, \citenamefont {Golovach}, \citenamefont {Katsaros}, \citenamefont {Stoffel}, \citenamefont {Fournel}, \citenamefont {Glazman}, \citenamefont {Schmidt},\ and\ \citenamefont {De~Franceschi}}]{ares2013nature}%
  \BibitemOpen
  \bibfield  {author} {\bibinfo {author} {\bibfnamefont {N.}~\bibnamefont {Ares}}, \bibinfo {author} {\bibfnamefont {V.~N.}\ \bibnamefont {Golovach}}, \bibinfo {author} {\bibfnamefont {G.}~\bibnamefont {Katsaros}}, \bibinfo {author} {\bibfnamefont {M.}~\bibnamefont {Stoffel}}, \bibinfo {author} {\bibfnamefont {F.}~\bibnamefont {Fournel}}, \bibinfo {author} {\bibfnamefont {L.~I.}\ \bibnamefont {Glazman}}, \bibinfo {author} {\bibfnamefont {O.~G.}\ \bibnamefont {Schmidt}},\ and\ \bibinfo {author} {\bibfnamefont {S.}~\bibnamefont {De~Franceschi}},\ }\bibfield  {title} {\bibinfo {title} {Nature of tunable hole g factors in quantum dots},\ }\href@noop {} {\bibfield  {journal} {\bibinfo  {journal} {Physical review letters}\ }\textbf {\bibinfo {volume} {110}},\ \bibinfo {pages} {046602} (\bibinfo {year} {2013})}\BibitemShut {NoStop}%
\bibitem [{\citenamefont {Liles}\ \emph {et~al.}(2021)\citenamefont {Liles}, \citenamefont {Martins}, \citenamefont {Miserev}, \citenamefont {Kiselev}, \citenamefont {Thorvaldson}, \citenamefont {Rendell}, \citenamefont {Jin}, \citenamefont {Hudson}, \citenamefont {Veldhorst}, \citenamefont {Itoh} \emph {et~al.}}]{liles2021electrical}%
  \BibitemOpen
  \bibfield  {author} {\bibinfo {author} {\bibfnamefont {S.}~\bibnamefont {Liles}}, \bibinfo {author} {\bibfnamefont {F.}~\bibnamefont {Martins}}, \bibinfo {author} {\bibfnamefont {D.}~\bibnamefont {Miserev}}, \bibinfo {author} {\bibfnamefont {A.}~\bibnamefont {Kiselev}}, \bibinfo {author} {\bibfnamefont {I.}~\bibnamefont {Thorvaldson}}, \bibinfo {author} {\bibfnamefont {M.}~\bibnamefont {Rendell}}, \bibinfo {author} {\bibfnamefont {I.}~\bibnamefont {Jin}}, \bibinfo {author} {\bibfnamefont {F.}~\bibnamefont {Hudson}}, \bibinfo {author} {\bibfnamefont {M.}~\bibnamefont {Veldhorst}}, \bibinfo {author} {\bibfnamefont {K.}~\bibnamefont {Itoh}}, \emph {et~al.},\ }\bibfield  {title} {\bibinfo {title} {Electrical control of the g tensor of the first hole in a silicon mos quantum dot},\ }\href@noop {} {\bibfield  {journal} {\bibinfo  {journal} {Physical Review B}\ }\textbf {\bibinfo {volume} {104}},\ \bibinfo {pages} {235303} (\bibinfo {year} {2021})}\BibitemShut {NoStop}%
\bibitem [{\citenamefont {Mauro}\ \emph {et~al.}(2024)\citenamefont {Mauro}, \citenamefont {Rodr{\'\i}guez-Mena}, \citenamefont {Bassi}, \citenamefont {Schmitt},\ and\ \citenamefont {Niquet}}]{mauro2024geometry}%
  \BibitemOpen
  \bibfield  {author} {\bibinfo {author} {\bibfnamefont {L.}~\bibnamefont {Mauro}}, \bibinfo {author} {\bibfnamefont {E.~A.}\ \bibnamefont {Rodr{\'\i}guez-Mena}}, \bibinfo {author} {\bibfnamefont {M.}~\bibnamefont {Bassi}}, \bibinfo {author} {\bibfnamefont {V.}~\bibnamefont {Schmitt}},\ and\ \bibinfo {author} {\bibfnamefont {Y.-M.}\ \bibnamefont {Niquet}},\ }\bibfield  {title} {\bibinfo {title} {Geometry of the dephasing sweet spots of spin-orbit qubits},\ }\href@noop {} {\bibfield  {journal} {\bibinfo  {journal} {Physical Review B}\ }\textbf {\bibinfo {volume} {109}},\ \bibinfo {pages} {155406} (\bibinfo {year} {2024})}\BibitemShut {NoStop}%
\bibitem [{\citenamefont {Bassi}\ \emph {et~al.}(2024)\citenamefont {Bassi}, \citenamefont {Rodr{\i}guez-Mena}, \citenamefont {Brun}, \citenamefont {Zihlmann}, \citenamefont {Nguyen}, \citenamefont {Champain}, \citenamefont {Abadillo-Uriel}, \citenamefont {Bertrand}, \citenamefont {Niebojewski}, \citenamefont {Maurand} \emph {et~al.}}]{bassi2024optimal}%
  \BibitemOpen
  \bibfield  {author} {\bibinfo {author} {\bibfnamefont {M.}~\bibnamefont {Bassi}}, \bibinfo {author} {\bibfnamefont {E.-A.}\ \bibnamefont {Rodr{\i}guez-Mena}}, \bibinfo {author} {\bibfnamefont {B.}~\bibnamefont {Brun}}, \bibinfo {author} {\bibfnamefont {S.}~\bibnamefont {Zihlmann}}, \bibinfo {author} {\bibfnamefont {T.}~\bibnamefont {Nguyen}}, \bibinfo {author} {\bibfnamefont {V.}~\bibnamefont {Champain}}, \bibinfo {author} {\bibfnamefont {J.~C.}\ \bibnamefont {Abadillo-Uriel}}, \bibinfo {author} {\bibfnamefont {B.}~\bibnamefont {Bertrand}}, \bibinfo {author} {\bibfnamefont {H.}~\bibnamefont {Niebojewski}}, \bibinfo {author} {\bibfnamefont {R.}~\bibnamefont {Maurand}}, \emph {et~al.},\ }\bibfield  {title} {\bibinfo {title} {Optimal operation of hole spin qubits},\ }\href@noop {} {\bibfield  {journal} {\bibinfo  {journal} {arXiv preprint arXiv:2412.13069}\ } (\bibinfo {year} {2024})}\BibitemShut {NoStop}%
\bibitem [{\citenamefont {Geyer}\ \emph {et~al.}(2024)\citenamefont {Geyer}, \citenamefont {Het{\'e}nyi}, \citenamefont {Bosco}, \citenamefont {Camenzind}, \citenamefont {Eggli}, \citenamefont {Fuhrer}, \citenamefont {Loss}, \citenamefont {Warburton}, \citenamefont {Zumb{\"u}hl},\ and\ \citenamefont {Kuhlmann}}]{geyer2024anisotropic}%
  \BibitemOpen
  \bibfield  {author} {\bibinfo {author} {\bibfnamefont {S.}~\bibnamefont {Geyer}}, \bibinfo {author} {\bibfnamefont {B.}~\bibnamefont {Het{\'e}nyi}}, \bibinfo {author} {\bibfnamefont {S.}~\bibnamefont {Bosco}}, \bibinfo {author} {\bibfnamefont {L.~C.}\ \bibnamefont {Camenzind}}, \bibinfo {author} {\bibfnamefont {R.~S.}\ \bibnamefont {Eggli}}, \bibinfo {author} {\bibfnamefont {A.}~\bibnamefont {Fuhrer}}, \bibinfo {author} {\bibfnamefont {D.}~\bibnamefont {Loss}}, \bibinfo {author} {\bibfnamefont {R.~J.}\ \bibnamefont {Warburton}}, \bibinfo {author} {\bibfnamefont {D.~M.}\ \bibnamefont {Zumb{\"u}hl}},\ and\ \bibinfo {author} {\bibfnamefont {A.~V.}\ \bibnamefont {Kuhlmann}},\ }\bibfield  {title} {\bibinfo {title} {Anisotropic exchange interaction of two hole-spin qubits},\ }\href@noop {} {\bibfield  {journal} {\bibinfo  {journal} {Nature Physics}\ }\textbf {\bibinfo {volume} {20}},\ \bibinfo {pages} {1152} (\bibinfo {year} {2024})}\BibitemShut {NoStop}%
\bibitem [{\citenamefont {Froning}\ \emph {et~al.}(2021)\citenamefont {Froning}, \citenamefont {Ran{\v{c}}i{\'c}}, \citenamefont {Het{\'e}nyi}, \citenamefont {Bosco}, \citenamefont {Rehmann}, \citenamefont {Li}, \citenamefont {Bakkers}, \citenamefont {Zwanenburg}, \citenamefont {Loss}, \citenamefont {Zumb{\"u}hl} \emph {et~al.}}]{froning2021strong}%
  \BibitemOpen
  \bibfield  {author} {\bibinfo {author} {\bibfnamefont {F.}~\bibnamefont {Froning}}, \bibinfo {author} {\bibfnamefont {M.}~\bibnamefont {Ran{\v{c}}i{\'c}}}, \bibinfo {author} {\bibfnamefont {B.}~\bibnamefont {Het{\'e}nyi}}, \bibinfo {author} {\bibfnamefont {S.}~\bibnamefont {Bosco}}, \bibinfo {author} {\bibfnamefont {M.}~\bibnamefont {Rehmann}}, \bibinfo {author} {\bibfnamefont {A.}~\bibnamefont {Li}}, \bibinfo {author} {\bibfnamefont {E.~P.}\ \bibnamefont {Bakkers}}, \bibinfo {author} {\bibfnamefont {F.~A.}\ \bibnamefont {Zwanenburg}}, \bibinfo {author} {\bibfnamefont {D.}~\bibnamefont {Loss}}, \bibinfo {author} {\bibfnamefont {D.}~\bibnamefont {Zumb{\"u}hl}}, \emph {et~al.},\ }\bibfield  {title} {\bibinfo {title} {Strong spin-orbit interaction and g-factor renormalization of hole spins in ge/si nanowire quantum dots},\ }\href@noop {} {\bibfield  {journal} {\bibinfo  {journal} {Physical Review Research}\ }\textbf {\bibinfo {volume} {3}},\ \bibinfo {pages} {013081} (\bibinfo {year} {2021})}\BibitemShut
  {NoStop}%
\bibitem [{\citenamefont {Schorling}\ \emph {et~al.}(2025)\citenamefont {Schorling}, \citenamefont {Vaidhyanathan}, \citenamefont {Schuff}, \citenamefont {Carballido}, \citenamefont {Zumb{\"u}hl}, \citenamefont {Milburn}, \citenamefont {Marquardt}, \citenamefont {Foerster}, \citenamefont {Osborne},\ and\ \citenamefont {Ares}}]{schorling2025meta}%
  \BibitemOpen
  \bibfield  {author} {\bibinfo {author} {\bibfnamefont {L.}~\bibnamefont {Schorling}}, \bibinfo {author} {\bibfnamefont {P.}~\bibnamefont {Vaidhyanathan}}, \bibinfo {author} {\bibfnamefont {J.}~\bibnamefont {Schuff}}, \bibinfo {author} {\bibfnamefont {M.~J.}\ \bibnamefont {Carballido}}, \bibinfo {author} {\bibfnamefont {D.}~\bibnamefont {Zumb{\"u}hl}}, \bibinfo {author} {\bibfnamefont {G.}~\bibnamefont {Milburn}}, \bibinfo {author} {\bibfnamefont {F.}~\bibnamefont {Marquardt}}, \bibinfo {author} {\bibfnamefont {J.}~\bibnamefont {Foerster}}, \bibinfo {author} {\bibfnamefont {M.~A.}\ \bibnamefont {Osborne}},\ and\ \bibinfo {author} {\bibfnamefont {N.}~\bibnamefont {Ares}},\ }\bibfield  {title} {\bibinfo {title} {Meta-learning characteristics and dynamics of quantum systems},\ }\href@noop {} {\bibfield  {journal} {\bibinfo  {journal} {arXiv preprint arXiv:2503.10492}\ } (\bibinfo {year} {2025})}\BibitemShut {NoStop}%
\bibitem [{\citenamefont {Gebhart}\ \emph {et~al.}(2023)\citenamefont {Gebhart}, \citenamefont {Santagati}, \citenamefont {Gentile}, \citenamefont {Gauger}, \citenamefont {Craig}, \citenamefont {Ares}, \citenamefont {Banchi}, \citenamefont {Marquardt}, \citenamefont {Pezze},\ and\ \citenamefont {Bonato}}]{gebhart2023learning}%
  \BibitemOpen
  \bibfield  {author} {\bibinfo {author} {\bibfnamefont {V.}~\bibnamefont {Gebhart}}, \bibinfo {author} {\bibfnamefont {R.}~\bibnamefont {Santagati}}, \bibinfo {author} {\bibfnamefont {A.~A.}\ \bibnamefont {Gentile}}, \bibinfo {author} {\bibfnamefont {E.~M.}\ \bibnamefont {Gauger}}, \bibinfo {author} {\bibfnamefont {D.}~\bibnamefont {Craig}}, \bibinfo {author} {\bibfnamefont {N.}~\bibnamefont {Ares}}, \bibinfo {author} {\bibfnamefont {L.}~\bibnamefont {Banchi}}, \bibinfo {author} {\bibfnamefont {F.}~\bibnamefont {Marquardt}}, \bibinfo {author} {\bibfnamefont {L.}~\bibnamefont {Pezze}},\ and\ \bibinfo {author} {\bibfnamefont {C.}~\bibnamefont {Bonato}},\ }\bibfield  {title} {\bibinfo {title} {Learning quantum systems},\ }\href@noop {} {\bibfield  {journal} {\bibinfo  {journal} {Nature Reviews Physics}\ }\textbf {\bibinfo {volume} {5}},\ \bibinfo {pages} {141} (\bibinfo {year} {2023})}\BibitemShut {NoStop}%
\bibitem [{\citenamefont {Tosato}\ \emph {et~al.}(2025)\citenamefont {Tosato}, \citenamefont {Elsayed}, \citenamefont {Poggiali}, \citenamefont {Stehouwer}, \citenamefont {Costa}, \citenamefont {Hudson}, \citenamefont {Esposti},\ and\ \citenamefont {Scappucci}}]{tosato2025qarpet}%
  \BibitemOpen
  \bibfield  {author} {\bibinfo {author} {\bibfnamefont {A.}~\bibnamefont {Tosato}}, \bibinfo {author} {\bibfnamefont {A.}~\bibnamefont {Elsayed}}, \bibinfo {author} {\bibfnamefont {F.}~\bibnamefont {Poggiali}}, \bibinfo {author} {\bibfnamefont {L.}~\bibnamefont {Stehouwer}}, \bibinfo {author} {\bibfnamefont {D.}~\bibnamefont {Costa}}, \bibinfo {author} {\bibfnamefont {K.}~\bibnamefont {Hudson}}, \bibinfo {author} {\bibfnamefont {D.~D.}\ \bibnamefont {Esposti}},\ and\ \bibinfo {author} {\bibfnamefont {G.}~\bibnamefont {Scappucci}},\ }\bibfield  {title} {\bibinfo {title} {Qarpet: A crossbar chip for benchmarking semiconductor spin qubits},\ }\href@noop {} {\bibfield  {journal} {\bibinfo  {journal} {arXiv preprint arXiv:2504.05460}\ } (\bibinfo {year} {2025})}\BibitemShut {NoStop}%
\bibitem [{\citenamefont {Dunn}(1974)}]{dunn1974well}%
  \BibitemOpen
  \bibfield  {author} {\bibinfo {author} {\bibfnamefont {J.~C.}\ \bibnamefont {Dunn}},\ }\bibfield  {title} {\bibinfo {title} {Well-separated clusters and optimal fuzzy partitions},\ }\href@noop {} {\bibfield  {journal} {\bibinfo  {journal} {Journal of cybernetics}\ }\textbf {\bibinfo {volume} {4}},\ \bibinfo {pages} {95} (\bibinfo {year} {1974})}\BibitemShut {NoStop}%
\bibitem [{\citenamefont {Ziegler}\ \emph {et~al.}(2023{\natexlab{b}})\citenamefont {Ziegler}, \citenamefont {Luthi}, \citenamefont {Ramsey}, \citenamefont {Borjans}, \citenamefont {Zheng},\ and\ \citenamefont {Zwolak}}]{ziegler2023automated}%
  \BibitemOpen
  \bibfield  {author} {\bibinfo {author} {\bibfnamefont {J.}~\bibnamefont {Ziegler}}, \bibinfo {author} {\bibfnamefont {F.}~\bibnamefont {Luthi}}, \bibinfo {author} {\bibfnamefont {M.}~\bibnamefont {Ramsey}}, \bibinfo {author} {\bibfnamefont {F.}~\bibnamefont {Borjans}}, \bibinfo {author} {\bibfnamefont {G.}~\bibnamefont {Zheng}},\ and\ \bibinfo {author} {\bibfnamefont {J.~P.}\ \bibnamefont {Zwolak}},\ }\bibfield  {title} {\bibinfo {title} {Automated extraction of capacitive coupling for quantum dot systems},\ }\href@noop {} {\bibfield  {journal} {\bibinfo  {journal} {Physical Review Applied}\ }\textbf {\bibinfo {volume} {19}},\ \bibinfo {pages} {054077} (\bibinfo {year} {2023}{\natexlab{b}})}\BibitemShut {NoStop}%
\bibitem [{\citenamefont {Ding}\ \emph {et~al.}(2022)\citenamefont {Ding}, \citenamefont {Zhang}, \citenamefont {Han},\ and\ \citenamefont {Ding}}]{ding2022scaling}%
  \BibitemOpen
  \bibfield  {author} {\bibinfo {author} {\bibfnamefont {X.}~\bibnamefont {Ding}}, \bibinfo {author} {\bibfnamefont {X.}~\bibnamefont {Zhang}}, \bibinfo {author} {\bibfnamefont {J.}~\bibnamefont {Han}},\ and\ \bibinfo {author} {\bibfnamefont {G.}~\bibnamefont {Ding}},\ }\bibfield  {title} {\bibinfo {title} {Scaling up your kernels to 31x31: Revisiting large kernel design in cnns},\ }in\ \href@noop {} {\emph {\bibinfo {booktitle} {Proceedings of the IEEE/CVF conference on computer vision and pattern recognition}}}\ (\bibinfo {year} {2022})\ pp.\ \bibinfo {pages} {11963--11975}\BibitemShut {NoStop}%
\bibitem [{\citenamefont {Czischek}\ \emph {et~al.}(2021)\citenamefont {Czischek}, \citenamefont {Yon}, \citenamefont {Genest}, \citenamefont {Roux}, \citenamefont {Rochette}, \citenamefont {Lemyre}, \citenamefont {Moras}, \citenamefont {Pioro-Ladri{\`e}re}, \citenamefont {Drouin}, \citenamefont {Beilliard} \emph {et~al.}}]{czischek2021miniaturizing}%
  \BibitemOpen
  \bibfield  {author} {\bibinfo {author} {\bibfnamefont {S.}~\bibnamefont {Czischek}}, \bibinfo {author} {\bibfnamefont {V.}~\bibnamefont {Yon}}, \bibinfo {author} {\bibfnamefont {M.-A.}\ \bibnamefont {Genest}}, \bibinfo {author} {\bibfnamefont {M.-A.}\ \bibnamefont {Roux}}, \bibinfo {author} {\bibfnamefont {S.}~\bibnamefont {Rochette}}, \bibinfo {author} {\bibfnamefont {J.~C.}\ \bibnamefont {Lemyre}}, \bibinfo {author} {\bibfnamefont {M.}~\bibnamefont {Moras}}, \bibinfo {author} {\bibfnamefont {M.}~\bibnamefont {Pioro-Ladri{\`e}re}}, \bibinfo {author} {\bibfnamefont {D.}~\bibnamefont {Drouin}}, \bibinfo {author} {\bibfnamefont {Y.}~\bibnamefont {Beilliard}}, \emph {et~al.},\ }\bibfield  {title} {\bibinfo {title} {Miniaturizing neural networks for charge state autotuning in quantum dots},\ }\href@noop {} {\bibfield  {journal} {\bibinfo  {journal} {Machine Learning: Science and Technology}\ }\textbf {\bibinfo {volume} {3}},\ \bibinfo {pages} {015001} (\bibinfo {year} {2021})}\BibitemShut {NoStop}%
\bibitem [{\citenamefont {Ziegler}\ \emph {et~al.}(2022)\citenamefont {Ziegler}, \citenamefont {McJunkin}, \citenamefont {Joseph}, \citenamefont {Kalantre}, \citenamefont {Harpt}, \citenamefont {Savage}, \citenamefont {Lagally}, \citenamefont {Eriksson}, \citenamefont {Taylor},\ and\ \citenamefont {Zwolak}}]{ziegler2022toward}%
  \BibitemOpen
  \bibfield  {author} {\bibinfo {author} {\bibfnamefont {J.}~\bibnamefont {Ziegler}}, \bibinfo {author} {\bibfnamefont {T.}~\bibnamefont {McJunkin}}, \bibinfo {author} {\bibfnamefont {E.}~\bibnamefont {Joseph}}, \bibinfo {author} {\bibfnamefont {S.~S.}\ \bibnamefont {Kalantre}}, \bibinfo {author} {\bibfnamefont {B.}~\bibnamefont {Harpt}}, \bibinfo {author} {\bibfnamefont {D.}~\bibnamefont {Savage}}, \bibinfo {author} {\bibfnamefont {M.~G.}\ \bibnamefont {Lagally}}, \bibinfo {author} {\bibfnamefont {M.}~\bibnamefont {Eriksson}}, \bibinfo {author} {\bibfnamefont {J.~M.}\ \bibnamefont {Taylor}},\ and\ \bibinfo {author} {\bibfnamefont {J.~P.}\ \bibnamefont {Zwolak}},\ }\bibfield  {title} {\bibinfo {title} {Toward robust autotuning of noisy quantum dot devices},\ }\href@noop {} {\bibfield  {journal} {\bibinfo  {journal} {Physical Review Applied}\ }\textbf {\bibinfo {volume} {17}},\ \bibinfo {pages} {024069} (\bibinfo {year} {2022})}\BibitemShut {NoStop}%
\end{thebibliography}%

%\appendix
\setcounter{section}{0}
\renewcommand{\thesection}{Appendix \Alph{section}}
\titleformat{\section}[block]
  {\normalfont\small\bfseries\centering}
  {\thesection:}
  {0.75em}
  {}

\renewcommand{\thesubsection}{\arabic{subsection}}

\setcounter{equation}{0}
\renewcommand{\theequation}{\Alph{section}\arabic{equation}}

\section{\label{app:clustering}Clustering}
Prior to $k$-means clustering, pruning is done recursively on positive classifications produced by the `Interdot CNN' with fewer than three neighbours within a neighbourhood of $8\sqrt{2}$ pixels. This corresponds to classifications up to two strides away. The Dunn index (DI) \cite{dunn1974well} is used to score the goodness of clustering, and is given by,
\begin{equation}
\mathrm{DI} = \frac{\underset{2 \leq i < j \leq k}{\mathrm{min}} \delta\left(C_i, C_j \right)}{\underset{2 \leq m \leq k}{\mathrm{min}} \Delta_m},
\end{equation}
where $\delta\left(C_i, C_j \right)$ is the inter-cluster distance between clusters $C_i$ and $C_j$, and $\Delta_m$ is the intra-cluster distance. A high DI signifies a clustering into $k$ clusters that have both high compactness and separation. Our routine computes the DI for $k \in [2, 40]$ and determines the optimal number of clusters as being the $k$ for which the Dunn index maximised. The centroids of this optimal clustering is used to define IDT locations.

\section{\label{app:connectivity}Connectivity}
After IDTs are found, we establish their connectivity by locating up to four nearest neighbours for each IDT. We constrain the search space for each neighbour by considering the physically reasonable bounds in which they can exist in the plunger voltage space. For scalable devices with constant gate pitch, it is reasonable to assume that a quantum dot will form nearest to the plunger gate that is intended to control the electrochemical potential of that dot. This implies that the capacitance between this dot and its dedicated plunger will always be greater than the cross-capacitance to a neighbouring plunger. Following this line of reasoning, the angles of transition lines in a CSD should always be less than 45\unit{\degree} as measured from the horizontal axes for vertical transition lines, and from the vertical axes for horizontal transition lines. In our case, vertical and horizontal transitions correspond to sweeping PL and PR, respectively. 

\begin{figure}[t]
\includegraphics[width=\linewidth]{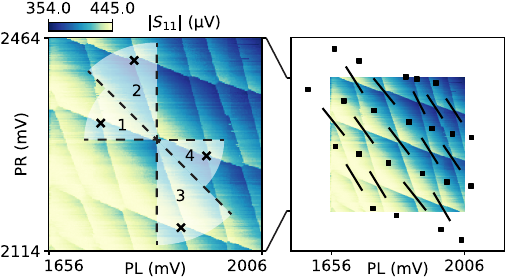}
\caption{\label{fig:fig_B1}\textbf{Procedure for finding connectivity and charge region midpoints.} Each IDT's neighbours are found by searching inside four sectors (left). Charge region centres are taken to be the midpoints between neighbours 1/2 and 3/4 (right). Pulsing amplitudes are calculated as a fraction of the distance to each midpoint.}
\end{figure}

With this, we confine the search space for each nearest neighbour to a 45\unit{\degree} sector emanating from each IDT, shown in white in Fig.~\ref{fig:fig_B1} (left) on an example IDT from the same voltage scan as in Fig.~\ref{fig:fig_2}(a) of the main text. The first IDT that is found in each sector is taken to be the nearest neighbour in that search direction, marked by a cross. Initial scan sizes used by the segmentation stage (Sec.~\ref{sec:segmentation}) are set to 70\% of the smallest nearest-neighbour distance for each IDT.

Knowing each IDT's set of nearest neighbours, and therefore their arrangement in plunger voltage space, the centre of each charge region can be approximated. These centres are given by the midpoints between neighbours labelled 1/2 and 3/4, producing the square points in Fig.~\ref{fig:fig_B1} (right). If only one of the neighbour pairs is complete (1/2 or 3/4), the missing midpoint is estimated by reflecting the midpoint determined from the existing pair. If both pairs are incomplete, an informed guess is made using the midpoint from a nearby charge region. 

Detuning amplitudes for relaxation and idling points (triangular and square markers in Fig.~\ref{fig:fig_1}(d)) are set to 40\% of the average distance to charge region midpoints either side of each IDT. The lines in Fig.~\ref{fig:fig_B1} (right) illustrate this amplitude, however, we note that their exact direction is given by the detuning axis, which is calculated only after potential meta-stable regions are found (Sec.~\ref{sec:segmentation})

\begin{figure}[t]
\includegraphics[width=\linewidth]{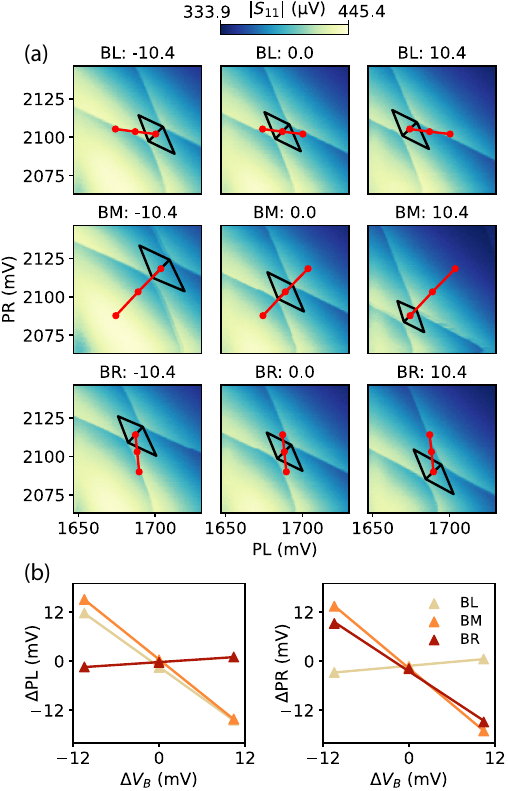}
\caption{\label{fig:fig_C1}\textbf{Barrier virtualisation procedure}. \textbf{(a)} IDT tracking in PL/PR voltage space for three equal voltage steps on each barrier gate. \textbf{(b)} IDT shifts decomposed into PL and PR components. Line slopes give virtual gate matrix elements.}
\end{figure}

\setcounter{equation}{0}
\section{\label{app:virtualisation}Virtualisation \& Fine Tuning}
Our virtualisation procedure is designed to maintain the degeneracy point between electrochemical potentials of two quantum dots ($\varepsilon = 0$). In doing so, IDT shifts are compensated in plunger voltage space during barrier tuning. We assume a linear response of the system to changes in gate voltage, which breaks down for larger tuning ranges. If an IDT were to originate from a spurious dot, this too would lead to starkly different gate couplings \cite{ziegler2023automated}. Hence, our virtualisation procedure is carried out each time a new IDT is visited.

Figure \ref{fig:fig_C1}(a) illustrates the virtualisation procedure about IDT 20 from the autonomous run of our routine presented in the main text. After the scan window is centred on the IDT, each barrier then makes three voltage steps in increments of 1/8 of the window size. Based on historical knowledge for devices with our specific gate layout, this translates to an expected shift of the IDT by up to approximately 1/4 of the scan window. Feeding scan images to Line and Angle CNNs enables the IDT's shifts to be decomposed into PL and PR voltage components. The deviations along each component are plotted in Fig.~\ref{fig:fig_C1}(b) as a function of the barrier gate step. A slope fitted to this data gives the compensating voltage needed on each plunger gate per unit voltage change on each barrier to maintain centring of the IDT.

\begin{figure*}[t]
\includegraphics[width=\linewidth]{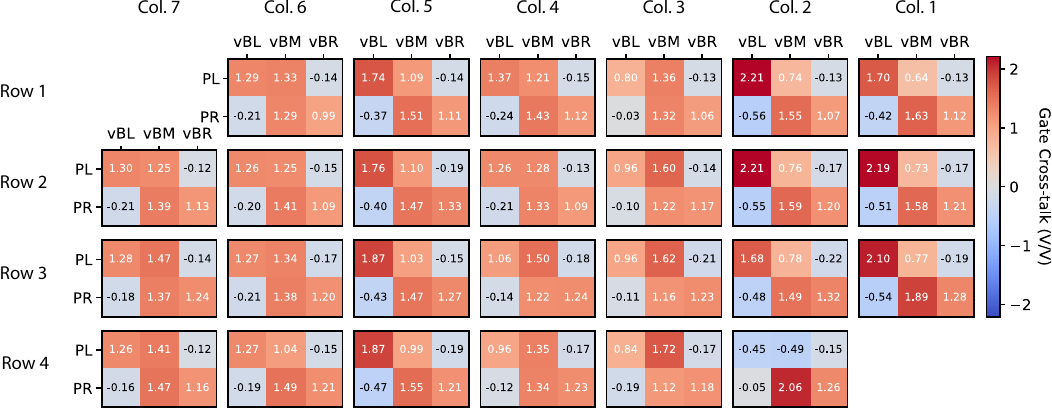}
\caption{\label{fig:fig_C2}\textbf{Automatically acquired barrier-plunger virtual gate matrices at each IDT.} Row and column indices correspond to the transitions line numbers of Fig.~\ref{fig:fig_4}, at whose intersections each IDT is located. The matrix in Col. 2, Row 4 (IDT 25), was miscalibrated.}
\end{figure*}

Figure \ref{fig:fig_C2} shows the virtual gate matrix between barrier and plunger gates for each IDT from the same run as in the main text. A large spread in the matrix entries motivates our procedure. We explain the negative entries between distant pairs of plungers and barriers (blue) by a change in Coulomb repulsion that outweighs gate cross-talk. This has the effect of lowering the electrochemical potential on the right/left dot when the voltage is reduced on the left/right barrier.

The sensing dot's electrochemical potential is maintained by virtualising all DQD gates against SP, thus creating a virtual sensor plunger gate vSP. Following the method of \cite{hickie2024}, which is well suited for long-range voltage scans, we establish the relationship,
\begin{equation}
\begin{split}
\mathrm{vSP} = \mathrm{SP} + ( 0.056 \mathrm{BL}& + 0.079 \mathrm{PL} + 0.193 \mathrm{BM} \\& + 0.089 \mathrm{PR} + 0.115 \mathrm{BR}). 
\end{split}
\end{equation}
This allows any change in BL/PL/BM/PR/BR to be compensated by adjusting SP by a scaled voltage that is opposite in magnitude, thus maintaining the set-point of vSP.

To maintain high sensitivity, SP is also routinely fine-tuned to probe the maximum slope of the sensing peak on the flank that is nearest to the current voltage setting of SP. This is done before each barrier virtualisation loop and before each image segmentation stage of the algorithm (Sec.~\ref{sec:segmentation}.). To ensure confident predictions by the Line CNN, the scan window is also rescaled and re-centred about the IDT with every barrier adjustment. Three attempts are given to have the IDT midpoint within 15\% of the scan window's centre, and for the scan size to be no greater than 5 times the IDT length. Initial scan sizes are determined uniquely for each IDT (Appendix \ref{app:connectivity}). Scan size requirements are relaxed during virtualisation and should be no greater than 6 times the IDT length.

\section{\label{app:references}Reference Acquisition}
Reference measurements provide a ``best guess'' for the means and standard deviations of readout signal distributions following our pulsing scheme (Fig.~\ref{fig:fig_1}d). We distinguish these reference quantities, $\mu_i^j$ and $\sigma_i^j$, by dropping the tilde ($~\tilde{}~$) from our notation, as it appears in Sec.~\ref{sec:evaluation} of the main text.

\begin{figure}[b]
\includegraphics[width=\linewidth]{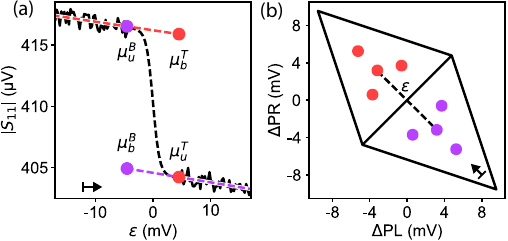}
\caption{\label{fig:fig_D1}\textbf{Diagrams of the reference acquisition procedure}. \textbf{(a)} Two traces are acquired in opposing directions along the detuning axis of a sample pair. At each readout point, unblocked reference signals are acquired. By extrapolating each trace (dashed coloured lines), blocked reference signals are obtained. The situation illustrated here resembles the signal levels show in Fig.~\ref{fig:fig_3}a of the main text. \textbf{(b)} Readout points plotted in plunger voltage space. Arrows show the same detuning point and direction in the space of the two diagrams.}
\end{figure}

If readout occurs with the system unblocked, the sensor signal will correspond to the ground-state charge configuration at which the readout voltage point is located. Hence, all of \{$\mu\mathrm{_u^T}, \mu\mathrm{_u^B}, \sigma\mathrm{_u^T}, \sigma\mathrm{_u^B}$\} are obtained simply by measuring the signal at the readout point, without pulsing, followed by a Gaussian fitting procedure. For consistency with pulsed measurements, all reference measurements also use 1000 shots, each with an integration time of 8\! \unit{\micro\second}. 

If readout occurs with the system blocked, either due to PSB or charge latching, the sensor signal will instead correspond to the charge state at which the idling point is located, with an additional cross-capacitance contribution. To emulate the sensor response, we linearly extrapolate traces acquired along the detuning axis of a sample pair, as shown in Fig.~\ref{fig:fig_D1}. Due to pairs of readout points being equidistant from $\varepsilon = 0$, this extrapolation is conveniently done over the same distance, yielding $\mu\mathrm{_b^T}$ and $\mu\mathrm{_b^B}$. Due to the inability to directly measure standard deviations of the blocked reference signals, we assume $\sigma\mathrm{_b^T} = \sigma\mathrm{_u^T}$ and $\sigma\mathrm{_b^B} = \sigma\mathrm{_u^B}$. 

All measurements, fits, and extrapolations for acquiring reference quantities are automatically built into our routine. We note that if a fast line were connected to SP, a.c.~compensation would be possible during pulsing. This would eliminate crosstalk-induced shifts on the sensing peak, and both blocked and unblocked references could be obtained from static reference measurements by the relationship: $\mu\mathrm{_u^T} = \mu\mathrm{_b^B}$, $\mu\mathrm{_u^B} = \mu\mathrm{_b^T}$, $\sigma\mathrm{_u^T} = \sigma\mathrm{_b^B}$ and $\sigma\mathrm{_u^B} = \sigma\mathrm{_b^T}$.

\setcounter{equation}{0}
\section{\label{app:MLE_methods}Maximum Likelihood Estimates}
Maximum likelihood estimation (MLE) is a statistical method used to estimate the parameters of a given probability distribution based on observed data. It works by maximising a likelihood function, such that the observed data is most probable under the assumed distribution.

\subsection{Calculating scores}
\label{app:score_calculation}
In our case, the probability distribution of interest is the Gaussian mixture model (GMM) given by Eq.~\ref{eq:GMM} of the main text.
The likelihood function of a GMM, $L^j$, given observed data, $X^j = \{x^j_1, x^j_2, ..., x^j_m\}$, is the product of individual probabilities,
\begin{equation}
L^j(\boldsymbol{w}^j, \tilde{\boldsymbol\mu}^j, \tilde{\boldsymbol\sigma}^j | X^j) = \prod_m p(x_m^j | \boldsymbol{w}^j, \tilde{\boldsymbol\mu}^j, \tilde{\boldsymbol\sigma}^j).
\end{equation}
If we let $\theta^j = \{\boldsymbol{w}^j, \tilde{\boldsymbol\mu}^j, \tilde{\boldsymbol\sigma}^j\}$, then MLE is the process of finding the distribution parameters, $\hat{\theta}^j$, for which $L^j$ is maximised. We find $\hat{\theta}^j$ using the L-BFGS-B algorithm as implemented in Python's SciPy package \cite{scipy}. However, since the number of observations $m$, i.e. pulsed measurement repetitions, is large, one encounters an underflow when calculating $L^j$. We overcome this common problem by instead maximising the log-likelihood, $\ell^j$:
\begin{equation}
    \ell^j(\boldsymbol{w}^j, \tilde{\boldsymbol\mu}^j, \tilde{\boldsymbol\sigma}^j | X^j) = \sum_m \sum_i w_i^j\mathcal{N}(x_m^j|\tilde{\mu}_i^j, \tilde{\sigma}_i^j),
\end{equation}
with $i=\{u, b\}$ indicating unblocked or blocked, and $w_u^j = 1 - w_b^j$. 

The superscript $j$ is kept as a reminder to the reader that this procedure is done for each sample, and repeated across sample pairs. We use Lorentzian barrier functions (chosen for smoothness) to constrain the MLE to bounds based on reference quantities (Sec.~\ref{app:references}:
\begin{subequations}
\begin{align}
    \begin{split}
        \left\vert\frac{(\tilde{\mu}_b^j - \tilde{\mu}_u^j) - (\mu_b^j - \mu_u^j)}{\mu_b^j - \mu_u^j}\right\vert &\leq 0.1,
    \end{split} \\
    \label{eq:const_1}
    \begin{split}
        \left\vert\frac{\tilde{\mu}_i^j - {\mu}_i^j}{\mu_b^j - \mu_u^j}\right\vert &\leq 0.5,
    \end{split} \\
    \begin{split}
        \left\vert\frac{\tilde{\sigma}_i^j - {\sigma}_i^j}{\sigma_i^j}\right\vert &\leq 0.2.
    \end{split}
    \label{eq:const_3}
\end{align}
\end{subequations}
In this way, we ensure that $\hat{\theta}^j$ stays near the expected distribution parameters, while still permitting minor deviations that may arise due to drift in the sensor signal.

The optimised blocked weights are assigned to $w_b^T = w^T$ and $w_b^B = w^B$, as denoted in the main text. Application of Eq.~\ref{eq:score_func} directly follows to calculate the PSB score.

\subsection{\label{app:outcome_methods}Assigning outcomes}
As stated in Sec.~\ref{subsec:criteria} of the main text, we use the Bayesian Information Criterion (BIC) to assign readout signal distributions of each sample pair to an outcome among `No PSB', `Top PSB', `Bottom PSB' and `Latching'. 

The BIC rewards outcomes with high likelihoods and mitigates over-fitting by penalising outcomes with more distribution parameters. The BIC is defined as,
\begin{equation}
\mathrm{BIC} = k\ln{(2m)} - 2\hat{\ell},
\label{eq:BIC}
\end{equation}
where $k$ is the number of distribution parameters estimated by an outcome (see Table \ref{tab:tab_1}) and $\hat{\ell}$ is the maximised value of the outcome's log-likelihood function. The outcome that is assigned to a sample pair is the one with the smallest BIC.

The $\hat{\ell}$ of each outcome is found by joining the maximised log-likelihoods of relevant distributions between paired samples as follows:

\begin{subequations}
\begin{align}
    \begin{split}
        \hat{\ell}_\mathrm{No\, PSB} &= \hat{\ell}_\mathrm{s}^T + \hat{\ell}_\mathrm{s}^B,
    \end{split} \label{eq:loglike-NoPSB} \\[4pt]
    \begin{split}
        \hat{\ell}_\mathrm{Top\, PSB} &= \hat{\ell}_\mathrm{d}^T + \hat{\ell}_\mathrm{s}^B,
    \end{split} \label{eq:loglike-TopPSB} \\[4pt]
    \begin{split}
        \hat{\ell}_\mathrm{Bottom\, PSB} &= \hat{\ell}_\mathrm{s}^T + \hat{\ell}_\mathrm{d}^B,
    \end{split} \label{eq:loglike-BotPSB} \\[4pt]
    \begin{split}
        \hat{\ell}_\mathrm{Latching} &= \hat{\ell}_\mathrm{d}^T + \hat{\ell}_\mathrm{d}^B.
    \end{split} \label{eq:loglike-latching}
\end{align}
\end{subequations}
Subscripts $s$ and $d$ indicate whether a single ($w^j = 0$) or double ($w^j \neq 0$) Gaussian is used in the MLE. Hence, determining the outcome of a sample pair boils down to determining whether the readout signal distribution for each sample is better described by a single or double Gaussian. This leads to the expanded forms of Eq.~\ref{eq:BIC} as follows:
\begin{subequations}
\begin{align}
    \begin{split}
        \mathrm{BIC}_\mathrm{No\, PSB} = 4\ln{(2m)} - 2\hat{\ell}_\mathrm{No\, PSB},
    \end{split} \\[4pt]
    \begin{split}
        \mathrm{BIC}_\mathrm{Top\, PSB} = 7\ln{(2m)} - 2\hat{\ell}_\mathrm{Top\, PSB},
    \end{split} \\[4pt]
    \begin{split}
        \mathrm{BIC}_\mathrm{Bottom\, PSB} = 7\ln{(2m)} - 2\hat{\ell}_\mathrm{Bottom\, PSB},
    \end{split} \\[4pt]
    \begin{split}
        \mathrm{BIC}_\mathrm{Latching} = 9\ln{(2m)} - 2\hat{\ell}_\mathrm{Latching},
    \end{split}
\end{align}
\end{subequations}
with $2m=2000$, due to each sample contributing 1000 data points.

\begin{table}[t]
\caption{\label{tab:tab_1}Parameters of each outcome model used in BIC calculations.}
\begin{ruledtabular}
\renewcommand{\arraystretch}{1.2}
\begin{tabular}{ccc}
Outcome & Model Parameters & $k$\\ 
\hline
No PSB & $\tilde{\mu}_u^T$, $\tilde{\mu}_u^B$, $\tilde{\sigma}_u^T$, $\tilde{\sigma}_u^B$ & 4 \\
Top PSB & $\tilde{\mu}_u^T$, $\tilde{\mu}_u^B$, $\tilde{\sigma}_u^T$, $\tilde{\sigma}_u^B$, $\tilde{\mu}_b^T$, $\tilde{\sigma}_b^T$, $w^T$ & 7 \\
Bottom PSB & $\tilde{\mu}_u^T$, $\tilde{\mu}_u^B$, $\tilde{\sigma}_u^T$, $\tilde{\sigma}_u^B$, $\tilde{\mu}_b^B$, $\tilde{\sigma}_b^B$, $w^B$ & 7 \\
Latching   & $\tilde{\mu}_u^T$, $\tilde{\mu}_u^B$, $\tilde{\sigma}_u^T$, $\tilde{\sigma}_u^B$, $\tilde{\mu}_b^T$, $\tilde{\sigma}_b^T$, $\tilde{\mu}_b^B$, $\tilde{\sigma}_b^B$, $w^B$ & 9 \\ 
\end{tabular}
\end{ruledtabular}
\end{table}

When calculating $\hat{l}_d^j$, we use the constraint $w_b^j \geq 0.1$ to penalise the MLE when $X^j$ follows a single Gaussian distribution. Without this constraint, a situation where $\hat{l}_d^j \sim \hat{l}_s^j$ may occur, making no outcome appear favourable.

Special care is also taken when calculating $\hat{\ell}_\mathrm{Latching}$ (Eq.~\ref{eq:loglike-latching}). Ideally, blocked weights for both samples in a pair should be equal, $w^T = w^B$, because the interdot tunnelling rate should be symmetric with respect to the pulsing direction. If this is not the case, $\hat{\ell}_\mathrm{Latching}$ is overestimated. To account for this, $\hat{l}_d^j$ is re-calculated for the sample with the smaller blocked weight, making it equal to the larger one. It is for this reason that the `Latching' outcome has 9, rather than 10 distribution parameters (see \ref{tab:tab_1}).

An example of the split in outcomes over sample pairs is shown in Fig.~\ref{fig:fig_4}(a) of the main text.

\subsection{Detecting anomalies}
Anomalous measurements may occur if a sample's readout voltage point lands within the thermally broadened range of the interdot charge transition, or if latching occurs from a nearby transition that is not the interdot charge transition, either during reference acquisition or during pulsing. To catch these instances, we conduct the following checks,
\begin{align}
    \begin{split}
        \frac{\underset{j}\max\, \sigma_u^j}{\underset{j}\min\, \sigma_u^j} &< 1.2
    \end{split} \\
    \begin{split}
        \frac{\underset{j}\max\, \left\vert\mu_b^j - \mu_u^j\right\vert}{\underset{j}\min\, \left\vert{\mu_b^j - \mu_u^j}\right\vert} &< 1.5
    \end{split} \\
    \begin{split}
        \frac{\left\vert\tilde{\mu}_b^j - \tilde{\mu}_u^j\right\vert}{\underset{j}\max\, \left\vert\mu_b^j - \mu_u^j\right\vert} &< 1.5 \quad \forall \quad \lbrace j\, |\, w_b^j \geq 0.1 \rbrace
    \end{split}
    \label{eq:pulsing_check}
\end{align}
where the parameters in Eq.~\ref{eq:pulsing_check} are found by unconstrained MLE. If any one check fails, the sample pair is excluded from the PSB score calculation and is assigned a `No PSB' outcome.

\setcounter{equation}{0}
\section{\label{app:optimisation}Optimisation Criteria}
Mirroring the stopping criteria in Sec.~\ref{subsec:criteria} of the main text, the criteria for optimisation are twofold. Priority is given to barrier voltage configurations that show oscillations, and will qualify for optimisation if $R^2 > 0.35$. If none exist, configurations may also qualify if their PSB score is at least 0.05 and half or more outcomes are either `Top PSB' or `Bottom PSB'. If either criterion is met for multiple configurations, the most promising candidate is taken to be the one with highest score.

The Nelder-Mead algorithm is used for optimisation with the PSB score as the objective function. Bounds of $\pm30$\! mV are used on each barrier gate, centred on the most promising candidate. The original barrier gate voltage bounds are disabled during optimisation, making it possible to evaluate configurations outside the initial barrier voltage search space. The algorithm iteratively updates vertices of a simplex using reflection, expansion, contraction, and shrinkage operations to converge upon vertices with high scores. A good initial simplex should be large enough to diversify the measured PSB scores, but needn't be much larger than the effective volume of the exploratory barrier configurations (see Supplementary Materials). With this in mind, we choose the following four initial simplex points:

\[
\begin{aligned}
    (\mathrm{BL}, \mathrm{BM}, \mathrm{BR}) \\
    (\mathrm{BL}, \mathrm{BM} + \delta_{\mathrm{BM}}, \mathrm{BR} + \delta_{\mathrm{BR}}) \\
    (\mathrm{BL} - \delta_{\mathrm{BL}}, \mathrm{BM}
    - \delta_{\mathrm{BM}}, \mathrm{BR}) \\
    (\mathrm{BL} + \delta_{\mathrm{BL}}, \mathrm{BM}, \mathrm{BR} - \delta_{\mathrm{BR}}).
\end{aligned}
\]
In the above, $\mathrm{BL}$, $\mathrm{BM}$ and $\mathrm{BR}$ refer to the barrier voltages of the most promising candidate, and $\delta_{\mathrm{BL/BM/BR}}$ correspond to twice the sampling resolution along each barrier dimension used during exploration (see Supplementary Materials). In the full run of our routine presented in the main text, this equates to 5.5/8.5/7\! mV. This optimisation loop led to PSB and oscillations being found at interdot 23, as highlighted in Fig.~\ref{fig:fig_5}b.

\begin{figure}[t]
\includegraphics[width=\linewidth]{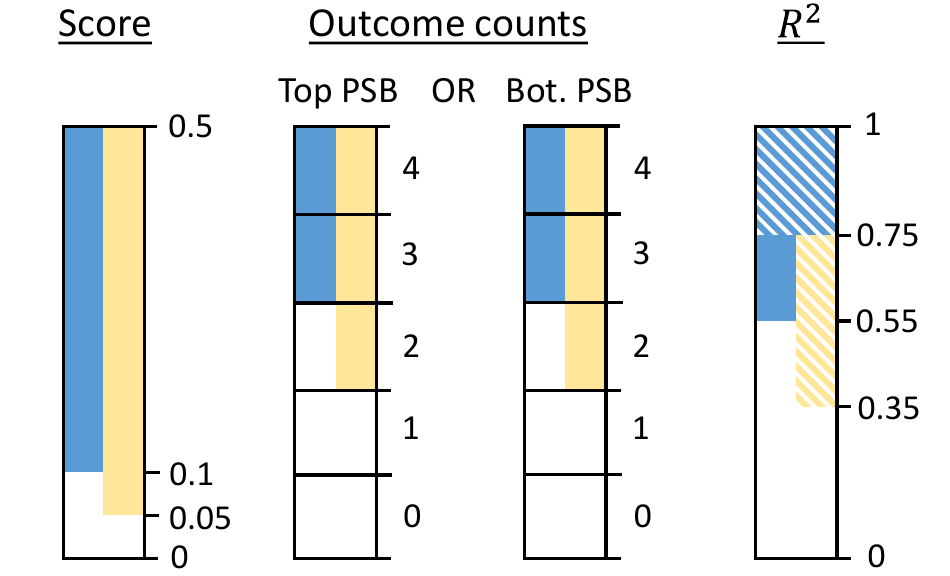}
\caption{\label{fig:fig_F1}\textbf{Overview of stopping and optimisation criteria.} Solid blue segments across bar plots visualise the range of values that each criterion must meet simultaneously in order for a qubit to be deemed present. Solid yellow segments visualise value ranges, also to be met simultaneously, for which a promising barrier voltage candidate would qualify for optimisation.  Hatched blue and yellow segments indicate ranges of $R^2$ that are stand-alone sufficient to satisfy qubit and optimisation conditions, respectively.}
\end{figure}

\clearpage

\onecolumngrid
\setcounter{section}{0}
\setcounter{equation}{0}
\setcounter{figure}{0}
\setcounter{table}{0}
\setcounter{page}{1}
\makeatletter
\renewcommand{\thesection}{S\! \Roman{section}}
\renewcommand{\thesubsection}{\Alph{subsection}}
\renewcommand{\theequation}{S\arabic{equation}}
\renewcommand{\thefigure}{S\arabic{figure}}
\renewcommand{\thetable}{S\arabic{table}}
\renewcommand{\bibnumfmt}[1]{[S#1]}
\renewcommand{\citenumfont}[1]{S#1}

\titleformat{\section}[block]
  {\normalfont\small\bfseries\centering}
  {\thesection.}
  {0.75em}
  {\uppercase}

{\centering
\large\bfseries 
Supplementary Materials: Automated All-RF Tuning \\ for
Spin Qubit Readout and Control\par
}

\section{Experimental Set-up}
A Leiden Cryogenics dry dilution cryostat was used to conduct the measurements. A Quantum Machines OPX+ was used to deliver all high frequency pulses. The high frequency lines for pulsing PL and PR were attenuated 38\! \unit{\dB}. To extend the OPX+ range of $\pm 0.5$\! \unit{\volt}, a Texas Instruments THS3491 Current Feedback Amplifier was placed on the OPX+ output with a gain of 5V/V. The reflectometry signal was only delivered to the device during the readout time with an amplitude of 20\! mVpp, followed by 64\! \unit{\dB} of cold attenuation. The $LC$ tank circuit is formed by a parasitic capacitance and an inductor mounted on the sample PCB. The reflectometry tone is routed by a MiniCircuits ZFDC-20-50-S+ directional coupler. The reflected signal is amplified by a CITLF3 cryogenic amplifier at the 4\! \unit{\kelvin} stage and homodyne demodulated by the OPX+. Rf-lever arms were calibrated using the method described in \cite{jirovec2021singlet}.

\section{Machine Learning Models}
\label{sup:ML_models}

\subsection{Architectures and Training}

The small input size of $20\times20$ pixels to the Interdot CNN lends itself to a very simple neural network architecture comprising just three convolutional layers and three fully-connected (FC) layers. The convolutional layers each use $3\times3$ kernels over increasing channel depths from 16 to 64. A kernel dilation of 2 is used in the first layer, $2\times2$ max-pooling is used after the second and third layers. The FC layers reduce the flattened feature vector to 128, then 32 features, and finally a single scalar output. ReLU activation functions are used in convolutional layers, Tanh in FC layers, and a sigmoid activation on the output. Drop-out probabilities of 0.15 and 0.7 are used between FC layers during training, with a learning rate set to 1e-3 and a weight decay of 2e-4. A batch size of 512 is used over 40 epochs with binary-cross-entropy (BCE) as the loss function. 

The Line CNN has a symmetric auto-encoder architecture. An encoder module maps interdot images into a latent space, while a decoder module extracts transition line features from this space. Four (transpose) convolutional layers are used in each, with equal numbers of channels. ReLU activations are used and a sigmoid activation is applied on the network's output. The encoder increases the depth of feature maps from 8 to 64, doubling each time. Larger kernel sizes were found to give the best performance, due to their ability to capture longer-range features as compared to standard $3\times3$ kernels \cite{ding2022scaling}. From the first through to the fourth encoder layer, kernel sizes of 13, 11, 9 and 5 are used, followed by $2\times2$ average pooling. The decoder uses exclusively $3\times3$ kernels and a stride of 2. Batch normalisation is used during training with a learning rate set to 1e-4 and a weight decay of 2e-4. A batch size of 32 is used over 70 epochs with BCE as the loss function.

The Angle CNN comprises three convolutional layers, with kernel sizes of 9, 7 and 5, respectively, followed by two FC layers. The channel depth increases from 16 to 32, each followed by $2\times2$ max pooling. The last convolutional layer, with 64 output channels, uses $4\times4$ max pooling. After flattening, the first FC layer outputs 64 features. All layers use ReLU as the activation function. Batch normalisation is used during training with a learning rate set to 2e-5 and a weight decay of 1.5e-4. A batch size of 32 is used over 50 epochs with mean-squared error (MSE) as the loss function

All training is done on a NVIDIA GeForce GTX 1080 Ti GPU with Adam as the optimiser.

\subsection{\label{sup:benchmarks}Benchmarking}
\begin{figure*}[b]
\includegraphics[width=\linewidth]{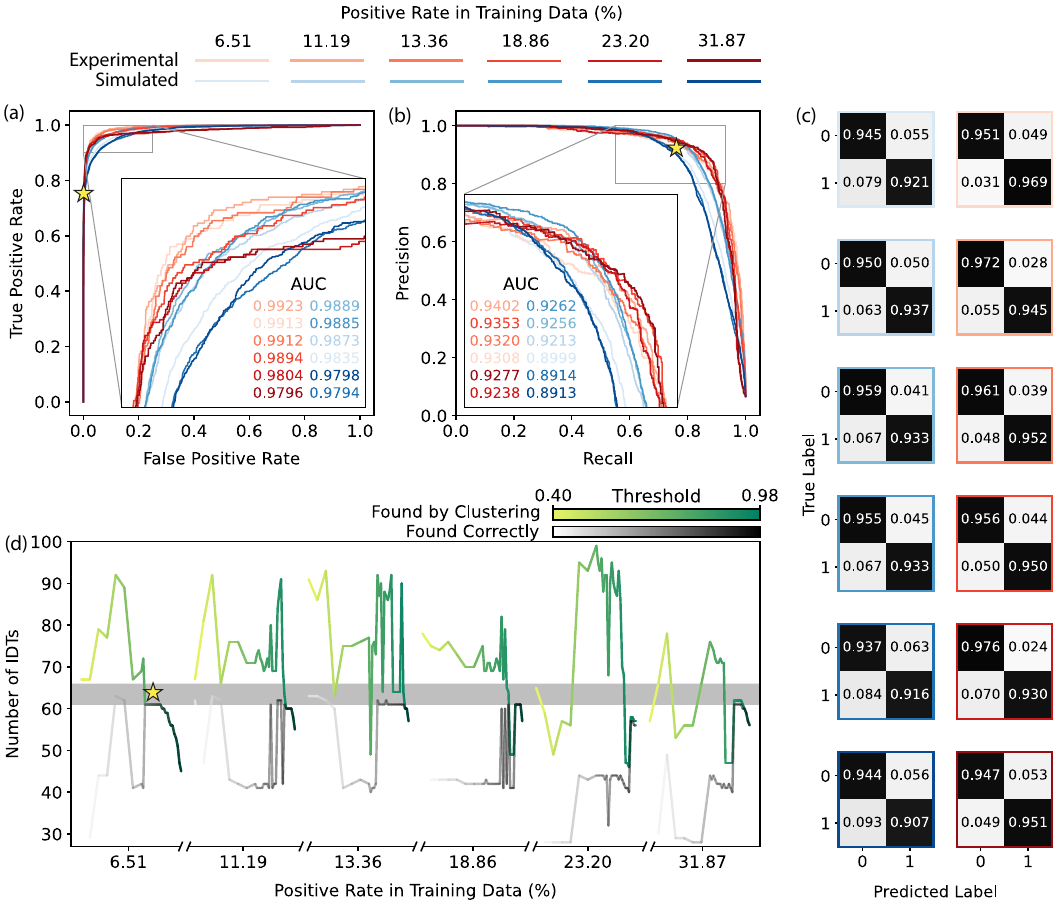}
\caption{\label{fig:fig_S1}\textbf{Performance of the Interdot CNN} \textbf{(a)} Receiver-Operator Characteristic (ROC) \textbf{(b)} Precision-recall curve. Stars indicate the performance of the model at the threshold at which the Interdot CNN was operated in experiments in the main text. \textbf{(c)} Normalised confusion matrices at the threshold for which the ROC curve locates furthest from the origin. \textbf{(d)} Clustering performance on positive classifications yielded by the Interdot CNN. Green curves indicate the total number of IDTs found as a function of the classification threshold. Grey curves indicate the total number of IDTs that were correctly located -- within 10 pixels of a true IDT location.}
\end{figure*}

\subsubsection{Interdot CNN}
\label{sup:benchmarks_classification}
Locating IDTs is an imbalanced classification problem, because the number of CSD patches that contain an IDT constitute a minority. This has been accounted for in a previous work by weighting towards patches with transition lines in the training data \cite{czischek2021miniaturizing}. Another well-known approach would be to add a weighting in the model's loss function that penalises predictions skewed toward the negative class. We combine these two approaches, but rather than weighting the loss function during training, we instead vary the classification threshold during inference, taking into account the clustering performance. Due to the added data redundancy about each IDT, we find that, perhaps unsurprisingly, using a threshold that produces a sub-optimal true-positive rate is preferred in order to form accurate clusters.

The stand-alone Interdot CNN is benchmarked on both simulated and hand-labelled experimental data using 50,000 and 10,200 samples, respectively. Both datasets contain a positive rate around 6.5\%. The receiver-operator characteristic (ROC) in Fig.~\ref{fig:fig_S1}a shows the true positive rate and false positive rate of the model's predictions at different thresholds. ROCs with a large area under their curve (AUC) are desirable. The confusion matrices in Fig.~\ref{fig:fig_S1}c are provided for the thresholds corresponding to the point in the upper most left of each ROC curve, where false positive and false negative rates are jointly minimised.

The first key observation is that the performance on experimental data is similar, if not better, than on simulated data in all cases. This underscores the ability of modern simulators \cite{van2024qarray} to produce charge stability diagrams that closely mimic those acquired in real experiments. Furthermore, this result also suggests that simulated data, if sufficiently realistic, can suffice in training machine learning models for automated quantum dot experiments. The small discrepancy in performance on simulated versus experimental data can be explained by the greater variety of noise and contrast levels contained in the simulated test dataset. This variety carries over to the training dataset, used to better generalise the Interdot CNN.

The second observation is that, for both simulated and experimental datasets, the AUC shows a marginal improvement for slight oversampling levels but quickly decreases when the oversampling becomes greater than $\sim 3\times$ the positive rate. This is also the case for the AUC in precision-recall plots shown in Fig.~\ref{fig:fig_S1}(b). The precision is defined as $TP/(TP + FP)$, and recall as $TP/(TP + FN)$, where $TP$, $FP$ and $FN$ are the true positive, false positive and false negative rates, respectively. 

\subsubsection{\label{sup:benchmarks_clustering}$k$-Means Clustering}
The combined performance of the Interdot CNN and clustering algorithm is presented in Fig.~\ref{fig:fig_S1}(d) for the experimental test data only. The total number of interdot transitions (IDTs) found as a function of the classification threshold is plotted in green. The number of IDTs found that are within 10 pixels of a true IDT position are plotted in gray. Accounting for IDTs located very close to CSD edges, we define an acceptable range of found IDTs between 61 and 66, given by the shaded area. 

Three main regimes can be identified from these plots. For low thresholds, the Interdot CNN's false positive rate is high, which, firstly, creates additional clusters away from the true IDT locations, and secondly, causes offsets in cluster centroids. This leads to a surplus and general inaccuracy in locating IDTs, evident from the large gap between green and gray lines. For moderate thresholds, spurious false positive classifications are reduced and this gap closes, making the clustering algorithm effective at both finding the correct number and location of IDTs. For high thresholds, the false negative rate increases, causing small IDT clusters to vanish and hence also the number of total found IDTs, however, these remaining IDTs are found accurately, as indicated by the overlap between green and grey lines. Abrupt jumps in the number of found IDTs, typically between the first and second regimes, is a consequence of the $k$-means algorithm combining several nearby positive classifications into one large cluster, or vice versa, partitioning a neighbourhood of positive classifications into small, more compact clusters. 

The combined performance is highest and most stable when the threshold range over which green and gray lines overlap in the shaded area is widest. This point exists at a threshold of 0.81 for the model trained on a 6.51\% rate of positive data, and is what we use in our routine and for the results presented in the main text. This operating point is marked with a star, with corresponding true positive rate, false positive rate, precision and recall values shown in Fig.~\ref{fig:fig_S1}a/b.

\newpage
\subsubsection{Angle and Line CNNs}
\label{sup:benchmarks_lines}
The Line CNN is trained on $1\times10^5$ simulated images of charge-sensed data \cite{van2024qarray}. The Angle CNN's training data, on the other hand, is generated in two ways. One subset includes predictions made by the trained Line CNN on $1\times10^5$ simulated images. A second subset includes an additional $5 \times 10^4$ line images generated retroactively from random angles. The second subset augments the first with the addition of highly local noise, introduced via dilation, blurring, affine transformations, warping and white noise operations, characteristic of line predictions made on non-ideal interdot images. Furthermore, control over the angle parametrisation allows unique slopes for all transition lines to be captured, which is not possible with a constant capacitor model. 

We first evaluate the overall accuracy when the two models, the Line CNN and Angle CNN, are applied in sequence. We compare simulated data, comprising 500 interdot images, with 308 experimentally measured images, each hand-labelled. The overall accuracy is summarised in Table \ref{tab:tab_S1} in terms of the mean absolute error (MAE) and mean squared error (MSE) of the predicted transition line angles. As noted in the previous section, the overall low errors, and similarity in performance between simulated and experimental datasets, highlight how the training data effectively captures the noise characteristics and interdot features present in real measurements. The higher MSE of simulated dataset can be attributed to a greater diversity of interdot images causing a handful of poorer predictions to inflate the error. 

\begin{table}[h]
\centering
\begin{minipage}{0.42\linewidth}
\centering
\caption{\label{tab:tab_S1}Errors in angles predicted by the Angle CNN on simulated and experimental data.}
\begin{ruledtabular}
\renewcommand{\arraystretch}{1.2}
\begin{tabular}{c|cc|cc}
 & \multicolumn{2}{c|}{Simulated} & \multicolumn{2}{c}{Experimental} \\
 Angle                 & MAE (\unit{\degree})    & MSE     & MAE (\unit{\degree})      & MSE      \\ \hline
$\alpha_1$        & 0.69          & 0.155          & 0.88            & 0.023           \\
$\alpha_2$        & 0.62          & 0.033          & 0.89            & 0.026           \\
$\beta_1$         & 0.78          & 0.142          & 0.87            & 0.031           \\
$\beta_2$         & 0.71          & 0.126          & 1.07            & 0.042           \\
$\gamma_1$        & 0.46          & 0.013          & 1.48            & 0.046           \\
$\gamma_2$        & 0.29          & 0.005          & 0.34            & 0.004           \\
$\gamma_3$        & 0.50          & 0.010          & 0.80            & 0.018           \\
$\gamma_4$        & 0.37          & 0.005          & 0.79            & 0.015           \\ \hline
Avg.    & 0.55 & 0.061 & 0.89   & 0.026 \\ 
\end{tabular}
\end{ruledtabular}
\end{minipage}
\end{table}

To decouple errors stemming from the Line CNN and Angle CNN, we now evaluate the accuracy of each model in isolation. To do this, we use the Jaccard index, which measures the similarity between ground truth line images and each model's predictions. The Jaccard index $J(\mathrm{\textbf{x}}, \mathrm{\textbf{y}})$ is defined as the intersection between the predicted image $\mathrm{\textbf{x}}$ and its target $\mathrm{\textbf{y}}$, divided by their union,
\begin{equation}
    J(\mathrm{\textbf{x}}, \mathrm{\textbf{y}}) = \frac{\sum_i x_i \land y_i}{\sum_i x_i \lor y_i},
\end{equation}
where $i$ runs over all pixel indices, $x_i$ is the binary output pixel value, $y_i$ is the target pixel value, $\land$ is the logical \texttt{AND} operator, and $\lor$ is the logical \texttt{OR} operator.

Since the output of the Line CNN is continuous, we binarise its pixels using a threshold of 0.5, followed by skeletonisation. For the Angle CNN, we feed it with the ground truth images themselves, which mimic a perfect prediction by the Line CNN, then reconstruct the line image from the returned angles. When the two models are evaluated together, the raw output from the Line CNN is given to the Angle CNN instead. 

\begin{table*}[t]
\caption{\label{tab:tab_S2}Jaccard index for line images constructed from Line and Angle CNN outputs on simulated and experimental data, and for varying dilations sizes applied to ground truth line images. Starred values correspond to the median sample shown in \ref{fig:fig_S2}}.
\begin{ruledtabular}
\renewcommand{\arraystretch}{1.2}
\begin{tabular}{l|llll||llll||llll}

       & \multicolumn{4}{l||}{Thresholded Line CNN}                               & \multicolumn{4}{l||}{Angle CNN}  
            & \multicolumn{4}{l}{Line CNN + Angle CNN} \\ \hline
            
Kernel & \multicolumn{2}{l|}{Simulated}      & \multicolumn{2}{l||}{Experimental} & \multicolumn{2}{l|}{Simulated}      & \multicolumn{2}{l||}{Experimental} & \multicolumn{2}{l|}{Simulated}      & \multicolumn{2}{l}{Experimental} \\
size   & Mean  & \multicolumn{1}{l|}{Med.} & Mean            & Med.          & Mean  & \multicolumn{1}{l|}{Med.} & Mean          & Med.     & Mean  & \multicolumn{1}{l|}{Med.}  & Mean          & Med.        \\ \hline

1      & 0.710 & \multicolumn{1}{l|}{0.747}  & 0.368           & 0.370   & 0.366 & \multicolumn{1}{l|}{0.360}  & 0.460         & 0.425        & 0.410 & \multicolumn{1}{l|}{0.414}  & 0.258         & $0.248^*$         \\

3      & 0.996 & \multicolumn{1}{l|}{1.000}  & 0.958           & 1.000   & 0.962 & \multicolumn{1}{l|}{1.000}  & 0.995         & 1.000           & 0.958 & \multicolumn{1}{l|}{1.000}  & 0.905         & $0.959^*$         \\

5      & 1.000 & \multicolumn{1}{l|}{1.000}  & 0.993           & 1.000   & 0.996 & \multicolumn{1}{l|}{1.000}  & 1.000         & 1.000           & 0.983 & \multicolumn{1}{l|}{1.000}  & 0.990         & $1.000^*$         \\

7      & 1.000 & \multicolumn{1}{l|}{1.000}  & 0.996           & 1.000   & 0.999 & \multicolumn{1}{l|}{1.000}  & 1.000         & 1.000           & 0.991 & \multicolumn{1}{l|}{1.000}  & 0.996         & $1.000^*$         \\ 

9      & 1.000 & \multicolumn{1}{l|}{1.000}  & 0.998           & 1.000   & 1.000 & \multicolumn{1}{l|}{1.000}  & 1.000         & 1.000           & 0.994 & \multicolumn{1}{l|}{1.000}  & 0.998         & $1.000^*$         \\ 
\end{tabular}
\end{ruledtabular}
\end{table*}

\begin{figure*}[t]
\includegraphics[width=0.65\linewidth]{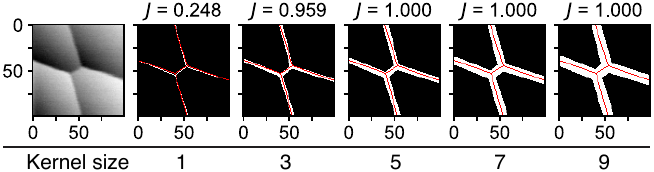}
\caption{\label{fig:fig_S2}Example Jaccard indexes for line images reconstructed using Line and Angle CNNs (red) on the median experimental IDT voltage scan (marked with an asterisk in Table \ref{tab:tab_S1}). The Jaccard index increases for dilated versions of the ground truth line image (white). Dilations are achieved by convolving the ground truth line image with square kernels of size 3, 5, 7, and 9, from left to right.}
\end{figure*}

The first row of Table \ref{tab:tab_S2} shows the Jaccard index for the above three cases on both simulated and experimental test datasets. For reference, the Jaccard index for two identical images is 1, for pure noise would be around 0.009, and for randomly placed transition lines would be around 0.011. 

A reduction in the Jaccard index on experimental data when Line and Angle CNNs are applied in sequence indicates a propagation of error. The performance gap between simulated and experimental data offers a clue to the dominant error source. Turning to the Line CNN, here the performance is inferior on experimental data. This is likely a result of transition line curvature, caused by tunnel coupling, which is absent from the simulated training data. Conversely, for the Angle CNN, the performance of simulated and experimental data is similar. This is consistent with the above claim, because both experimental and simulated inputs to the Angle CNN are images with straight lines. Finally, when the two models are evaluated together, the mean and median Jaccard indices are again higher for the simulated dataset. This leads to the conclusion that the Line CNN bears a greater contribution to the overall error when run on experimental data.

To give physical meaning to the above results, we re-calculate the Jaccard index over dilated versions of the ground truth line images. An example of this procedure is shown in Fig.~\ref{fig:fig_S2}. Under a dilation by a symmetric $n \times n$ kernel with $n \in \{3, 5, 7, 9\}$, an output pixel is deemed as intersecting a target pixel within a tolerance of $n-2$ pixels. For increasing kernel sizes, a convergence in the Jaccard index between Line and Angle CNNs can be observed. For the case where $n\geq5$, we measure a median Jaccard index of 1 across both datasets. In other words, a majority of interdot voltage scans will have their transition lines placed within 2 pixels of the true transition lines. For typical scan ranges of $35 \times 35$\! mV, this translates to a precision of 0.7\! mV. 

\newpage

\section{\label{sup:out_of_distribution}Out-of-distribution Detection}
\begin{figure*}[b]
\includegraphics[width=0.75\linewidth]{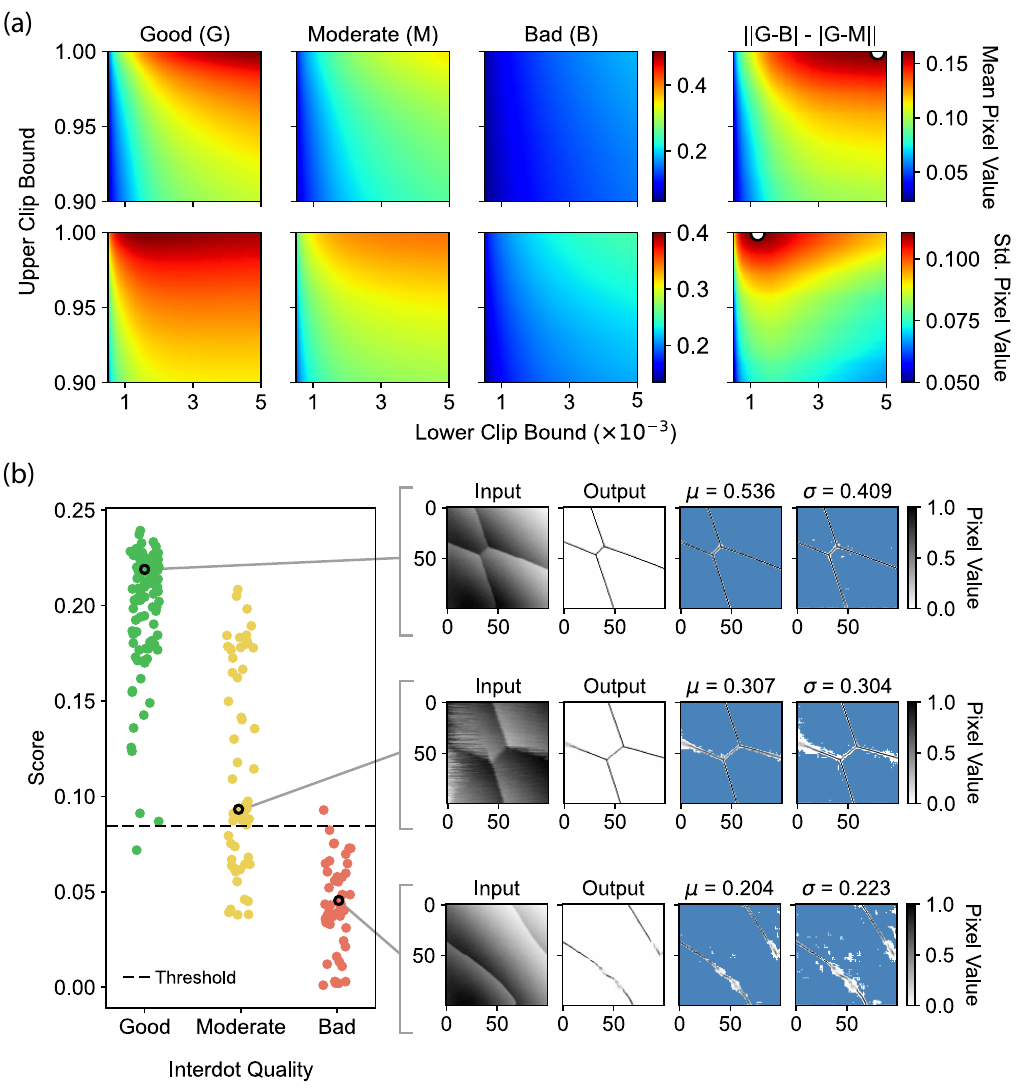}
\caption{\label{fig:fig_S3}\textbf{(a)} Means (top) and standard deviations (bottom) of pixel values for sets of `good', `moderate' and `bad' interdot images when clipped between different lower and upper bound values. Each point represents averages by set. White markers indicate the clipping values for which good and moderate images are best distinguished from bad images. \textbf{b)} Interdot quality score for each image alongside examples of clipped pixels from images in each group in blue (insets).}
\end{figure*}

As mentioned in Sec.~\ref{sec:segmentation} of the main text, the two-stage framework we use for parametrising transition lines -- applying the Line CNN followed the Angle CNN -- enabling poor interdot (IDT) scans to be detected. Consider the following scenarios: (i) the navigation stage incorrectly locates an IDT where none is present, (ii) a voltage adjustment tunes out of a well-defined double-dot regime, (iii) imperfect virtualisation causes the IDT to fall outside the voltage scan range. The quality of inference by a neural network reflects the similarity between a measurement and the expected input \cite{ziegler2022toward}. Hence, by analysing the pixel value distribution returned by the Line CNN, we develop a proxy for whether the aforementioned scenarios have occurred.

To develop this proxy, we analyse 101, 55 and 46 images of `good', `moderate' and `bad' IDT scans, respectively. In Fig.~\ref{fig:fig_S3}(a), we present the mean (top) and standard deviations (bottom) of pixel values for clipped outputs of the Line CNN for each set. For good scans, the pixel distribution is heavily gapped with a majority of pixels falling below $\sim0.01$ and above $\sim0.8$. Conversely, bad scans have appreciable counts of pixels between these values, due to a lack of confidence in assigning transition line locations. Hence, when clipping pixels outside bounds that lay near these values, we find that the mean of `good' scans tend to be skewed toward higher-valued pixels, while `bad' scans by lower-valued pixels. The high density of middle-valued pixels in `bad' scans also reduces their standard deviation. The white circles in (a) mark the clipping bounds for which the means and standard deviations between `good' and `moderate' scans are maximally different from `bad' scans. We define our proxy score as the product of the mean and standard deviation at these clipping bounds.

Our quality score for each image is shown in Figure \ref{fig:fig_S3}b. We run our routine at a threshold of 0.085, above which all but one good image passes from the analysis set, and all but one bad image does not pass. On occasion, imperfect virtualisation by our routine would cause scan images to feature just a single transition line, or pair of IDTs, both of which we found to also return high scores. To handle this, we required exactly one transition line to be detected along each border of the scan. If this additional check is passed, and the score is sufficient, only then would the IDT image continue through the segmentation stage (Sec.~\ref{sec:segmentation} in main text), and the barrier voltage configuration would be evaluated (Sec.~\ref{sec:evaluation} in main text).  

\vfill
\section{\label{sup:tuning_path}Tuning Path}
In this section, we substantiate our choice to use Latin hypercube sampling to explore the barrier voltage space, and explain why using 40 sampling points is sufficient in the majority of cases. We compare our approach with purely random sampling, which we will now treat. 

We consider a simplified picture of a PSB volume, $v$, that is continuous, without voids, and that lives inside the larger barrier voltage search space $V$. Inside $v$, we assume that our stopping criteria would be met. In this case, the probability that our tuning loop exits for a single sample would be $v/V$. For $s$ samples, the probability that at least one sample occupies $v$ is given by,
\begin{equation}
    P_{\mathrm{success}}= 1 - \left(1 - \frac{v}{V} \right)^s
\end{equation}
To meet a minimum confidence level $P_{\mathrm{success}} \geq p$, re-arranging the above yields,
\begin{align}
    %\left(1 - \frac{v}{V} \right)^s \geq 1 - p, \\[10pt]
    s \geq \frac{\ln{(1-p})}{\ln{(1-v/V)}}
\end{align}
Since $v \ll V$, we can approximate $\ln{(1-v/V)} \approx -v/V$, giving the final result
\begin{equation}
    s \geq \left(\frac{V}{v}\right) \ln{\frac{1}{1-p}}.
\label{eq: random-search}
\end{equation}

Latin hypercube sampling works by partitioning each dimension of the search space into $s$ equal intervals and evenly distributing the $s$ samples among them. For small $s$, this improves the efficiency with which the search space is explored by making the distance between samples more uniform. We verify this through a simple experiment in which we vary the number of samples in a tuning loop about IDTs that are known to exhibit PSB, followed by Monte-Carlo simulations.

\begin{figure}[ht]
\includegraphics[width=0.5\linewidth]{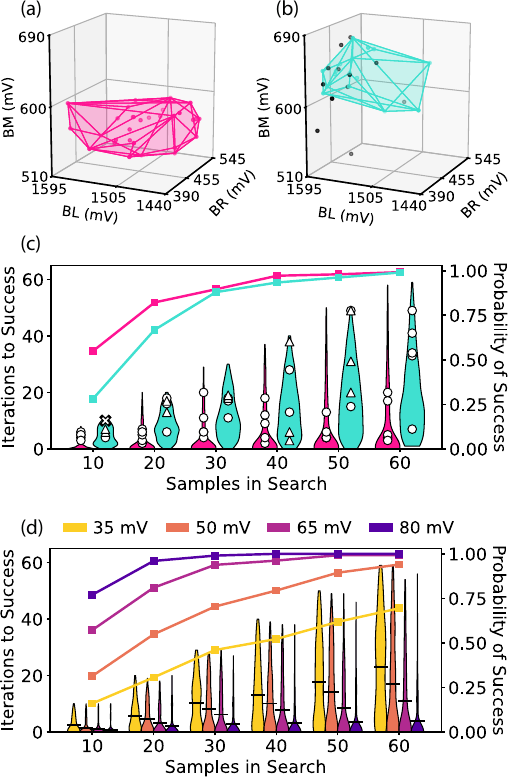}
\caption{\label{fig:fig_S4}\textbf{Monte-Carlo simulations of single-IDT tuning loops.} \textbf{(a-b)} Locations in barrier space where PSB and oscillations were found experimentally over 30 runs on (a) interdot 1 and (b) interdot 23. Every five runs, the number of sampling points was incremented by 10, starting at 10. Black markers indicate most promising candidates in runs that did not find PSB. These are depicted using triangles in panel (c). \textbf{(c)} Violin plots for the number of iterations to find PSB in 500 simulated searches. PSB volumes and distances are fixed and extracted from the experimental runs in (a-b). The probability of finding the PSB volume is plotted on the secondary y-axis. Data from (a-b) is plotted in white. Triangles mark iterations at which a most promising candidate was found in runs that did not meet the stopping criteria. The cross at 10 samples is a run where neither PSB nor a promising candidate was found. \textbf{(d)} Violin plots for the simulated number of iterations to find PSB for four different PSB volumes. Each violin plot is a distribution over 500 runs where the distance to the volume is randomised. The colour legend corresponds to the cube root of the PSB volume. All simulations use search bounds of $[-20,+90]$/$[-80,+90]$/$[-60,+80]$\! mV on BL/BM/BR.}
\end{figure}

In Fig.~\ref{fig:fig_S4}(a-b) we plot the convex hulls that enclose points at which PSB was found over 30 single-IDT tuning runs for (a) interdot 1 and (b) interdot 23. The bounds of the barrier voltage search space are the same as those presented in the main text. We take a conservative estimate of the PSB volume for each interdot by finding the largest parallelepiped that is contained within each hull. This yields $v^{1/3} = 60$\! mV and $v^{1/3} = 58$\! mV, respectively. We use these values to simulate tuning paths that search for hulls of the same volume and distance from the starting barrier configuration. 

The number of simulated iterations needed to find $v$ in each case are shown in Fig.~\ref{fig:fig_S4}(c). The trend in the number of iterations to reach success, as well as the success probability (secondary y-axis), is consistent with results from experimental tuning runs, overlayed in white. On average, interdot 23 requires more iterations to reach success, due to its PSB volume locating further away. This is a natural consequence of tracing a path through the sampled configurations that minimises large voltage jumps, causing regions near the starting configuration to be explored sooner. This path is determined by a greedy algorithm. Furthermore, the number of tuning loops that do not end in success, depicted by non-circular markers, is more frequent for interdot 23, where the PSB volume is marginally smaller. This is reflected in the lower simulated probability of success.

Figure \ref{fig:fig_S4}(d) shows simulated tuning runs for four different PSB volumes with randomised distances from the starting configuration. Here we learn that the probability of success scales non-linearly with the addition of more samples and competes with the average number of iterations to find $v$. 40 samples offer a good middle ground between these two variables, yielding overall high probabilities in a moderate number of iterations, which would keep measurement overheads low. Even for very small volumes of $v^{1/3} = 35$\! \unit{mV}, which amounts to $v/V = 0.01$ in our simulations, 40 Latin hypercube samples would be sufficient to find $v$ in a majority of tuning runs. By comparison, using Eq.~\ref{eq: random-search}, a random search with $p \geq 0.5$ would require a minimum of 70 samples for the same volume ratio.

\vfill

\newpage

\newpage

\section{Limitations of the routine}
\label{sup:limitations}
Despite the overall good performance of our routine, evidenced by its ability to autonomously find PSB and qubit oscillations at multiple interdot charge transitions, we discuss a few limitations. 

\subsection{Readout}
Firstly, our routine assumes a high enough sensitivity to achieve single-shot spin readout. If this is not possible, an adapted score function and set of likelihood functions based solely on averaged multi-shot readout signals is expected to work just as well. In this case, the extent of blockade would be given by the position of the (now single) Gaussian mean with respect to reference means. The score function then becomes a function of the distance of this mean from the unblocked reference mean in each pulsing direction. As for MLE calculations, the likelihoods of single and double Gaussian distributions would translate to the comparison of two single Gaussian distributions, albeit with different means and standard deviations. All other parts of our routine would be unchanged.

\subsection{Stopping criteria}
Regarding our ensemble of stopping criteria (Sec.~\ref{sec:evaluation} of the main text), there exist three special cases that may yield false negatives: 
\begin{enumerate}
    \item $0.55 < R^2 \leq 0.75$ but $\mathrm{score} \leq 0.1$ and/or the majority of outcomes are not PSB. 
    \item $R^2 \leq 0.55$ but  $\mathrm{score} > 0.1$ and the majority of outcomes are PSB.
    \item Decoherence occurs faster than the oscillation frequency or the period of the oscillations is larger or comparable to the maximum $t_{\mathrm{idle}}$.
\end{enumerate} 
\noindent 

In all of the above cases, the stopping criteria are not met. The first case could result from initialisation and readout errors caused by Landau-Zener transitions at the $ST_-$ anti-crossing. The second case could result from a highly asymmetric latching probability or, again, low state preparation or readout fidelities, leading to poor visibility. The third case is self-explanatory.  

In addition to these cases, false negatives may arise if the voltage configurations at which PSB emerge fall outside the barrier search space. Since the exact voltage coordinates of PSB are unknown \textit{a priori}, we have made an educated guess of the search space based on the barrier ranges over which our DQD confinement is maintained. For a fully automated tuning pipeline, this could be estimated algorithmically \cite{schuff2024fully}. In general, larger search spaces reduce the risk of missing relevant PSB coordinates, however this comes at the cost of measuring at more barrier voltage configurations in order to maintain a fine-grained voltage exploration.

Finally, it is important to realise that, while most sensitive to barrier gate voltages \cite{katiraee2025unified}, the emergence of PSB and singlet-triplet oscillations depends on a variety of other experimental parameters, such as pulsing amplitudes, pulsing directions, ramp times, magnetic field strengths and field orientations, which we set for this experiment. 

\newpage
\section{\label{sup:flowchart}Flowchart}
\begin{figure*}[b]
\includegraphics[width=0.79\linewidth]{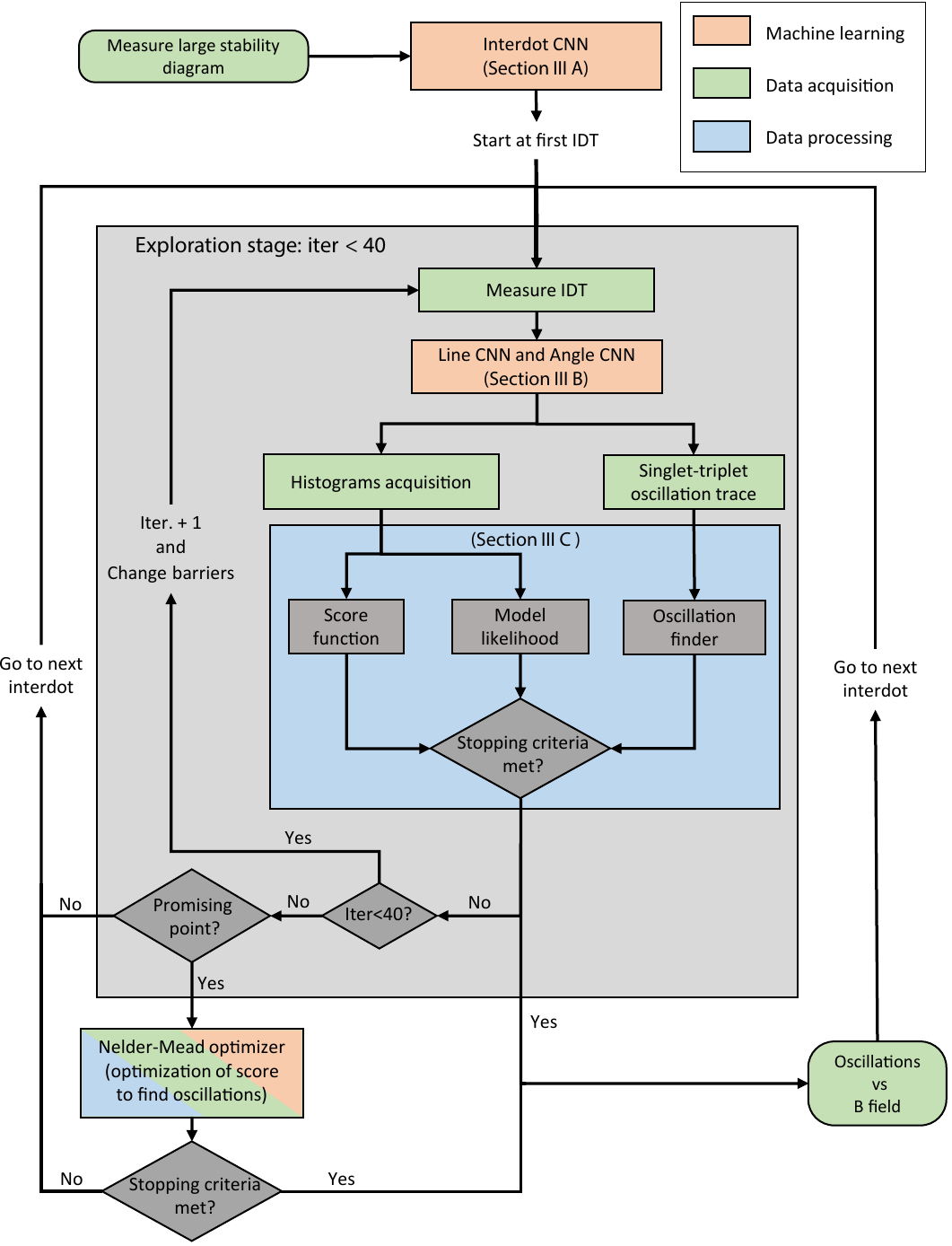}
\caption{\label{fig:fig_S5}\textbf{Flowchart of our routine for tuning spin qubits}. Colour codes highlight the function of each step and indicate parts of the routine that are enhanced by machine learning. The Optimisation Stage, if entered, follows the same sequence of measurements as the Exploration Stage, but barrier adjustments are informed by the PSB score rather than following a pre-defined path.}
\end{figure*}

An overview of our full routine's flow is presented in Fig.~\ref{fig:fig_S5}.

\clearpage

\section{Further Data}

\subsection{\label{sup:osc_fits}Oscillation Fits}
Below we present fits to $A\cos(2\pi f t +\phi)\exp(-t/T_2^\ast)$ for the data presented in Figure 5e. For each interdot, we plot oscillation traces that yielded the largest (top) and smallest (bottom) $T_2^\ast$. Oscillations at the electrostatic configuration for which the stopping criteria were met are shown in middle plots. We note that for interdots 13 and 16, a charge switch occurred in the device between the calibration of the voltage readout point and magnetic field sweep.

\begin{figure}[htbp!]
    \centering
    \includegraphics[width=0.86\linewidth]{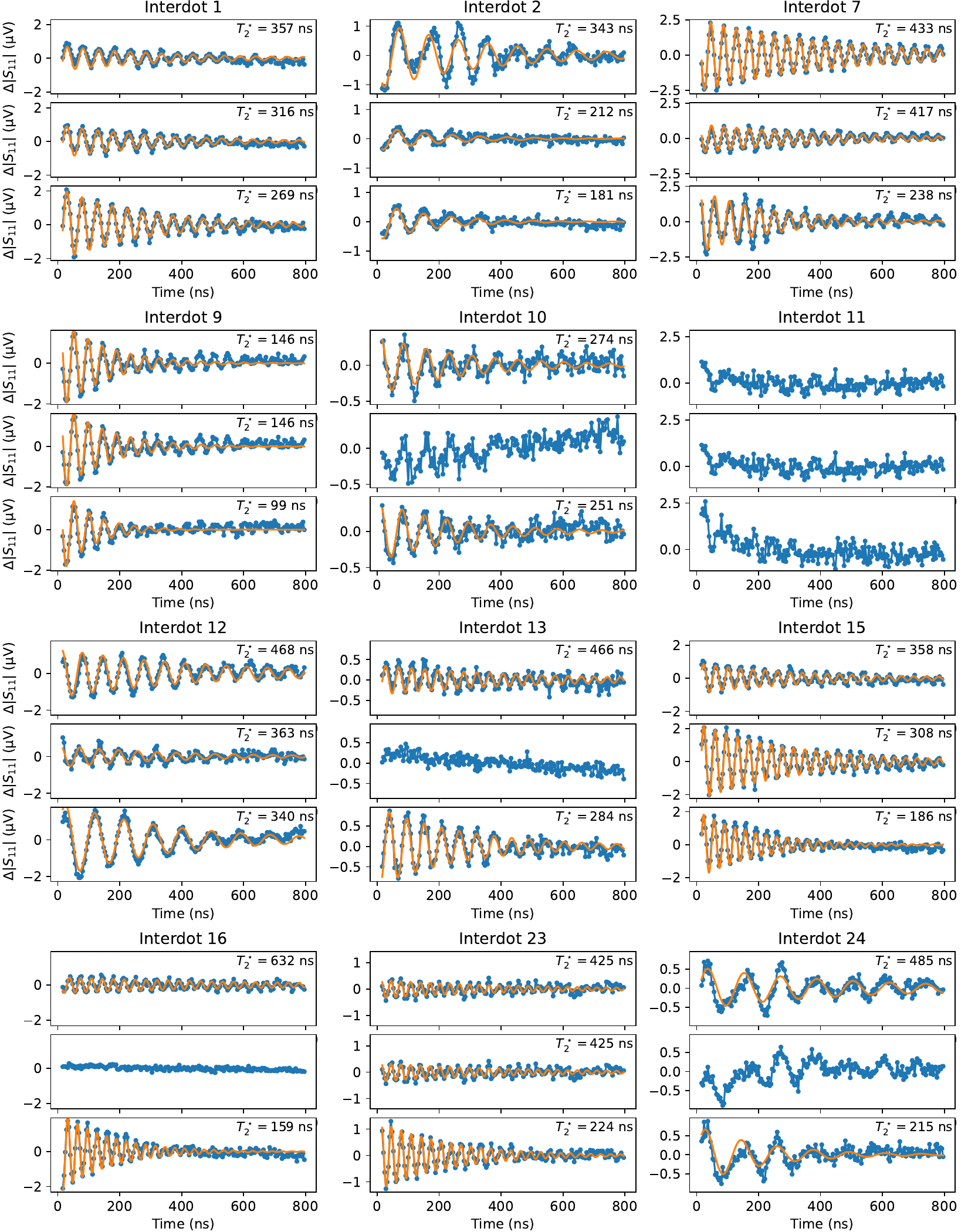}
\end{figure}

\subsection{\label{sup:magnetic_sweeps}Magnetic Field Sweeps}

Below we present raw data for the oscillations automatically acquired by our routine, each paired with their Fourier transform. The middle row of figure panels for each interdot, comprising both time and frequency domain data, corresponds to the electrostatic configuration at which the stopping criteria were met. Panel columns, from left to right, show data for variations in the voltage on BL, BM, and BR, respectively. On the top of each panel, the variation in the respective gate is indicated. The average signal along each time trace has been subtracted.

\begin{figure}
\includegraphics[width=0.8\linewidth]{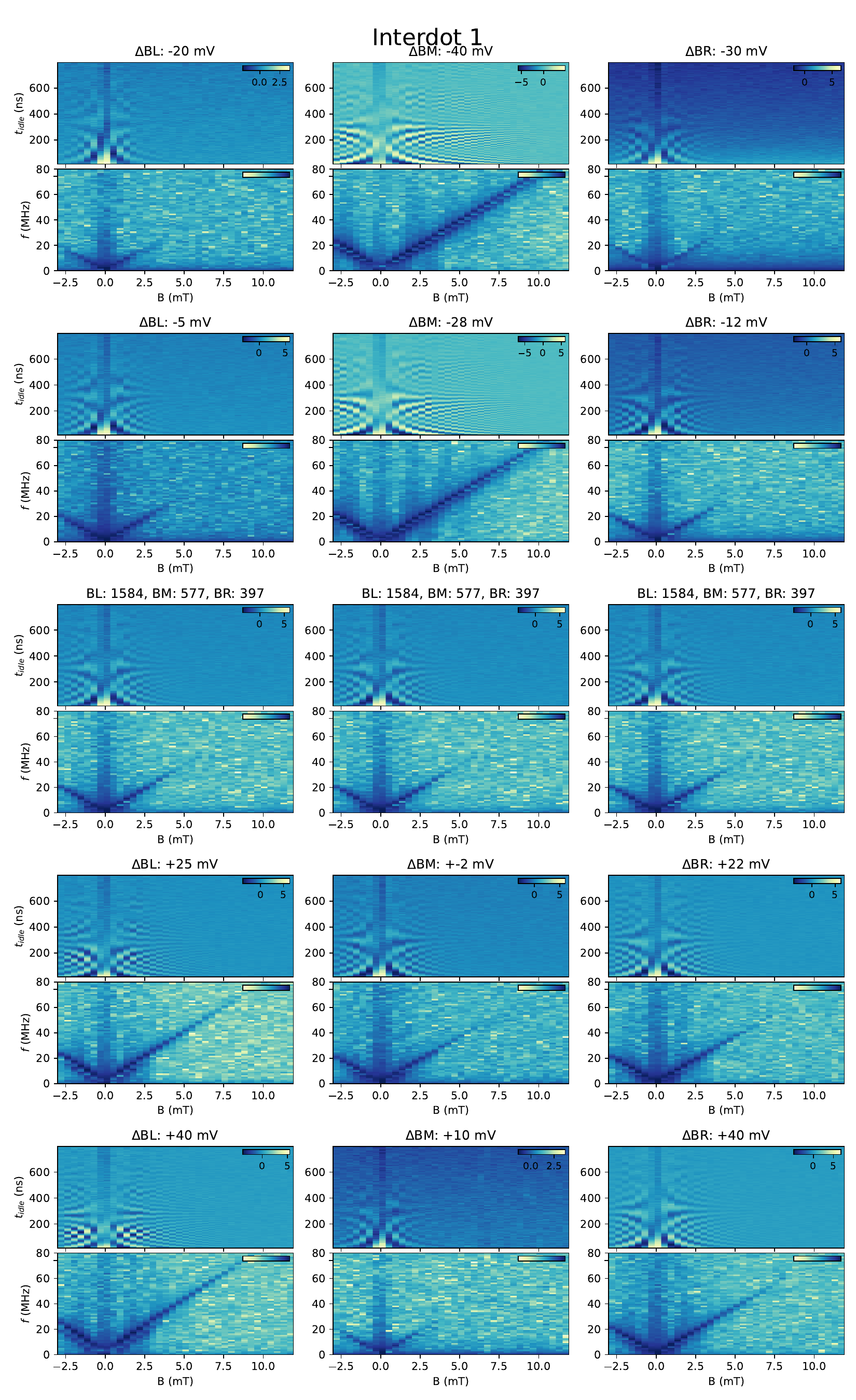}
\end{figure}

\begin{figure}
\includegraphics[width=0.8\linewidth]{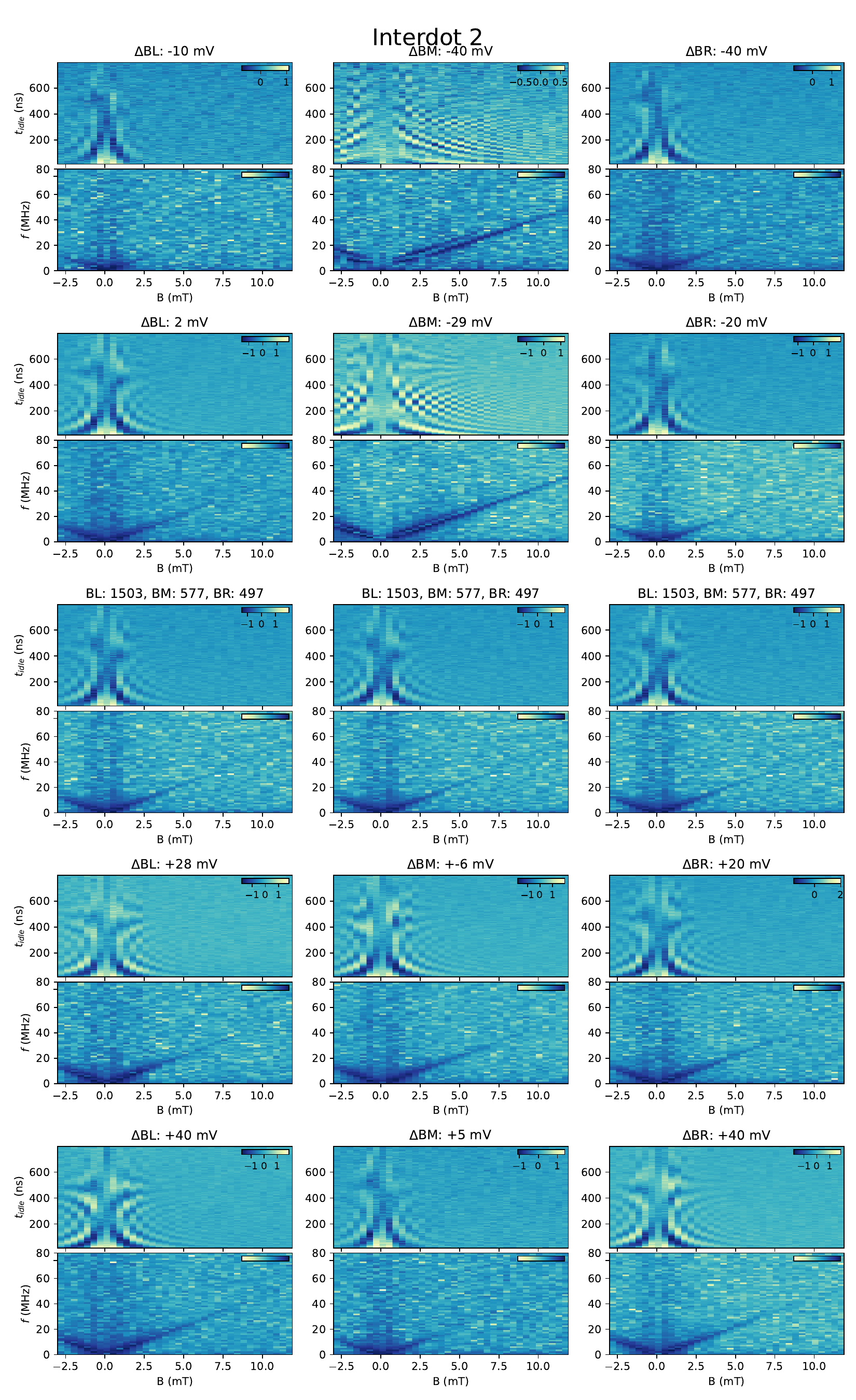}
\end{figure}

\begin{figure}
\includegraphics[width=0.8\linewidth]{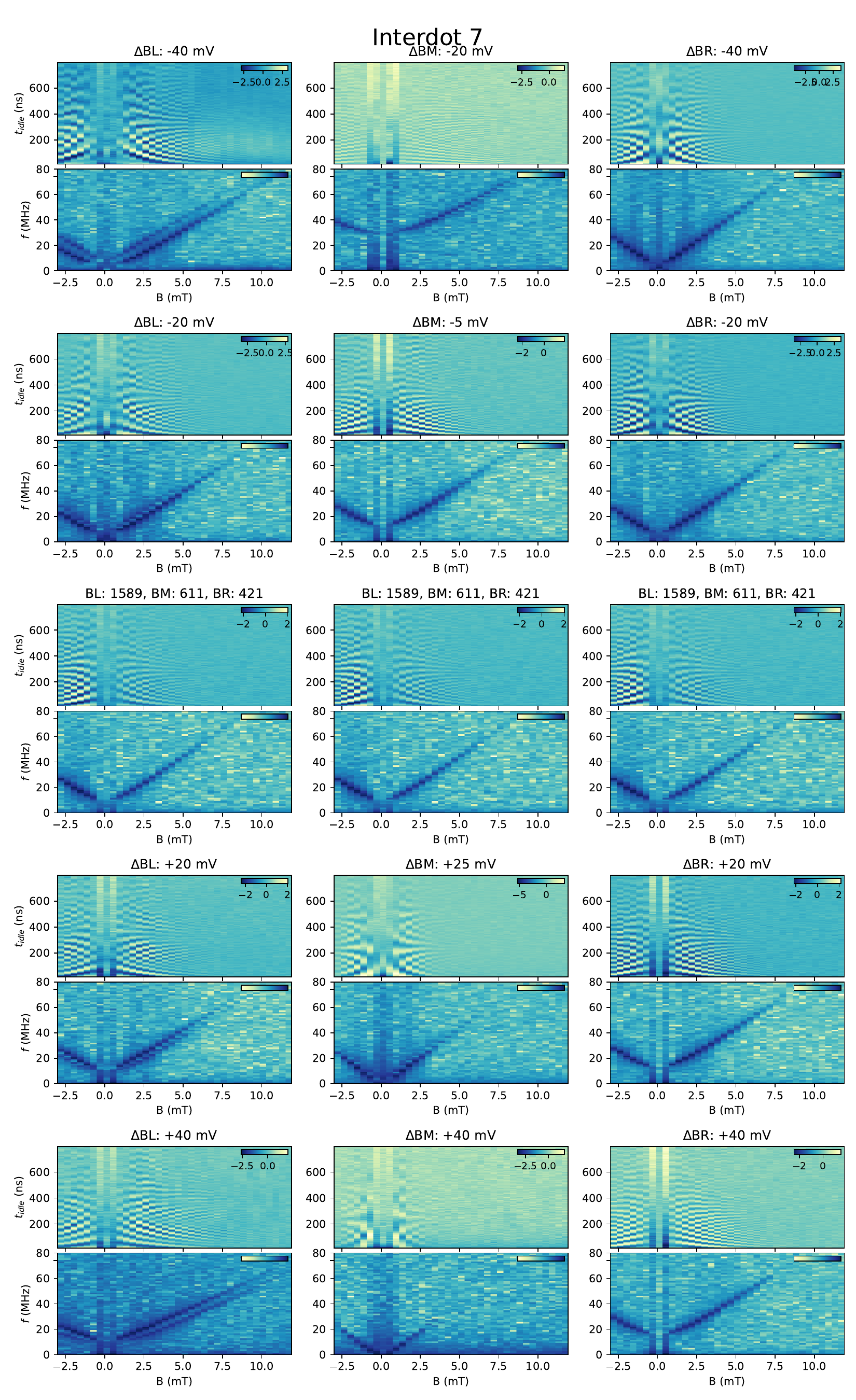}  
\end{figure}

\begin{figure}
\includegraphics[width=0.8\linewidth]{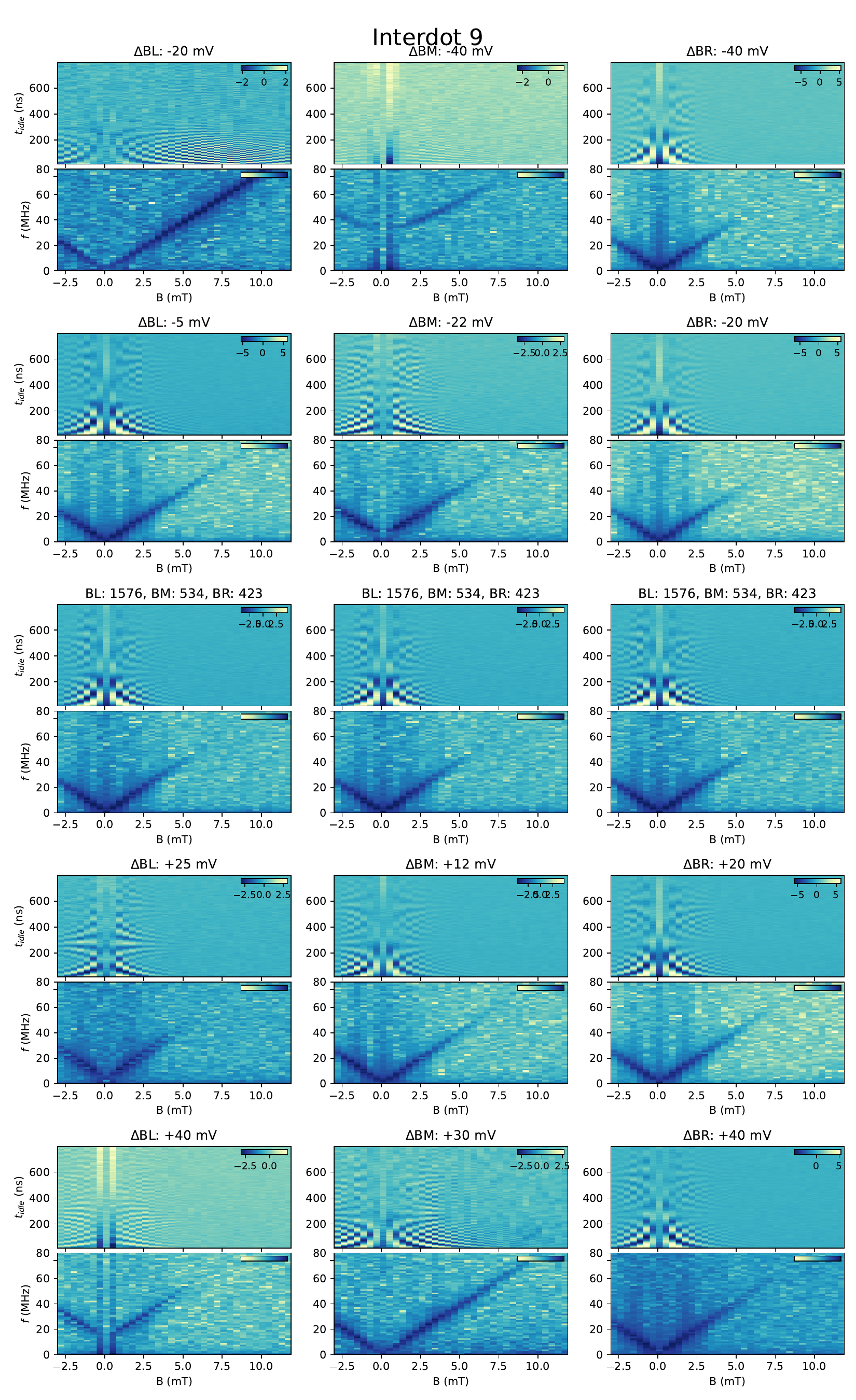}  
\end{figure}

\begin{figure}
\includegraphics[width=0.8\linewidth]{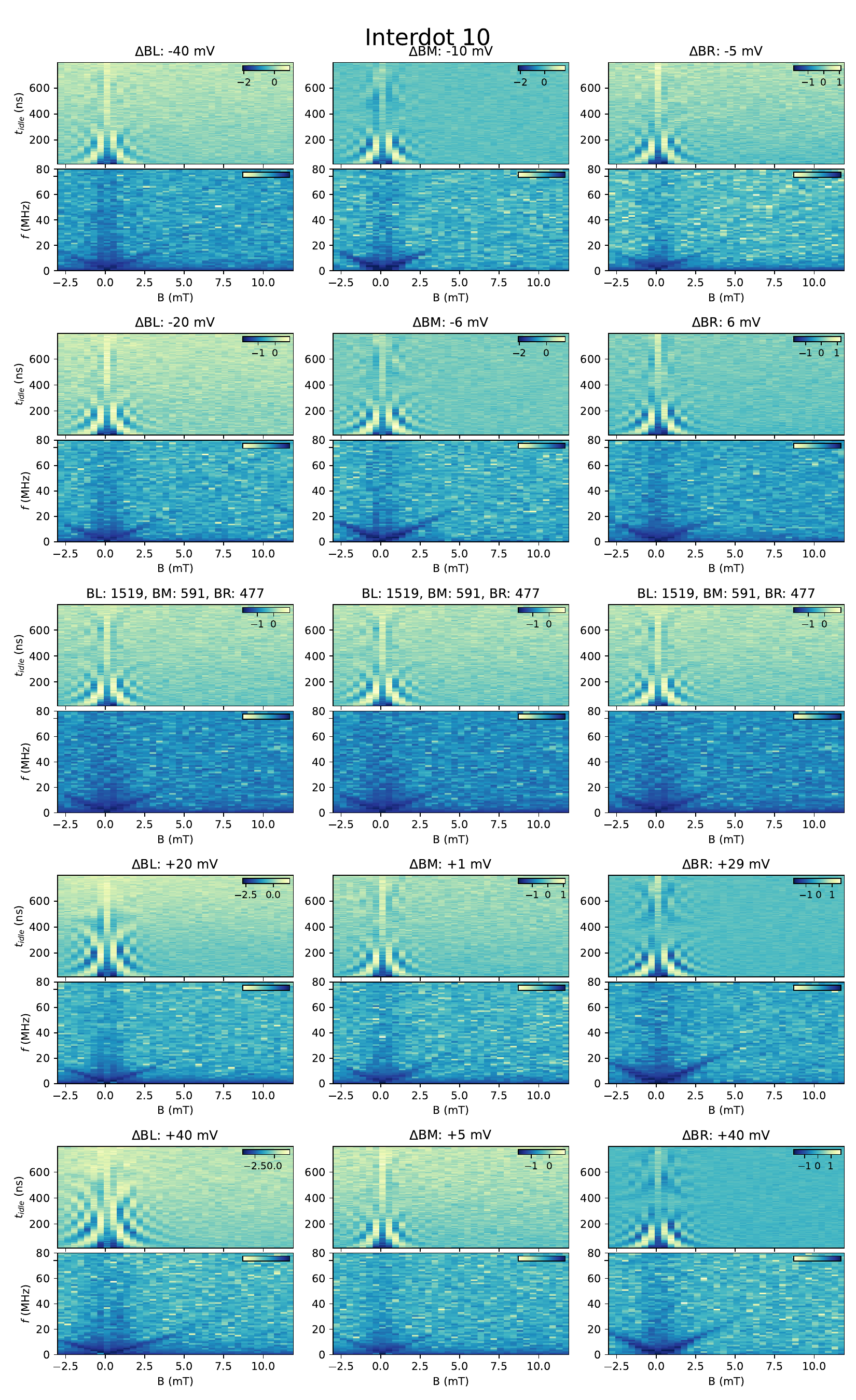}
\end{figure}

\begin{figure}
\includegraphics[width=0.8\linewidth]{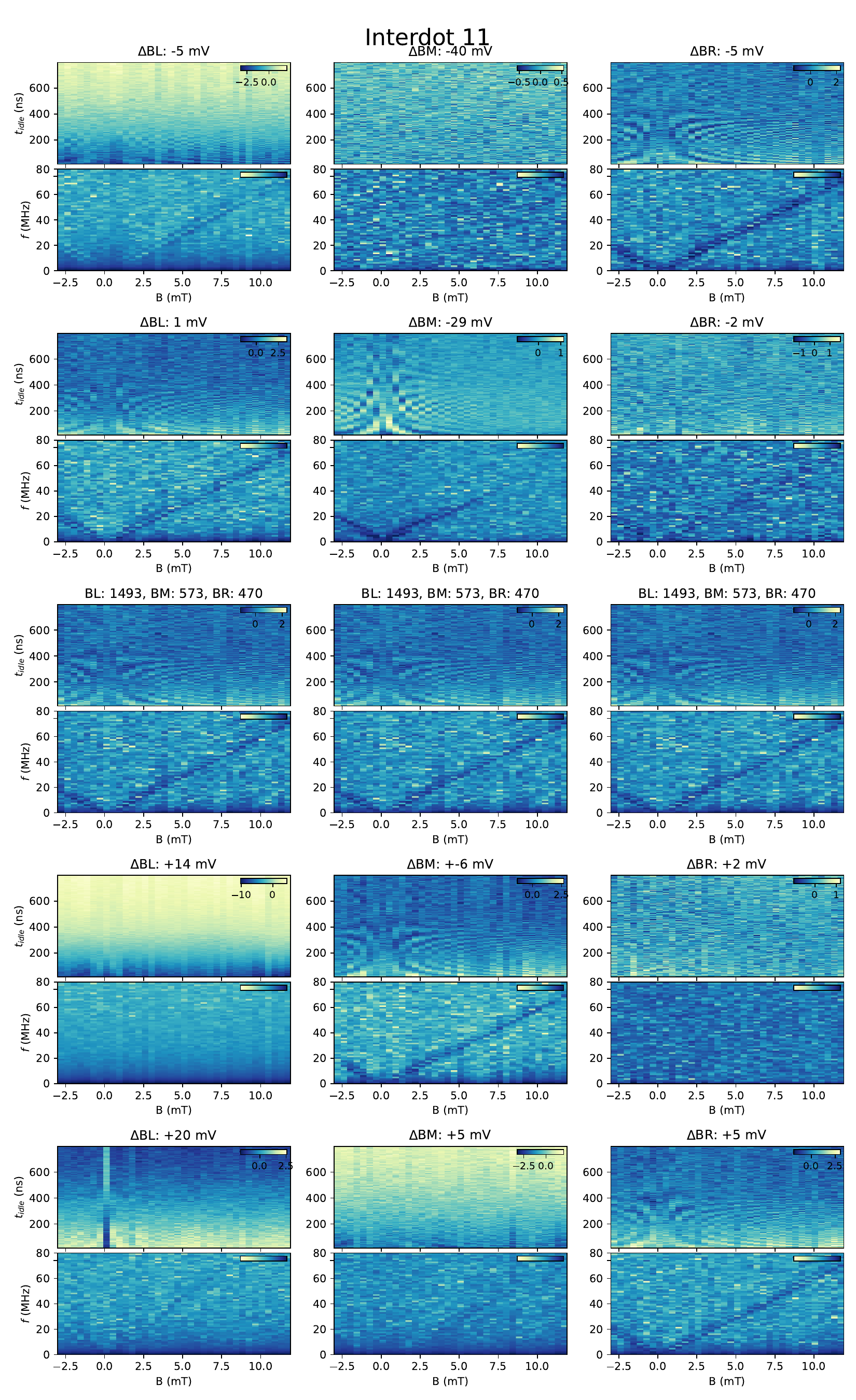}
\end{figure}

\begin{figure}
\includegraphics[width=0.8\linewidth]{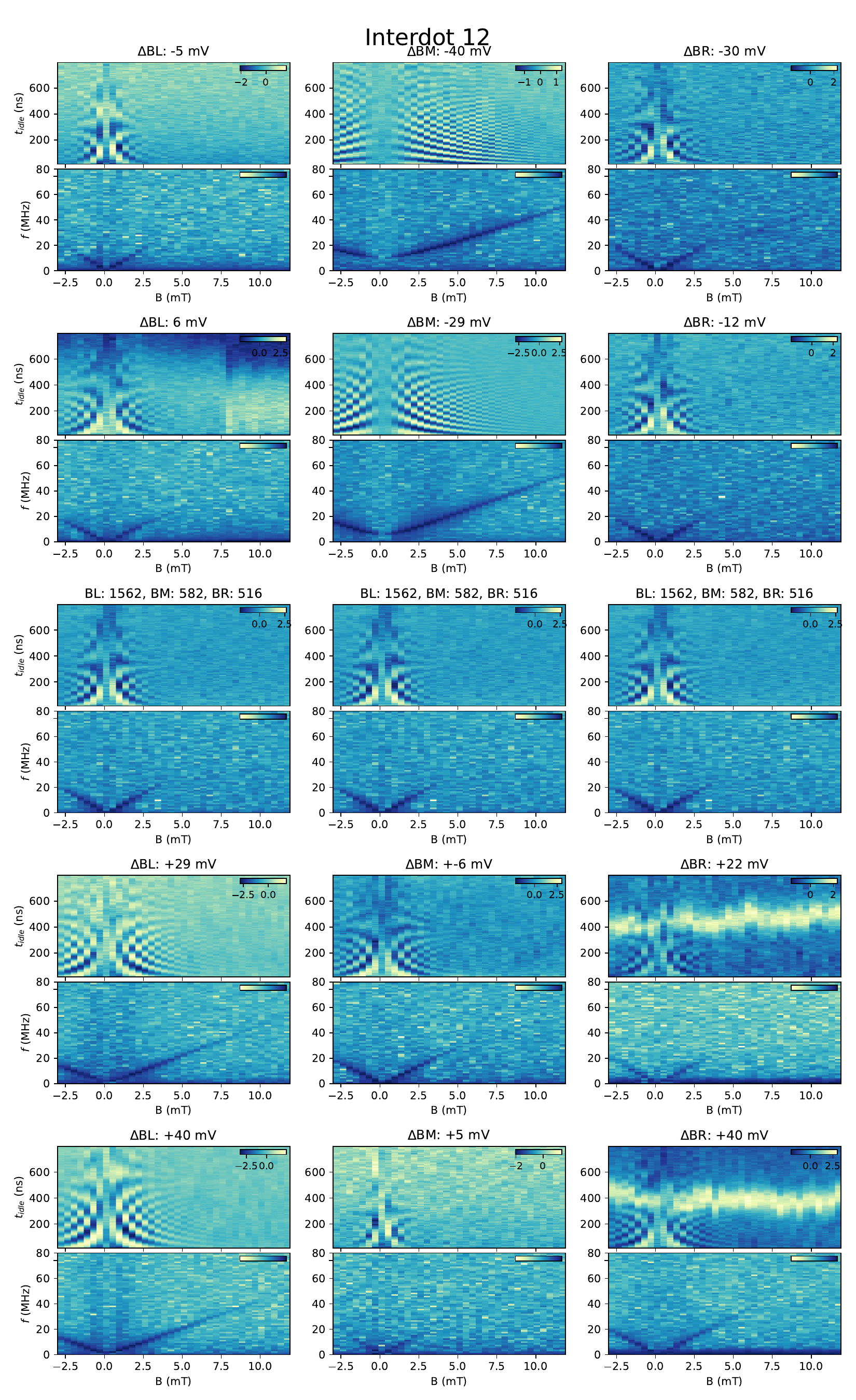}
\end{figure}

\begin{figure}
\includegraphics[width=0.8\linewidth]{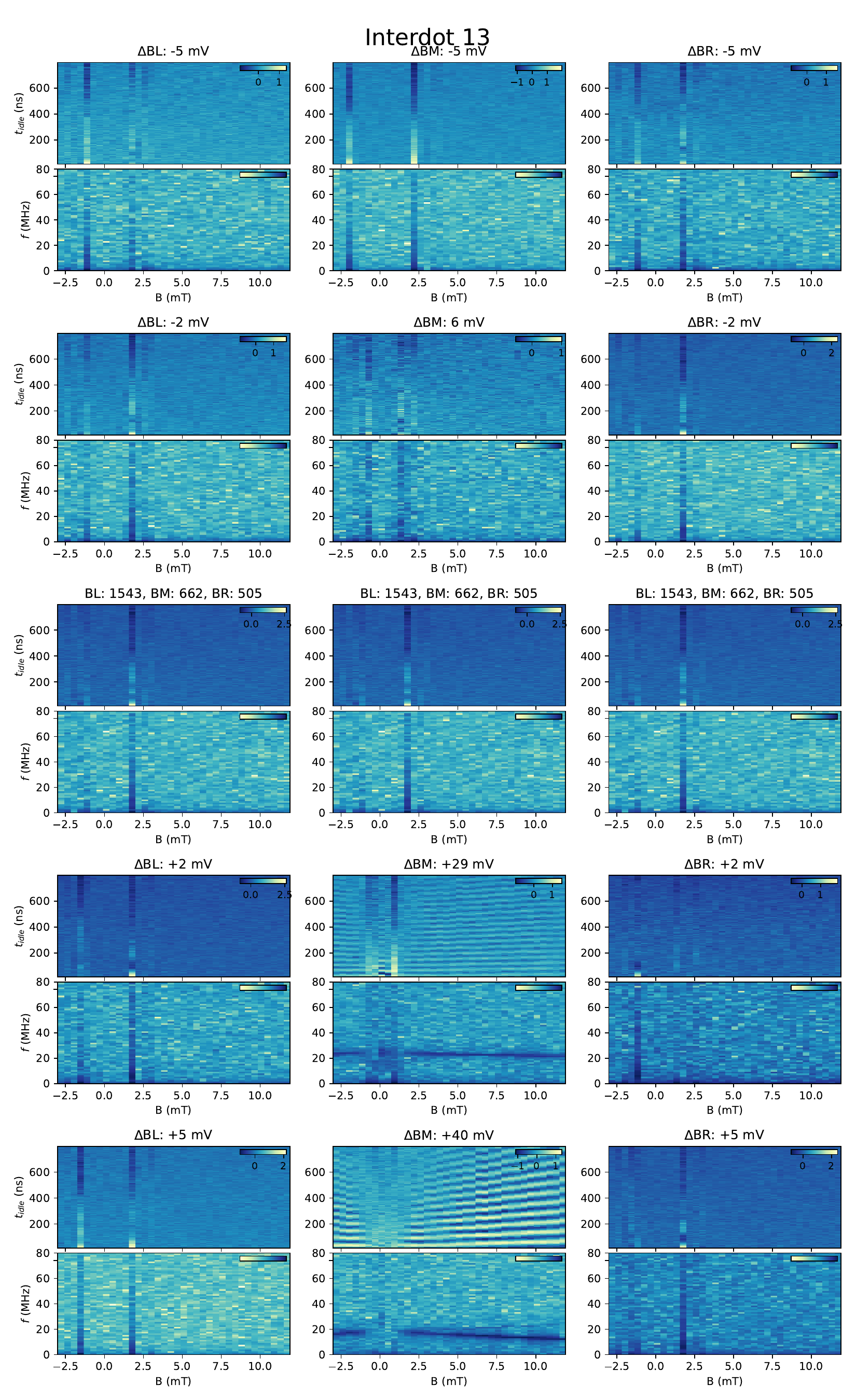}
\end{figure}

\begin{figure}
\includegraphics[width=0.8\linewidth]{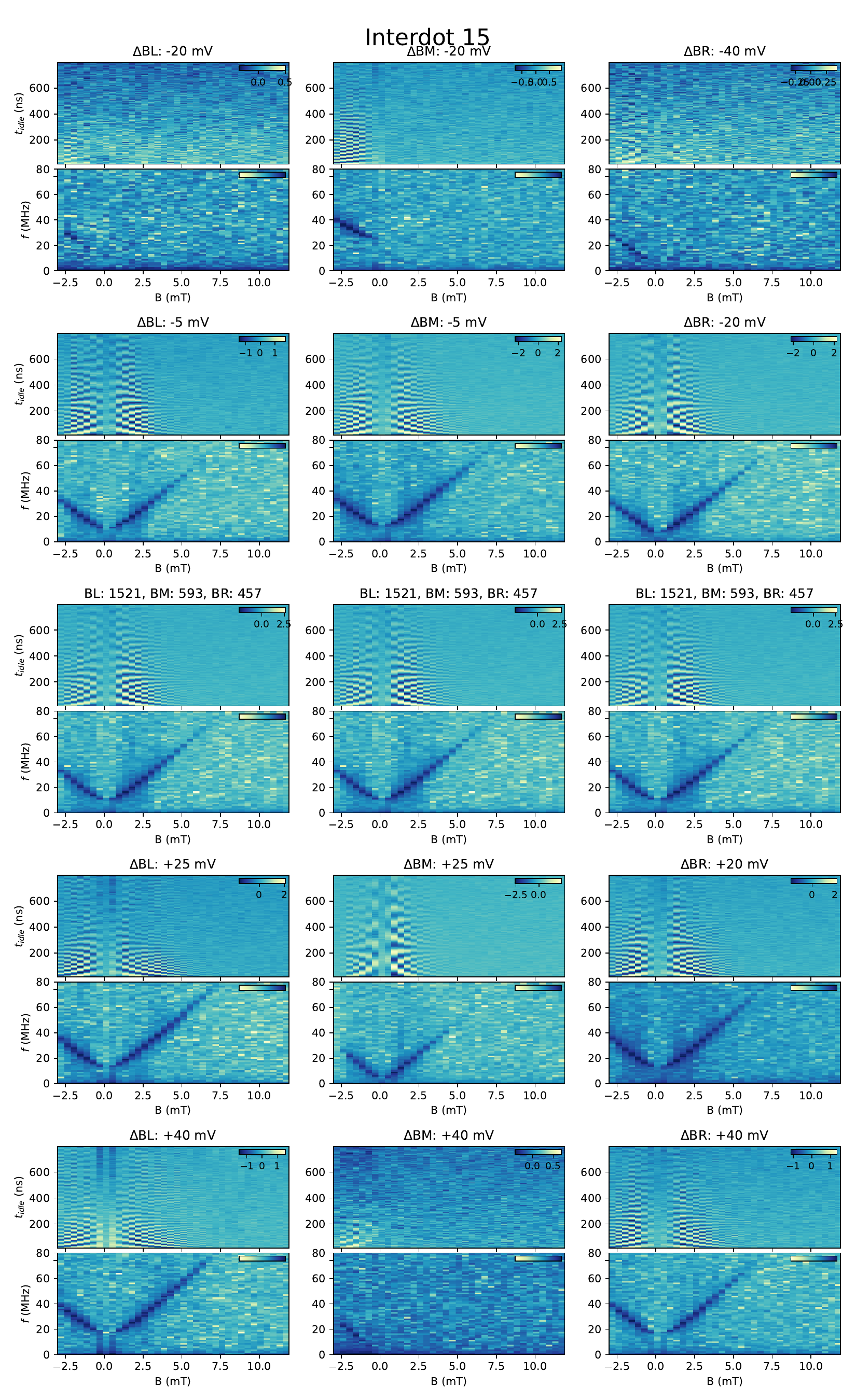} 
\end{figure}

\begin{figure}
\includegraphics[width=0.8\linewidth]{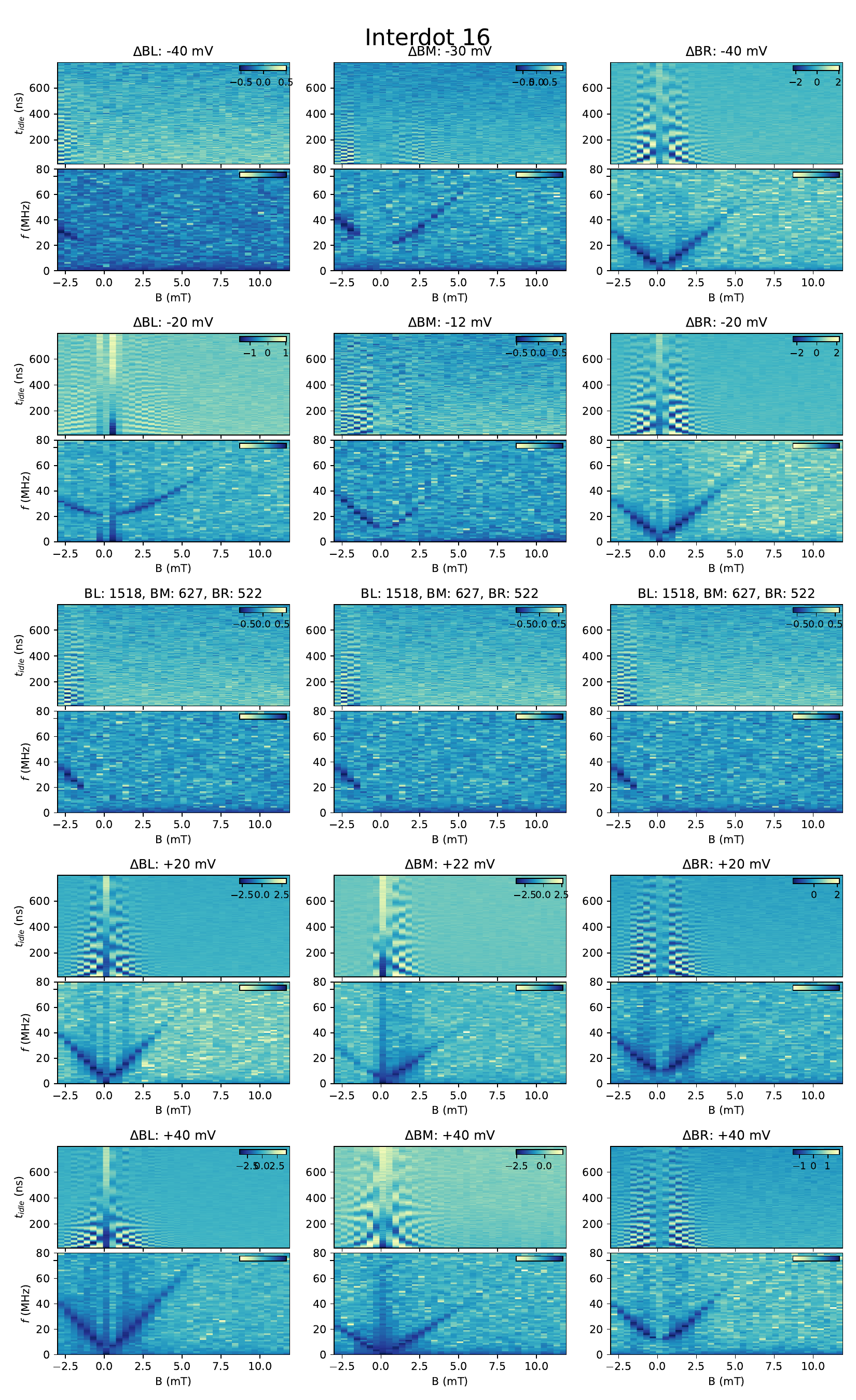}
\end{figure}

\begin{figure}
\includegraphics[width=0.8\linewidth]{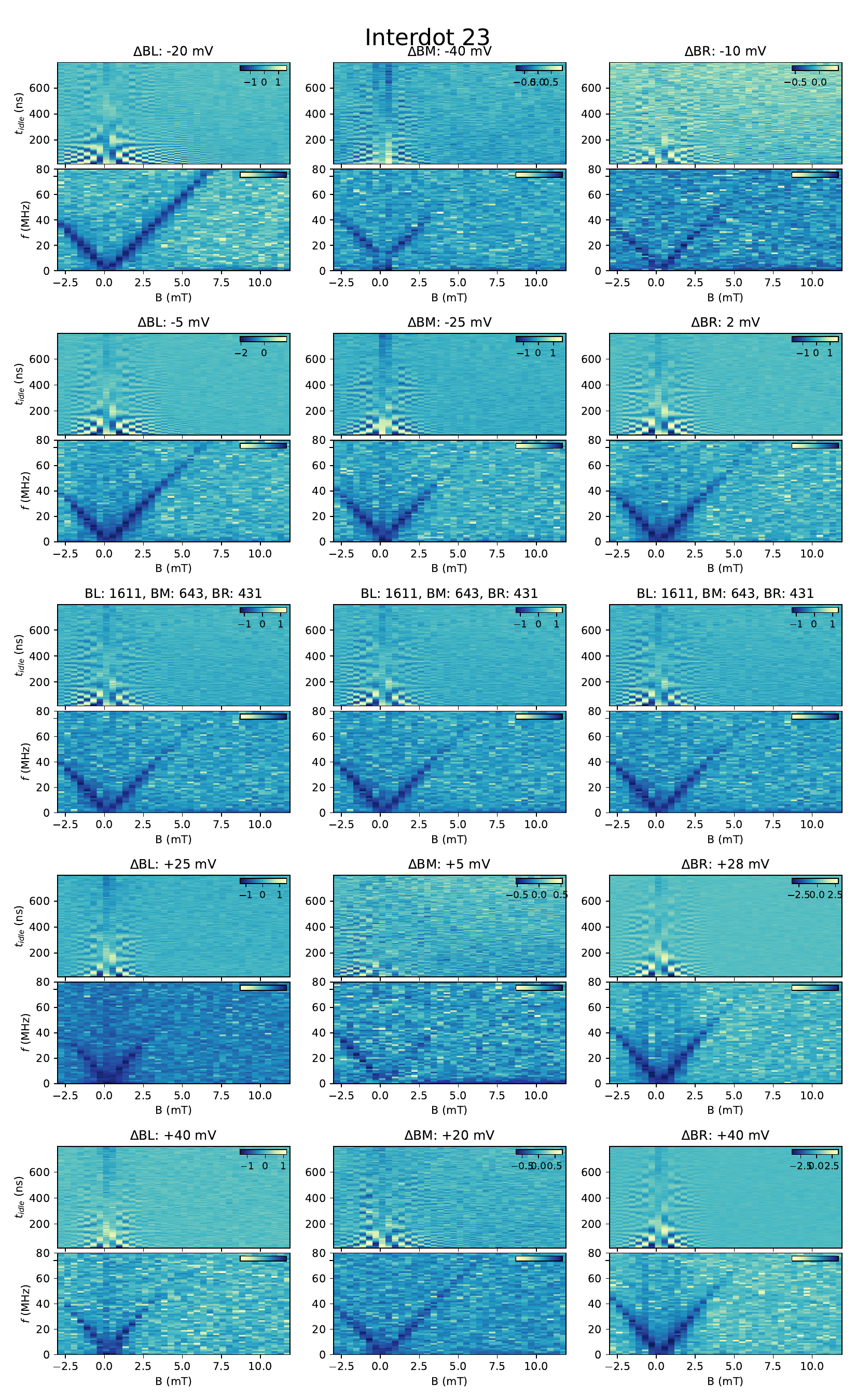}
\end{figure}

\begin{figure}
\includegraphics[width=0.8\linewidth]{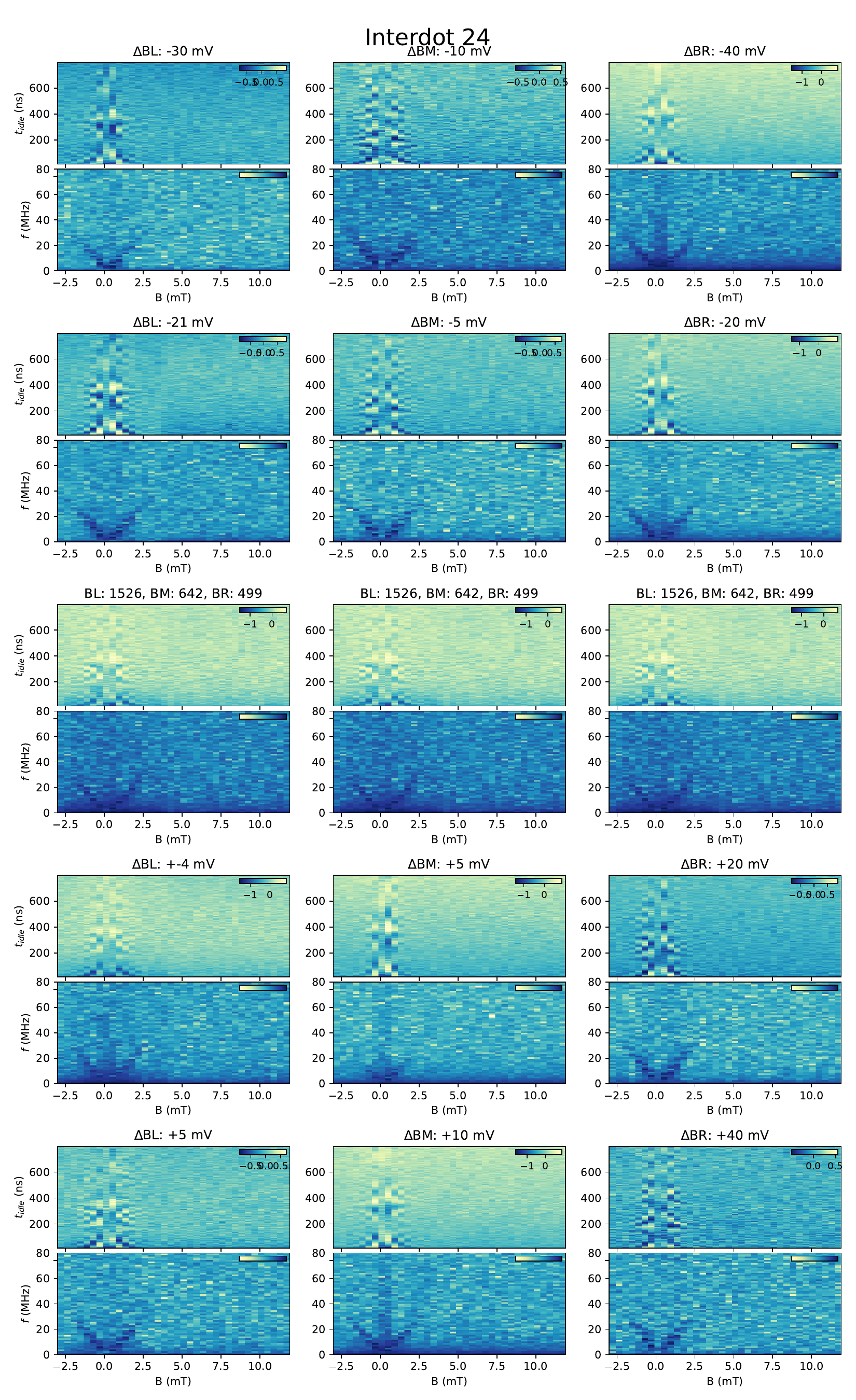}
\end{figure}

\newpage

\subsection{\label{sup:spurious_lines}Spurious Transition Lines}

\begin{figure}[ht]
\includegraphics[width=0.35\linewidth]{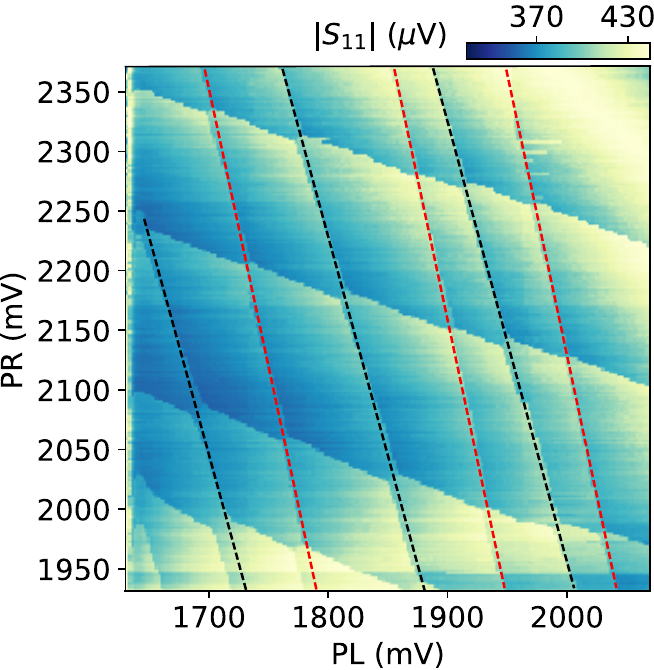}
\caption{\label{fig:fig_S6}Charge stability diagram with vertical transition lines traced in black and red. Traces of the same colour share the same slope. The differing slope of red and black traces is a strong indicator for the presence of a spurious dot.}
\end{figure}

In the main text, we propose spurious dot formation as the most likely cause for PSB emerging at interdot transitions that break a conventional ``checkerboard'' pattern. Spurious dots can be identified by additional sets of transition lines that differ in slope or separation from the underlying double quantum dot charge stability diagram. In Fig.~\ref{fig:fig_S6}, we provide a measurement taken during the initial tune-up of our device where both of these features are apparent.

\end{document}